\documentclass[prx, twocolumn, amsmath, amssymb, notitlepage, longbibliography, showpacs]{revtex4-1}

\usepackage{dsfont}
\usepackage[english]{babel}
\usepackage{graphicx}
\usepackage[usenames, dvipsnames]{color}
\usepackage{rotating}
\usepackage{hyperref}
\usepackage{enumerate}

\graphicspath{ {./Pictures/} }

\newcommand{\bra}[1]{\ensuremath{\left\langle #1\r|}}
\newcommand{\ket}[1]{\ensuremath{\left|#1\r\rangle}}

\newcommand{\mean}[1]{\ensuremath{\left\langle #1\r\rangle}}

\newcommand{\cc}{^{\ast}}							
\newcommand{\hc}{^{\dagger}}							

\newcommand{\op}[1]{\hat{#1}}							
\newcommand{\ee}{\mathrm{e}}						
\newcommand{\ii}{i}			             					
\renewcommand{\H}[0]{H}  							
\renewcommand{\L}[0]{\mathcal{L}}  						

\newcommand{\s}{k}  

\newcommand{\comm}[2]{\left[ #1, #2 \right]} 				
\newcommand{\nn}{\nonumber}							
\newcommand{\abs}[1]{\ensuremath{ \left| #1 \right| }}		
\newcommand{\abss}[1]{\ensuremath{ \left| #1 \right|^{2} }}	
\newcommand{\diss}[1]{\mathcal{D}[ #1 ]}					
\renewcommand{\l}[0]{\left}
\renewcommand{\r}[0]{\right}

\newcommand{\Tr}{\text{Tr}}

\def\ct{\ensuremath{\cos^{2}\theta}}						
\def\st{\ensuremath{\sin^{2}\theta}}						

\newcommand{\eq}[1]{\eqref{#1}}

\newcommand{\eqn}[1]{Eq.~\eq{#1}}
\newcommand{\fig}[1]{Fig.~\ref{#1}}

\begin{document}

\title{	
	Keldysh meets Lindblad: Correlated Gain and Loss in Higher-Order Perturbation Theory 
	}

\author{Clemens M\"uller}
\email{c.muller2@uq.edu.au}
\affiliation{ARC Centre of Excellence for Engineered Quantum Systems, School of Mathematics and Physics, The University of Queensland, Saint Lucia, Queensland 4072, Australia}

\author{Thomas M. Stace}
\email{stace@physics.uq.edu.au}
\affiliation{ARC Centre of Excellence for Engineered Quantum Systems, School of Mathematics and Physics, The University of Queensland, Saint Lucia, Queensland 4072, Australia}

\date{\today}

\begin{abstract}
	Motivated by correlated decay processes producing gain, loss and lasing in driven semiconductor quantum-dots~\cite{Liu:PRL:2014, Liu:S:2015, Gullans:PRL:2015}, 
	we develop a theoretical technique using Keldysh diagrammatic perturbation theory to derive a Lindblad master equation that goes beyond the usual second order perturbation theory.
	We demonstrate the method on the driven dissipative Rabi model, including terms up to fourth order in the interaction between the qubit and both the resonator and environment. 
	This results in a large class of Lindblad dissipators and associated rates which go beyond the terms that have previously been proposed to describe similar systems.  
	All of the additional terms contribute to the system behaviour at the same order of perturbation theory. 
	We then apply these results to analyse the phonon-assisted steady-state gain of a microwave field driving a double quantum-dot in a resonator. 
	We show that resonator gain and loss are substantially affected by dephasing-assisted dissipative processes in the quantum-dot system. These additional processes, which go beyond recently proposed polaronic theories, are in good quantitative agreement with experimental observations.
\end{abstract}


\maketitle

\section{Introduction}

	Microwave-driven double-quantum dots (DQD) have demonstrated a rich variety of quantum phenomena, including population inversion~\cite{pet04, Stace:PRL:2005, Petta:S:2005,Stace:PRL:2013, Colless:NC:2014}, gain~\cite{Liu:PRL:2014, Kulkarni:PRB:2014, Stockklauser:PRL:2015}, 
	masing~\cite{Marthaler:PRL:2011, Jin:RPP:2012, Liu:S:2015, Liu:PRA:2015, Marthaler:PRB:2015, Karlewski:PRB:2016} and Sysiphus thermalization~\cite{Gullans:PRL:2016}. 
	These processes are well understood in quantum optical systems, however mesoscopic electrostatically-defined quantum dots exhibits additional complexity not typically seen in their optical counterparts, arising from coupling to the phonon environment.  
	
	A notable experimental example of this, which motivates our work, is an electronically open, DQD system coupled to a driven resonator, as illustrated in~\fig{fig:dissipation}(a),(b)~\cite{Liu:PRL:2014}.  
	Substantial gain in the resonator field  was observed when the DQD is blue-detuned with respect to the resonator, and capacitively biased to induce substantial population inversion.  
	The observed gain is attributed to correlated emission of a resonator photon and a phonon into the semiconductor medium in which the DQD system is defined.  
	This process ensures conservation of energy, since the phonon carries the energy difference, $\hbar(\omega_q-\omega_r)$, between the energy of the qubit  and the energy of the resonator phonon. 
	
	In some experimental regimes of~\cite{Liu:PRL:2014}, the observed gain is well described by a theory  based on a canonical transformation to a polaron frame~\cite{Gullans:PRL:2015}.  
	In this frame, conventional quantum optics techniques and approximations (Born-Markov, secular etc) are used to derive dissipative Lindblad superoperators that are quadratic in both the qubit-phonon bath coupling strength, $\beta_j$, and  in the qubit-resonator coupling strength, $g$.    
	However the same theory fails to describe substantial loss (sub-unity gain) in other experimental regimes, which strongly suggests that there are additional dissipative processes that are not captured in the polaron frame.
	
	One problem with relying on a canonical transformation as the basis for a perturbative expansion is that it is tailored to a specific frame, which emphasises some processes over others. 
	It is therefore not guaranteed to find all dissipative processes that occur at a given order in perturbation theory.
	
	In this paper we present a systematic approach to derive a Lindblad-type master equation at higher  perturbative orders, which  contains all relevant correlated dissipative processes at a given  order. 
	We use the Keldysh real-time diagrammatic technique to calculate the perturbative series, and explicitly evaluate all fourth order diagrams that generate correlated dissipation.  In doing so, we make explicit the link between the Keldysh self-energy and the Lindblad superoperators.  
	We then apply this technique to the experimental situation described above, showing that the additional terms we find quantitatively explain anomalous gain and loss profiles in driven DQD-resonator systems.
	
	This paper is structured as follows: 
	
	Section \ref{Sec:System} introduces the model system that motivates this work, namely a single artificial two-level system ({qubit}) coupled to a resonator as well as a dissipative environment \cite{Liu:PRL:2014, Liu:S:2015}.

	Section \ref{Sec:Keldysh} derives the Keldysh self-energy superoperator in terms of irreducible Keldysh diagrams, 
	and decomposing it into a sum of Lindblad superoperators to form the Keldysh-Lindblad master equation. 
	The complete set of fourth-order Keldysh-Lindblad dissipators, presented in \eqn{eq:4thAll}, is the central theory result of this paper. 

	Section \ref{Sec:Steady} introduces resonator driving, and solves the steady-state Keldysh-Lindblad master equation for a driven, DQD-resonator system in a mean-field approximation. 

	Section \ref{sec:Results} shows that the additional Lindblad dissipators that arise in the Keldysh analysis of the Dyson series quantitatively account 
	for the  of gain and loss in a driven DQD-resonator system, across all experimental regimes of Ref.~\onlinecite{Liu:PRL:2014}.	

	We finish in Section \ref{Sec:Discussion} with a discussion of our results, specifically the correlated decay rates we obtain, that describe the dynamics of qubit and resonator.
	
	The Keldysh and Lindblad formalisms are each well used in their respective research communities, 
	however the link between them is usually not made explicit.
	Equations of Lindblad type are the generators of completely-positive, trace-preserving dynamics~\cite{Lindblad:CMP:1976} and a wide variety of techniques have been developed to analyse the dynamics of open quantum systems under the actions of Lindblad master equations, 
	including the input-output~\cite{QuantumOptics} and cascaded `SLH' formalisms~\cite{baragiola2012n}, quantum trajectories and measurement~\cite{QuantumMeasurement, stace2004parity, barrett2006continuous, kolli2006all, barrett2006two}, and phase-space methods~\cite{QuantumNoise}.	
	There are widely used, but somewhat heuristic techniques in quantum optics to derive Lindblad master equations at second order in the bath coupling~\cite{QuantumNoise}.  
	Going beyond this approximation requires a well formulated perturbative approach.
	
	The Keldysh diagrammatic technique~\cite{Keldysh:JETP:1965} is widely used in condensed matter physics as a tool to calculate Greens functions of non-equilibrium systems~\cite{Schoeller:PRB:1994}.
	It has been previously applied with great success for calculating the properties of mesoscopic quantum systems~\cite{Makhlin:JCP:2004, Torre:PRA:2013, Schad:PRB:2014, Marthaler:PRB:2015, Marthaler:PRB:2016}. 
	However in these previous cases the resulting master equations are not typically expressed explicitly in the Lindblad form.  
	Furthermore, these derivation are either limited to second order or include only a small class of diagrams for which closed form expressions are available. 
	
	In contrast here we develop a systematic approach to account for \emph{all} Keldysh diagrams at a particular order, with which to derive a master equation that is explicitly in Lindblad form. 
			
	This paper therefore has the subsidiary objective of making explicit the connection between the Keldysh and Lindblad approaches to open quantum systems.   
	As such, we have provided extensive appendices giving the technical details of our calculations. 
	As we show, Keldysh perturbation theory is a powerful approach to deriving consistent Lindblad dissipators at arbitrary order. It is especially well suited for the treatment of  open quantum systems 
	in which correlated dissipative process, which go beyond the usual second order in perturbation theory, are significant.
	As such, in addition to the specific problem of correlated gain and loss in a driven DQD-resonator, we foresee immediate applications of this technique in numerous context, like co-tunneling in open quantum-dots~\cite{Aghassi:APL:2008}, 
	charge transport through chains of Josephson junctions~\cite{Cedergren:PRB:2015} or the calculation of correlated decay rates in multi-level circuit-QED systems~\cite{Slichter:PRL:2012}.

\section{Driven dissipative Rabi Model\label{Sec:System}}

	\begin{figure}
		\begin{center}
		\includegraphics[width=\columnwidth]{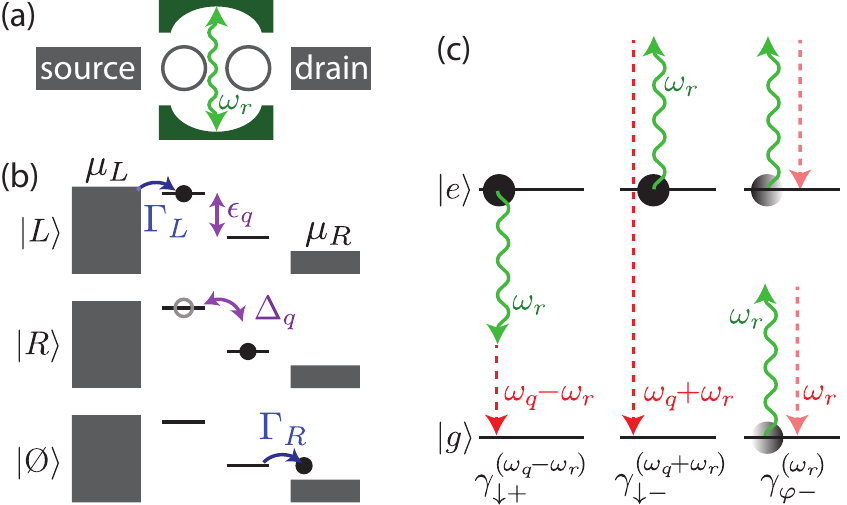}
		\caption{(Color online) (a) Schematic showing  a double quantum dot (DQD) coupled to a cavity resonator mode, and to source/drain leads. 
			(b) An electron  (black dot) tunnels from the source lead to the DQD state $\ket{L}$, which couples to the DQD state $\ket{R}$ with matrix element $\Delta_q$, and then tunnels to the drain, leaving the DQD in the empty state $\ket{\O}$. 
	 		Also shown are inter-dot bias, $\epsilon_q$, and coupling rates $\Gamma_{L,R}$ to metallic leads, with chemical potentials $\mu_{L,R}$.   (c) Dissipative processes due to phonon emission (dashed  arrows) responsible for rates in~\eqn{eq:CorrelatedRates}.  
			Each processes is correlated with resonator photon  creation (downward  wiggly arrows) or annihilation (upward wiggly arrows); $\gamma_{\downarrow\pm}^{(\omega_q\pm\omega_r)}$ correspond to DQD relaxation from $\ket{e}$ to $\ket{g}$, 
			whilst  $\gamma_{\varphi-}^{(\omega_r)}$ corresponds to DQD dephasing, leaving the  populations of $\ket{e}$ and $\ket{g}$ unchanged.
		} 
		\label{fig:dissipation}
		\end{center}
	\end{figure}

	The Rabi Hamiltonian describe a single two-level system or qubit coupled to a harmonic mode. 
	It is one of the main workhorses of modern quantum physics and describes such diverse situations as natural atoms interacting with visible light~\cite{QuantumOptics}, superconducting artificial atoms coupled to microwave resonators~\cite{Blais:PRA:2004} or 
	semiconductor double quantum-dots coupled to superconducting resonators~\cite{Petersson:N:2012}.
	Here we additionally consider the situation where the two-level system is coupled to an environment, inducing dissipation in the dynamics.
	Here we adopt the language of a semiconducting open double quantum-dot, but the technique presented here is applicable to many other situations that are described by the Rabi-model.
	
	The Hamiltonian of a DQD coupled to a superconducting resonator, expressed in the DQD 
	position basis $\{{\ket{R}},{\ket{L}}\}$, is
	\mbox{$H=H_S+H_B+H_I$}, with
	\begin{eqnarray}
		H_{S} &=& \omega_{r} a\hc a -  \epsilon_{q} \sigma_{z}^{(p)}/2 +  \Delta_{q} \sigma_{x}^{(p)}/2, \\
		H_B&=&{\sum}_{\textrm{modes }j}\omega_{j} b_{j}\hc b_{j},\\
		H_I&=&g \sigma_{z}^{(p)} ( a + a\hc )/2+ \sigma_{z}^{(p)} X /2,
	\end{eqnarray}
	where $\sigma_{z}^{(p)} ={ \ket L}{ \bra L} - {\ket R}{ \bra R}$ is the dipole operator of the DQD (also referred to as a qubit), \mbox{$\sigma_x^{(p)}={\ket R }{\bra L} + {\ket L}{ \bra R}$} induces transitions between its two charge states and $a$ is an annihilation operator for microwave photons in the cavity mode at frequency $\omega_{r}$.
	Here the position states indicate the presence of a single charge either on the right or on the left dot.
	The DQD has an asymmetry energy between its position states of $\epsilon_{q}$ and they are tunnel-coupled with strength $\Delta_{q}$.
	The coupling between DQD and resonator is described by the dipolar interaction, with coupling strength $g$. Here, as in the rest of the paper we use the convention $\hbar = 1$.
	
	The DQD couples to a bath of environmental modes, $b_j$, with eigenfrequencies $\omega_{j}$ via the system-bath coupling term 
	\begin{equation}
		X=\sum_{j}\beta_{j} ( b_{j} + b_{j}\hc ) .
	\end{equation}
	For descriptive purposes we will assume the environment to be a bath of phonons as this is generally the dominant source of dissipation for charge-based quantum-dots~\cite{Stace:PRL:2005}. 
	In general, the DQD environment could include other electronic fluctuations, such as charged intrinsic two-level defects~\cite{Simmonds:PRL:2004, Mueller:PRB:2015} or charge traps on surfaces~\cite{Choi:PRL:2009}.
	
	In addition to the coupling of the DQD to a bath, the microwave resonator will also be coupled to a dissipative environment leading to loss and spurious excitation of resonator photons.
	Also, in the experiments motivating this work, the quantum-dot was subject to an external bias, leading to charge transport through the DQD as illustrated in Fig.~\ref{fig:dissipation}(b).
	Both these processes will be included in the master equation and we describe them in sections~\ref{Sec:ResDecay} and~\ref{Sec:QDBias}.
	
	We write the interaction Hamiltonian $\H_{\text I}$ as a sum of the intra-system interaction $\H_{\text{I,S}}$ (i.e. between the qubit and resonator), and the interaction between the system and phonon environment $\H_{\text{I,B}}$,
	\begin{align}
		\H_{\text I} ={}& \H_{\text{I,S}} + \H_{\text{I,B}},\nn\\
			={}& \frac12 g \left( \cos\theta\: \sigma_{z} + \sin\theta\: \sigma_{x}  \right) \left( a+ a\hc \right) \nn\\
			& + \frac12 \left( \cos\theta\: \sigma_{z} + \sin\theta\: \sigma_{x} \right) \sum_{j}\beta_{j} \left( b_{j} + b_{j}\hc \right) \,,
		\label{eq:HI}
	\end{align}
	where we have decomposed $\sigma_{z}^{(p)}= \cos\theta\: \sigma_{z} + \sin\theta\: \sigma_{x}$ into the eigenoperators, $\sigma_{x}={\ket{e}}{\bra{g}}+{\ket{g}}{\bra{e}}$ and \mbox{$\sigma_{z}={\ket{e}}{\bra{e}}-{\ket{g}}{\bra{g}}$} of the bare system-bath Hamiltonian
	\begin{align}
		H_0&=H_S+H_B,\\
			&= \omega_{r} a\hc a - \omega_{q} \sigma_{z}/2 + \sum_{j}\omega_{j} b_{j}\hc b_{j},
	\end{align}
	where the bare qubit level splitting is $\omega_{q} = ({\epsilon_{q}^{2} + \Delta_{q}^{2}})^{1/2}$ and the qubit mixing angle $\tan{\theta} = \Delta_{q} / \epsilon_{q}$, with the convention $0\leq\theta<\pi$. 
	The qubit eigenstates are given by 
	\begin{align}
		\ket{g} &= \cos{\frac{\theta}{2}} \ket{L} - \sin{\frac{\theta}{2}} \ket{R}, \nn\\
		\ket{e} &= \sin{\frac{\theta}{2}} \ket{L} + \cos{\frac{\theta}{2}} \ket{R}.\nn
	\end{align}		
		
	In the following we  perform perturbation theory in the interaction Hamiltonian $\H_{\text I}$.
	We are thus assuming that the qubit resonator coupling is weak compared to their detuning, $g<|\omega_{q} -\omega_{r}|$. 
	Additionally we make the usual weak coupling approximation between the system and the phonon environment.
	The standard situation in experiments is $\omega > g \gg \beta_{j}$, where $\omega$ represents all relevant system and bath frequencies.

\section{Lindblad Super-Operators from Keldysh Self-Energy\label{Sec:Keldysh}}

	The objective of this paper is to derive a Lindblad master equation 
	\begin{align}
		\dot{ \rho} &= \sum_{m} \mathcal L_{m} \rho,\nn\\
			&=\mathcal L_{2} \rho+\mathcal L_{4} \rho+\mathcal L_{\textrm{res}} \rho+ \L_{\text{leads}}\rho\,,	
		\label{eqn:ME}	
	\end{align}
	for the evolution of the (slowly varying) system density matrix, $\rho$, where the index $m$ labels orders of perturbation theory. 	
	In particular, $\mathcal L_{2}$ includes dispersive and dissipative terms at second order in $H_\textrm{I}$, $\mathcal L_{4}$ includes dissipative terms at fourth order in $H_\textrm{I}$, $\mathcal L_{\textrm{res}}$ accounts for the resonator decay 
	and $\L_{\text{leads}}\rho$ describes a transport bias across the DQD, as in Ref.~\onlinecite{Liu:PRL:2014}.  
	
	The superoperators $\mathcal L_m$ are composed of linear combination of coherent evolution processes, expressed as commutators with some effective Hamiltonian, and dissipative evolution, expressed in the form of Lindblad dissipators 
	\begin{align}
		\diss{\op o}\rho = \op o \rho \op o\hc -  (\op o\hc \op o \rho + \rho \op o\hc \op o )/2\,.
		\label{eqn:lind}
	\end{align}
	For the purposes of this paper, we will refer to any term of the form in \eqn{eqn:lind} as a Lindblad dissipator, though we note that at higher order, we find such terms with negative coefficients. 

	Keldysh perturbation theory provides a systematic way to evaluate self-energy terms in the Dyson equation, in order to derive the dynamical master equation for the system density matrix.  
	To determine the system evolution, we evaluate the Keldysh self-energy superoperator $\Sigma$, which is written as a perturbation series  in the interaction Hamiltonian.  
	Terms in this series are expressed in the graphical language of  Keldysh diagrams, in order to keep track of all relevant contributions and avoid double counting. 
	
	To begin the analysis, the Keldysh master equation for the density matrix in the Schr\"odinger picture, $\rho^{(S)}(t)$,  is given by
	\begin{align}
		\dot\rho^{(S)}(t) =&- \ii [H_0, \rho^{(S)}(t)]+ \int_{t_{0}}^{t} dt_{1}\: \rho^{(S)}(t_{1}) \Sigma (t_{1},t) ,
		\label{eq:KeldyshMEsp}
	\end{align}
	where the superoperator $\Sigma$ is the self-adjoint self-energy, which  acts on the state from the right.
	A derivation of~\eqn{eq:KeldyshMEsp} can be found in Appendix~\ref{App:Keldysh}.  In the interaction picture defined by $H_0$, this becomes
			\begin{align}
		\dot\rho^{(I)}(t) =& \int_{t_{0}}^{t} dt_{1}\: \rho^{(I)}(t_{1}) \Sigma (t_{1},t).
		\label{eq:KeldyshMEip}
	\end{align}
	Hereafter, we drop the interaction picture label ${}^{(I)}$.
	In what follows, we seek to express the self-energy superoperator on the RHS of~\eqn{eq:KeldyshMEip} in the form of dispersive terms (i.e. commutators with the Hamiltonian) and dissipative terms (i.e. Lindblad superoperators), as in~\eqn{eqn:ME}.

\subsection{Keldysh self-energy}

	As described in Appendix~\ref{App:Keldysh}, the self-energy superoperator, $\Sigma(t_1,t)$, is given by the sum of irreducible superoperator-valued Keldysh diagrams, each of which evolves the state from an early time $t_1$ to the current time, $t$. 
	We evaluate $\Sigma(t_1,t)$ to a given order of perturbation theory by truncating the sum at that order.

	Keldysh superoperators are represented diagrammatically as two parallel lines representing the time-evolution of the forward (\bra{\cdot}) and reverse (\ket{\cdot}) components of the density matrix.  
	Vertices placed on the lines correspond to the action of the interaction Hamiltonian, so at $m^\textrm{th}$ order of perturbation theory, there are $m$ interaction vertices.  
	These act at specific interaction times, $t_j$, over which we integrate. Tracing over bath degrees of freedom corresponds to contracting vertices together, indicated by dashed lines. 
	
	To evaluate the time integral appearing in the Keldysh master equation,~\eqn{eq:KeldyshMEip}, we note that each Keldysh diagram takes the form of a convolution kernel with oscillatory time-dependence, 
	so that the Laplace transform in the time-domain becomes a product of terms in Laplace space, i.e.\ $\rho(t)\rightarrow \rho_{s}$.  
	In Laplace space, \eqn{eq:KeldyshMEip} becomes
	\begin{align}
		s \rho_s-\rho(0) =&  \sum_{\omega_j} \rho_{s+i\omega_j} \bar\Sigma_s \,,
		\label{eq:KeldyshMElap}
	\end{align}
	where $\omega_j$ labels a set of system frequencies.  	
	
	One approach to the solution of  these equations was presented in  Ref.~\onlinecite{Stace:PRL:2013} starting from the ansatz
	\begin{align}
		\rho_{s} = \frac{\bar\rho_{0}}{s} + \sum_{\omega_j} \frac{\bar\rho_{j}}{s+\ii \omega_{j}} \,.
		\label{eq:RhoAnsatz1}
	\end{align}
	In brief, the unknown residues in  \eqn{eq:RhoAnsatz1} are found by comparing the poles on both sides of \eqn{eq:KeldyshMElap}, and imposing a self-consistency condition thereupon.  
	This self-consistent solution ultimately truncates the summation in \eqn{eq:RhoAnsatz1}, which corresponds to a secular or rotating-wave approximation. 
	 Further details are given in Appendix \ref{App:Laplace}.	
	
	In the experiments that motivate this work, the steady-state response of the system determines the phenomenology, 
	so we  focus here on the slow dynamics of the system. 	
	Retaining only the pole at $s=0$ corresponds to the conventional secular approximation made in deriving the quantum optical master equation~\cite{Stace:PRL:2013}, so we use the quasi-static ansatz 
	\begin{equation}
		\rho_{s} = {\bar\rho}/{s}\,,
		\label{en:ansapprox}
	\end{equation}
	so that $\bar\rho$ captures the quasi-static evolution associated with the weak coupling to the bath.

	We now describe the results of evaluating $\mathcal{L}_m$ at each order.
	
\subsection*{First order}

	There are  no dissipative terms surviving the perturbative treatment at  first order in perturbation theory, since for any thermal state of the environment one finds $\langle b_j\rangle_\textrm{th}=\langle b_j^\dagger\rangle_\textrm{th}=0$~\cite{QuantumOptics, QuantumNoise}.  
	This is generally true for any odd moments of bath operators.
	
	In our case, the perturbative interaction Hamiltonian $\H_{\text I}$ also contains coherent terms between the qubit and resonator, i.e.\ terms that couple two parts of the system to each other. 
	In this case the above reasoning no longer applies and some terms might survive even in first order of perturbation theory.  However these terms carry a rapidly rotating time dependence.
	Making a rotating-wave approximation (consistent with our approximate ansatz,~\eqn{en:ansapprox}) eliminates these terms also.
	It follows that $\mathcal L_{1} \bar \rho = 0$.

\subsection*{Second order}
	
	\begin{figure}
		\begin{center}
			\includegraphics[width=\columnwidth]{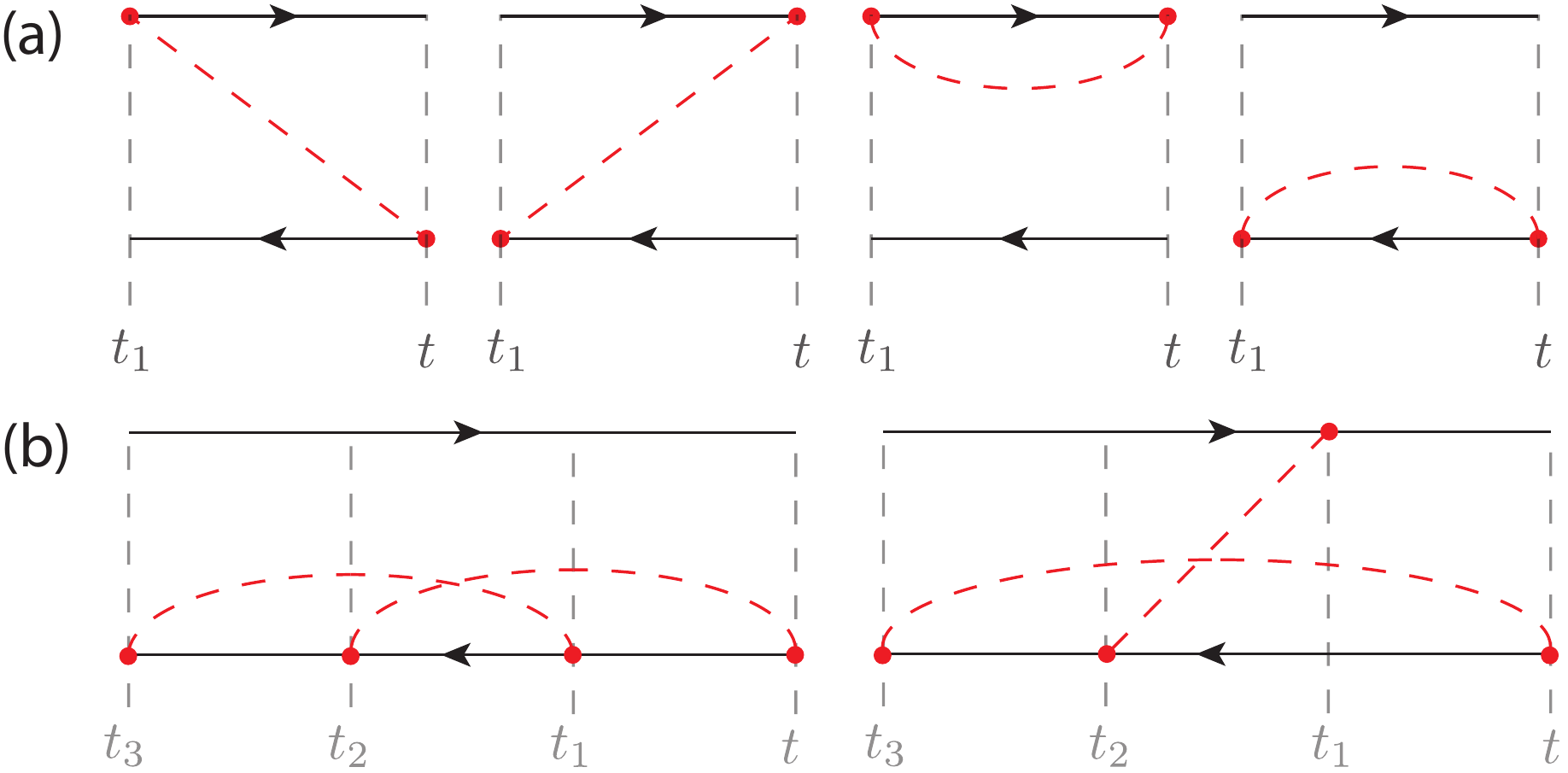}
			\caption{(Color online) (a) Diagrams representing all 2nd order terms in the Keldysh self-energy.  
				The red dashed line indicates that bath operators in the vertices connected by it are contracted together to form a bath spectral function.
				(b) Two examples of irreducible fourth-order Keldysh diagrams. 
				All 32 irreducible diagrams at this order can be generated from the two shown by swapping sub-sets of interaction vertices between the lower and upper lines. 
			} 
			\label{fig:keldysh}
		\end{center}
	\end{figure}

	In second order perturbation theory there are four classes of diagrams to consider, depicted in Fig.~\ref{fig:keldysh}(a). 
	Evaluating these diagrams generates a dissipative contribution arising from system-phonon interactions, as well as a coherent contribution from the dispersive, coherent terms arising from qubit-resonator coupling. We provide worked examples evaluating these diagrams in Appendix \ref{App:2ndK}.  
	At second order we write,
	\begin{align}
		\mathcal L_{2}\bar \rho = \mathcal L_{2,\text{diss}} \bar \rho -\ii \comm{\H_{2}}{\bar \rho},\label{eqn:L2}
	\end{align}
	where we specify $\mathcal L_{2,\text{diss}}$ and $H_2$ below.

\subsubsection*{2nd order dissipative terms}

	If the vertices in Fig.~\ref{fig:keldysh}(a) represent the interaction with the underlying phonon bath, $\H_{I,B}$, 
	the diagrams yield the usual dissipative terms in the master equation for a two-level system coupled to an environment~\cite{QuantumOptics, MesoscopicQuantumOptics}. 
	They evaluate to
	(c.f. Appendix~\ref{App:2ndKdiss})
	\begin{align}
		\mathcal L_{2,\text{diss}}\bar \rho = \gamma_{\downarrow,2} \diss{\sigma_{-}} \bar\rho + \gamma_{\uparrow,2} \diss{\sigma_{+}} \bar\rho + \gamma_{\varphi,2} \diss{\sigma_{z}} \bar\rho ,
		\label{eq:ME2nd}
	\end{align}
	where the second-order rates are 
	\begin{align}
		\gamma_{\downarrow,2} &=  \sin^{2}{\theta}\: C(\omega_{q}) /2,\nn\\
		 \gamma_{\uparrow,2} &=  \sin^{2}{\theta}\: C(-\omega_{q})/2,\nn\\
		\gamma_{\varphi,2} &=  \cos^{2}{\theta}\: C(0) /2,
		\label{eq:QDRates0}
	\end{align}
	and in which the spectral function of the phonon environment is
	\begin{align}
		C(\omega) &=  \int_{-\infty}^{\infty} dt\: \ee^{\ii \omega (t_{1} - t_{2})} \langle{\hat X(t_{1}) \hat X(t_{2})}\rangle \,,\nn\\
			&= \frac12 J(\omega) \l( n_{\text{th}} (\omega) + \theta(\omega) \r) ,\nn
	\end{align}
	where $\hat X(t) = \sum_{j} \beta_{j} ( b_{j}\: \ee^{-\ii\omega_{j }t} + b_{j}\hc\: \ee^{\ii \omega_{j} t} )$ is the bath coupling operator in the interaction picture,  
	$J(\omega)$ is the spectral density, and $\langle{b_j\hc b_j}\rangle = n_{\text{th}}(\omega_j) = (\ee^{\beta\omega_j} - 1)^{-1}$ is the Bose distribution with $\beta=1/k_BT$. 
	$\theta(\omega)$ is the step-function.
	See Appendix~\ref{App:SpectralFunction} for more details.

\subsubsection*{2nd order coherent terms}

	If the vertices in Fig.~\ref{fig:keldysh}(a) represent dot-resonator interactions from $\H_{\text{I,S}}$, most diagrams carry fast oscillatory terms, so that they vanish in a rotating wave approximation. 
	The residual, stationary terms contribute to the imaginary part of the self-energy, leading to a renormalisation of the Hamiltonian. We find (c.f. Appendix~\ref{App:2ndKdisp})
	\begin{align}
		\H_{2} = \chi \frac{\omega_{q}}{\omega_{q} + \omega_{r}} \l( \sigma_{z} + 2a\hc a \sigma_{z} \r) \,,
		\label{eq:HDispersive}
	\end{align}
	where the dispersive shift is \mbox{$\chi = g^{2}\sin^{2}\theta / (4\omega_{q} - 4\omega_{r})$}, similar to the standard treatment of the Jaynes-Cummings model in the dispersive regime~\cite{Blais:PRA:2004}.
	The difference between our result Eq.~\eqref{eq:HDispersive} and the usual Jaynes-Cummings dispersive shift lies in the inclusion of counter-rotating terms in the system-bath coupling Eq.~\eqref{eq:HI}, leading to the additional ratio of system frequencies in Eq.~\eqref{eq:HDispersive}.
	Our result does not make the rotating-wave approximation in this coupling and thus includes both the usual dispersive as well as the Bloch-Siegert shift~\cite{Bloch:PR:1940, FornDiaz:PRL:2010} in one analytic expression.

\subsection*{Third order}
	
	Third order diagrams can again contribute due to the system-system coupling terms in our perturbative interaction Hamiltonian. 
	Their contributions can be at order $\sim g^{3}$ or $~\sim g \beta^{2}$, i.e. they are at even order in $\H_{\text{I,B}}$ and odd order in $\H_{\text{I,S}}$. 
	These terms contribute Hamiltonian corrections, i.e. small energy shifts and changes of the bare system parameters, so we ignore them here. 	
	Instead, we assume these contributions are implicitly included in renormalised  experimental parameters. Thus we take $\mathcal L_{3}\bar \rho = 0$.

\subsection*{Fourth order}

	In fourth order perturbation theory, there is a total of 32 irreducible diagrams. Two examples are depicted in Fig.~\ref{fig:keldysh}(b) and the full set is shown in Appendix~\ref{App:4thDiag}. 
	These 32 diagrams all have 4 vertices, each of which represents terms in $\H_{\text{I}}=\H_{\text{I,S}}+\H_{\text{I,B}}$. 
	When evaluating diagrams, we first expand each one into all possible combinations of $\H_{\text{I,S}}$ and $\H_{\text{I,B}}$ over the vertices.  
	Within this expansion, terms with an odd number of instances of $\H_{\text{I,B}}$ vanish, since odd moments of bath operators vanish.  
	Terms with an even number of instance of both $\H_{\text{I,B}}$ and $\H_{\text{I,S}}$ are further classified into three distinct categories of diagrams:
	\begin{enumerate}
		\item{Correlated photon-phonon decay processes: Diagrams in which two vertices come from the intra-system interaction Hamiltonian $\H_{\text{I,S}}$ 
			and the remaining two are from the system-bath interaction $\H_{\text{I,B}}$.
		}
		\item{Coherent dispersive photon-photon processes: Diagrams in which all four vertices originate from  the coherent intra-system interaction $\H_{\text{I,S}}$.
		}
		\item{Two phonon decay processes: Diagrams in which all four vertices come from the system-bath interaction term $\H_{\text{I,B}}$
		}
	\end{enumerate}
	We thus write
	\begin{align}
		\mathcal L_{4}\bar \rho = \mathcal L_{4,\text{corr}} \bar\rho -\ii \comm{\H_{4}}{\bar\rho} + \mathcal L_{4,2\text{phonon}}\bar\rho\,,
	\end{align}
	where the order of the terms corresponds to the numbered list above.
	
	In this paper we are only concerned with calculating the self-energy diagrams leading to correlated photon-phonon decay processes, $\mathcal L_{4,\text{corr}}$. 
	We therefore evaluate neither $H_4$ nor  $L_{4,2\text{phonon}}$ here, but leave this to future work. 
	In passing, we note that $H_4$ will renormalise the bare Hamiltonian terms, and $L_{4,2\text{phonon}}$ will renormalise the rates in \eqn{eq:QDRates0}, and potentially include dispersive shifts as well.

\subsubsection*{4th order correlated photon-phonon decay terms}
	
	Evaluating  correlated photon-phonon decay self-energy diagrams at fourth order results in a total of 21 individual dissipators, each associated with rates $\Gamma_j$. 	
	
	An example diagram at fourth order is evaluated in Appendix~\ref{App:4thKdiss} for pedagogical purposes.  
	After evaluating all the relevant fourth-order Keldysh diagrams, we are left with a large number of superoperators that contribute to $\mathcal L_{4,\text{corr}} \bar\rho$.  
	Collecting these into Lindblad dissipators requires careful analysis, which we describe in Appendix~\ref{App:Diss}.
	
	The full set of correlated photon-phonon decay terms that appear at fourth order of  Keldysh-Lindblad perturbation theory are
	\begin{align}
		\mathcal L_{4,\text{corr}} \bar \rho   =&\hphantom{{}+{}} \Gamma_{\downarrow +} \diss{\sigma_{-}a\hc}\bar\rho + \Gamma_{\downarrow -}\diss{\sigma_{-}a}\bar\rho \nn\\
			&+ \Gamma_{\uparrow +}\diss{\sigma_{+}a\hc}\bar\rho + \Gamma_{\uparrow-}\diss{\sigma_{+}a}\bar\rho \nn\\
			&+ \Gamma_{\varphi +} \diss{\sigma_{z} a\hc}\bar\rho + \Gamma_{\varphi - }\diss{\sigma_{z}a} \bar\rho \nn\\
			&+ \Gamma_{-} \diss{a}\bar\rho + \Gamma_{+} \diss{a\hc}\bar\rho \nn\\
			&+ \Gamma_{\uparrow\downarrow} \l( \diss{\sigma_{+}}\bar\rho +\diss{\sigma_{-}}\bar\rho \r) \nn\\
			&+ \Gamma_{n} \l( \diss{a\hc a}\bar\rho - \diss{\sigma_{z}+a\hc a}\bar \rho \r) + \Gamma_{\varphi,4}\diss{\sigma_{z}}\bar\rho \nn\\
			&+ \Gamma_{\pm,\varphi\pm} \l( \diss{a+ \sigma_{z}a}\bar\rho + \diss{a\hc + \sigma_{z}a\hc}\bar\rho \r) \nn\\
			&+ \Gamma_{\varphi n} \l(\diss{\sigma_{z} a\hc a}\bar\rho - \diss{\sigma_{z} + \sigma_{z}a\hc a}\bar \rho \r) \nn\\
			&+ \Gamma_{\downarrow n} \l( \diss{\sigma_{-} a\hc a}\bar\rho - \diss{\sigma_{-} + \sigma_{-}a\hc a }\bar\rho\r)\nn\\
			&+ \Gamma_{\uparrow n}\l(\diss{\sigma_{+} a\hc a}\bar\rho -\diss{\sigma_{+} + \sigma_{+}a\hc a }\bar\rho\r) \,.
		\label{eq:4thAll}
	\end{align}
	Each of the rates $\Gamma_{j}$ are sums of  terms depending on the bath spectral function, $C(\omega)$, evaluated at frequencies 
	$\omega \in \l\{0, \pm\omega_{q}, \pm \omega_{r}, \pm (\omega_{q} - \omega_{r}), \pm (\omega_{q}+\omega_{r})\r\}$, c.f. Table \ref{tab:AllRates}. 
	Additionally there are contributions to the $\Gamma_j$ that are proportional to the derivative of the bath spectral function, $C'(\omega)$, evaluated at frequency $\omega=\pm \omega_{q}$. 
	Those originate from diagrams where the time evolution involves degenerate frequencies, leading to double poles in Laplace space, as has been noticed before~\cite{Shnirman:2016}.
	Their physical interpretation is uncertain. One possibility is that they are first order terms in a Taylor expansion of the bath spectral function and as such indicative of an implicit renormalisation of system frequencies.
	We have verified that the derivative terms do not contribute significantly in the experimental parameter regime we consider below and leave their detailed understanding to future work.

	Analysing all 21 dissipators above in detail is not the main concern of this paper.  
	However, the first six dissipators in~\eqn{eq:4thAll} are easy to interpret, and have dominant contributions, $\gamma_j$, to the corresponding rates $\Gamma_j$, that can be expressed simply.  
	These are
	\begin{align}
		\mathcal L_{4,\text{corr}}^{(0)} \bar \rho = &\hphantom{{}+{}} \gamma_{\downarrow +}^{(\omega_q-\omega_r)} \diss{\sigma_{-}a\hc}\bar\rho + \gamma_{\downarrow -}^{(\omega_q+\omega_r)} \diss{\sigma_{-}a}\bar\rho \nn\\
			&+ \gamma_{\uparrow +}^{(-\omega_q-\omega_r)} \diss{\sigma_{+}a\hc}\bar\rho + \gamma_{\uparrow-}^{(-\omega_q+\omega_r)} \diss{\sigma_{+}a}\bar\rho \nn\\
			&+ \gamma_{\varphi -}^{(\omega_{r})} \diss{\sigma_{z}a}\bar\rho + \gamma_{\varphi +}^{(-\omega_{r})} \diss{\sigma_{z} a\hc}\bar\rho\,,
		\label{eq:4thME}
	\end{align}
	where $\uparrow / \downarrow$ denote qubit flipping processes, $\varphi$ indicates qubit dephasing, and $+/-$ photon excitation and loss.
 	The rates are 
	\begin{align}
		\gamma_{\downarrow+}^{(\omega_q-\omega_r)} &= \frac12 g^{2}\cos^{2}\theta\ \frac{\omega_q^2 \sin^{2}\theta}{\omega_{r}^{2}(\omega_{q} - \omega_{r})^{2}}\: C(\omega_{q} - \omega_{r})\,, \nn\\
		\gamma_{\downarrow-}^{(\omega_q+\omega_r)} &=  \frac12 g^{2}\cos^{2}\theta \frac{\omega_q^2 \sin^{2}\theta}{\omega_{r}^{2}(\omega_{q} + \omega_{r})^{2}}\: C(\omega_{q} + \omega_{r}) \,,\nn\\
		\gamma_{\varphi-}^{(\omega_{r})} &= \frac12 g^{2}\sin^{2}\theta \frac{\omega_{q}^{2}\sin^{2}\theta}{(\omega_{q} - \omega_{r})^{2}(\omega_{q}+\omega_{r})^{2}} C(\omega_{r}) .
		\label{eq:CorrelatedRates}
	\end{align}

	The remaining three rates in \eqn{eq:4thME} are related by thermal occupation factors, 
	\begin{align}
		\gamma_{\uparrow -}^{(-\omega_q+\omega_r)}&= \gamma_{\downarrow +}^{(\omega_q-\omega_r)} e^{-\beta(\omega_{q} - \omega_{r})},\nn\\ 
		\gamma_{\uparrow +}^{(-\omega_q-\omega_r)}&= \gamma_{\downarrow -}^{(\omega_q+\omega_r)} e^{-\beta(\omega_{q} + \omega_{r})},\nn\\
		\gamma_{\varphi +}^{(-\omega_r)}&= \gamma_{\varphi -}^{(\omega_r)} e^{-\beta \omega_{r} }.\nn
	\end{align}  	
	In section~\ref{sec:Results}, we analyse in detail the contributions from Eqs.~\eqref{eq:CorrelatedRates} to gain and loss in the parameter regime of recent experiments
	in semiconductor double quantum-dots coupled to a microwave resonator. 
	There we show that the six dissipators in Eq.~\eqref{eq:CorrelatedRates} quantitatively explain recent experimental results~\cite{Liu:PRL:2014}, 
	whereas including the full set of rates from Eq.~\eqref{eq:4thAll} does not change the picture qualitatively apart from a change in numerical parameters.

\subsection{Resonator decay\label{Sec:ResDecay}}
	
	The resonator couples weakly to the external electromagnetic environment~\cite{Reagor:APL:2013}, and we include this as an additional decay process in the usual way~\cite{QuantumNoise}
	through the Lindblad dissipators
	\begin{align}
		\mathcal{L}_{\textrm{res}}\bar\rho= \kappa_{-,r}\diss{a}\bar\rho + \kappa_{+,r}\diss{a\hc}\bar\rho \,,
	\end{align}
	with the resonator relaxation and excitation rates $\kappa_{\mp,r}$ and the bare resonator linewidth $\kappa = \kappa_{-,r} - \kappa_{+,r}$.

\subsection{External quantum-dot bias\label{Sec:QDBias}}
	In the experiments that motivate this work, the  open DQD coupled to external leads, inducing a charge-discharge transport cycle, illustrated in~\fig{fig:dissipation}(b).  
	To model this process, we extend the DQD  basis to include the empty state $\ket{\O}$, in which the DQD is uncharged. 
	As electrons tunnel between the leads and the DQD, it passes transiently through the empty state.    
	This process is described by the incoherent Lindblad superoperator~\cite{Lindblad:CMP:1976, QuantumOptics, Stace:PRL:2005, stace2004parity,barrett2006continuous}
	\begin{equation}
		\L_{\text{leads}}\rho= \Gamma_{L} \diss{\,{\ket L}{\bra \O}\,}\rho + \Gamma_{R} \diss{\,{\ket \O}{\bra R}\,}\rho.
	\end{equation}
	For simplicity, we will assume $\Gamma_R = \Gamma_{L} = \Gamma$ in the following. 
	Depending on the sign of $\epsilon_q$ and the strength of $\Gamma$, the DQD population may become inverted in steady state.

\section{Driven resonator\label{Sec:Steady}}
	 
	The additional dissipators will tend to drive the system to its ground state.  
	To see non-trivial behaviour arising from the master equation, the system needs to be driven out of equilibrium.  
	Depending on the specific system, there are a variety of ways to achieve this.  Here we consider resonator driving, in which an external microwave field is imposed on the resonator.  
	To this end we introduce an additional resonator driving term in the system Hamiltonian, $H = H_{\text{S}} + H_{D}$, where
	\begin{align}
		H_D =  \epsilon_{d} \l( a\hc \ee^{ i \omega_{d} t} + a \ee^{- i \omega_{d} t} \r)/2 \,,\label{eqn:hd}
	\end{align}
	with the amplitude of the drive $\epsilon_{d}$ and  frequency $\omega_{d}$. In writing Eq.~\eqref{eqn:hd} we have assumed a rotating wave approximation, 
	relevant to near-resonant driving $\omega_{d} \approx \omega_{r}$. 
	
	Under driving, we anticipate that the resonator will tend to a steady-state which is close to a coherent state, $\ket{\alpha}$.  We therefore apply a displacement transformation on the resonator modes~\cite{Gambetta:PRA:2008, Slichter:NJP:2016}, $a\rightarrow \tilde a+\alpha$.  
	This transformation forms the basis of a semi-classical, mean-field approximation for dissipative steady-state equations for qubit and resonator, from which we establish a self-consistency condition that determines $\alpha$.
	
	The canonical unitary for the displacement transformation is 
	\begin{align}
		D(\alpha) = \exp{ \{ \alpha a\hc - \alpha\cc a \}} \,,
	\end{align}
	so that $D(\alpha)\hc a D(\alpha) = \tilde a + \alpha$. 
	The Hamiltonian in the displaced frame is transformed as
	\begin{align}
		\tilde H =& D(\alpha)\hc H D(\alpha) - \ii D(\alpha)\hc \dot D(\alpha)\nn\\
		=& D(\alpha)\hc H D(\alpha) - \ii (\tilde a\hc \dot\alpha - \tilde a \dot\alpha\cc ) \,,
	\end{align}
	where $\alpha$ may be time-dependent. 
	
	When applying the displacement operation, we have to take care to include dissipative processes containing resonator operators in the transformation. For example, when applying the transformation to the dissipator describing direct photon decay $\sim \diss{a}\rho$ we find
	\begin{align}
		\diss{D\hc a D} \bar\rho =&{} \diss{\tilde a}\bar\rho - \ii \l[i{ ( \alpha\cc \tilde a - \alpha \tilde a\hc )/2},{\bar\rho}\r] \,,
	\end{align}
	where the second term describes a coupling between the residual mode $\tilde a$ and the coherent field amplitude $\alpha$. 
	This term, and similar ones from other dissipators, contributes to a self-consistent equation for the coherent field amplitude $\alpha$. 

	Correlated dissipators transform similarly. For example 
	in the displaced frame, $\diss{\sigma_{-} a\hc}\rho$ transforms as
	\begin{align}
		\diss{D\hc\sigma_{-}a\hc D}\bar\rho ={}& \diss{\sigma_{-} \tilde a\hc}\bar\rho + \abss{\alpha}\diss{\sigma_{-}}\bar\rho \nn\\
			&\hspace*{-0mm}+ \alpha \sigma_{-}\tilde a\hc \bar\rho \sigma_{+} + \alpha\cc \sigma_{-} \bar\rho \tilde a \sigma_{+} \nn\\
			&\hspace*{-0mm}-\frac12 \big( \alpha \sigma_{+}\sigma_{-} \tilde a\hc \bar\rho + \alpha\cc \tilde a\sigma_{+}\sigma_{-}\bar\rho \nn\\
			&\hspace*{-0mm}\quad +\alpha \bar\rho\sigma_{+}\sigma_{-} \tilde a\hc + \alpha\cc \bar\rho \tilde a\sigma_{+}\sigma_{-} \big) \,.
		\label{eq:corrDiss}
	\end{align}

\subsection{Separable and semi-classical approximations}
	
	In principle, we could solve the steady-state (or dynamical behaviour) of the master equation, \eqn{eqn:ME}, accounting fully for correlations between the resonator and qubit.  
	However, to make analytical progress we now make a simplifying mean-field approximation that enables us to calculate the qubit and resonator steady-states.
	
	That is, we assume the resonator to be in the semi-classical regime, so that it is well approximated by a coherent state, $\ket{\alpha}$. We will self-consistently determine the coherent amplitude $\alpha$ by finding a displaced resonator frame which is effectively undriven. 
	
	In this approximation, the resonator and qubit are separable, 
	\begin{align}
		\bar\rho = \bar\rho_{r} \otimes \bar\rho_{q} \,.
		\label{eqn:separation}
	\end{align}
	Under this approximation, \eqn{eqn:ME} decomposes into two coupled {mean-field} equations, one for the resonator state, $\bar\rho_{r}={\ket{\alpha}}{\bra{\alpha}}$, and one for the qubit state, $\bar\rho_{q}$, in which each sub-system experiences the mean field of the other. We find
	\begin{align}
		\dot{ \bar\rho}_r &=\text{Tr}_{q}\Big\{ \sum_{m} \mathcal D^\dagger L_{m} D \bar\rho\Big\},
		\label{eqn:mer}\\
		\dot{ \bar\rho}_q &=\text{Tr}_{r}\Big\{ \sum_{m} \mathcal D^\dagger L_{m} D \bar\rho\Big\}.
		\label{eqn:meq}
	\end{align}
	Evaluating these traces simplifies correlated dissipators.  
	For example, tracing over the qubit in \eqn{eq:corrDiss} contributes a term on the RHS of~\eqn{eqn:mer}:
	\begin{align}
		& \text{Tr}_{q}\l\{\diss{D\hc\sigma_{-}a\hc D}\bar\rho\r\}  \nn\\
		&\,\,= P_{e}\l (\diss{\tilde a\hc} \bar\rho_{r} - i [{i (\alpha \tilde a\hc- \alpha\cc \tilde a)/2},{\bar\rho_{r}}]
		\r),
	\end{align}
	where $P_{i} = \text{Tr}_{q}\l\{\bar\rho{\ket{i}}{\bra{i}}\r\}$, so that $\mean{\sigma_{+} \sigma_{-}}=P_e$.
	The first term describes incoherent photon generation originating from the correlated decay process $\diss{\sigma_{-}a\hc}\rho$, while the second and third terms represent effective resonator driving, which  depends on the qubit population and the resonator field expectation value $\alpha$.
	
	Finally, tracing~\eqn{eq:corrDiss} over the resonator degrees of freedom results in a contribution to the qubit master equation,~\eqn{eqn:meq}
	\begin{equation}
		\text{Tr}_{r}\l\{\diss{D\hc\sigma_{-}a\hc D}\bar\rho\r\} = \langle{(\tilde a\!+\!\alpha)(\tilde a\hc\!\!+\!\alpha^*)}\rangle  \diss{\sigma_{-}} \bar\rho_{q} \,.
		\label{eq:corrDissQubit}
	\end{equation}

\subsection{Coupled steady-state equations}

	In the application to gain and loss in a driven DQD-resonator, we are interested in steady-state properties of the system, so we solve the above equations subject to $\dot{\bar\rho}=0$, and $\dot\alpha=0$.  
	The mean-field approximation for the resonator field 
	then yields coupled steady-state equations for the two sub-systems. 
	In a frame rotating at the resonator drive frequency $\omega_{d}$, we find the  steady-state of the resonator master equation, \eqn{eqn:mer}, becomes
	\begin{align}
		0 = -\ii [{\tilde H_r},{\bar\rho_{r}}] + \kappa_{-} \diss{\tilde a}\bar\rho_{r} + \kappa_{+}\diss{ \tilde a\hc} \bar\rho_{r} \,,
		\label{eq:resME}
	\end{align}
	with the renormalized photon loss and generation rates 
	\begin{align}
		\kappa_{\pm} = \kappa_{\pm,r} + \kappa_{\pm,4}\,,
		\label{eq:kappapm}
	\end{align}
	and where the contributions from the fourth-order Keldysh perturbation theory are given by
	\begin{align}
		\kappa_{-,4} =&  \gamma_{\downarrow -}^{(\omega_q+\omega_r)} P_{e} + \gamma_{\uparrow -}^{(-\omega_q+\omega_r)} P_{g} + \gamma_{\varphi -}^{(\omega_{r})} (1\!-\!P_{\O} ) \nn\\
			&+  \kappa_{\varphi}^{(\omega_{q})} \mean{\sigma_{z}}+  \kappa^{(\omega_{q})} P_{e}-  \kappa^{(-\omega_{q})} P_{g}\,,\nn\\
		\kappa_{+,4} =&  \gamma_{\downarrow +}^{(\omega_q-\omega_r)} P_{e}+  \gamma_{\uparrow +}^{(-\omega_q-\omega_r)}  P_{g}+ \gamma_{\varphi +}^{(-\omega_{r})} (1\!-\!P_{\O} ) \nn\\
			&+ \kappa_{\varphi}^{(\omega_{q})} \mean{\sigma_{z}}-  \kappa^{(\omega_{q})} P_{e}+\kappa^{(-\omega_{q})} P_{g} \,.\label{eq:kappa}
	\end{align}
	Here $\mean{\sigma_{z}} = P_{g} - P_{e}$ is the expectation value of the qubit dipole operator and $\kappa_{\pm,r}$ are the original resonator decay and excitation rates. $P_{\O}$ is the population of the empty state, when no charge is on either island.
	For the parameter regime we consider below, $P_{\O}\ll1$.
	The full expressions for the rates $\kappa_{\varphi}^{(\omega_{q})}$ and $\kappa^{(\pm \omega_{q})}$ are somewhat cumbersome and are given in appendix~\ref{App:SteadyRates} where we also give their relationship to the $\Gamma_{j}$ in Eq.~\eqref{eq:4thAll}.
	
	The effective Hamiltonian for the resonator in the displaced frame is 
	\begin{align}
		\tilde H_r =& \big(\delta\omega_{r} + 2 \tilde\chi \langle{\sigma_{z}}\rangle\big) \tilde a\hc \tilde a \nn\\
				&+ \tilde a\hc  \big( 
				 \epsilon_{d}/2 + \alpha ( \delta\omega_{r} + 2\tilde\chi \mean{\sigma_{z}} - \ii \kappa' /2 )	\big) \nn\\ 
				&+ \text{h.c} \,,			
		\label{eq:HDisp}
	\end{align}
	where $\tilde\chi = \chi\: \omega_{q} / (\omega_{q}+\omega_{r})$ is the effective dispersive shift. 
	The resonator linewidth is renormalized by the interaction with the qubit to
	\begin{align}
		\kappa' =& \kappa_{-} - \kappa_{+} \,,
		\label{eq:KappaFull}
	\end{align}
		
	The term in~\eqn{eq:HDisp} proportional to $\tilde a^\dagger$ (and its hermitian conjugate) describes effective driving of the transformed resonator mode $\tilde a$, which would lead to a non-zero expectation value of the resonator operators $\mean{\tilde a}$.
	We now self-consistently require that the coefficient of $\tilde a^\dagger$ should vanish. 	Since the effective driving terms depend on $\alpha$, this choice leads to the condition 
	\begin{align}
		0 =   \epsilon_{d} /2+ \alpha \big(& \delta\omega_{r} + 2 \tilde\chi \langle\sigma_{z}\rangle - {\ii}\kappa'/2 \big).
		\label{eq:AlphaEigen}
	\end{align}
	In this case, the effective driving terms in~\eqn{eq:HDisp} are zero and~\eqn{eq:resME} describes the residual quantum dynamics of an un-driven resonator subject only to relaxation at rate $\kappa_{-}$ and excitation at rate $\kappa_{+}$ 
	In the displaced frame with the appropriately chosen value of $\alpha$, the displaced resonator mode $\tilde a$ will be in a thermal state, with its effective temperature $T_{r}$ depending on the ratio of its relaxation and excitation rates 
	like $\kappa_{+} / \kappa_{-} = \exp{\{ -\omega_{r} / k_{B} T_{r} \}}$.
	In this case, we find $\langle{\tilde a}\rangle = \langle{\tilde a\hc}\rangle =0$.
	In the following we additionally assume a small effective temperature of the resonator $T_{r}$ such that $\mean{\tilde a\hc \tilde a} = \frac{\kappa_{+}}{\kappa_{-} - \kappa_{+}} \ll 1$.

	The  steady-state of the qubit master equation,~\eqn{eqn:mer}, then becomes	
	\begin{align}
		0 =& -\ii [{\tilde H_q},{\bar\rho_{q}}] + \gamma_{\downarrow}\diss{\sigma_{-}}\bar\rho_{q} + \gamma_{\uparrow} \diss{\sigma_{+}}\bar\rho_{q}\nn\\
			& + \gamma_{\varphi} \diss{\sigma_{z}} \bar\rho_{q} + \mathcal{L}_{\text{leads}} \rho_{q} \,,
		\label{eq:qDotME}
	\end{align}
	where the effective  qubit Hamiltonian is
	\begin{align}
		\tilde H_q = - \omega_{q} \sigma_{z}/2 + \tilde\chi (1+2\abss{\alpha}) \sigma_{z} \,,
	\end{align}
	and the correlated qubit dissipation rates are given by $\gamma_{\uparrow}= \gamma_{\uparrow,2}+\gamma_{\uparrow,4}$, $\gamma_{\downarrow}= \gamma_{\downarrow,2}+\gamma_{\downarrow,4}$ and $\gamma_{\varphi}= \gamma_{\varphi,2}+\gamma_{\varphi,4}$ 
	with the additional rates from correlated fourth-order processes 
	\begin{align}
		\gamma_{\downarrow,4} =& \abss\alpha \gamma_{\downarrow-}^{(\omega_q+\omega_r)} + \big({\abss\alpha} +1 \big) \gamma_{\downarrow+}^{(\omega_q-\omega_r)} \nn\\
			&+ \gamma_{\downarrow}^{(\omega_{q})} + \gamma_{\downarrow}^{(-\omega_{q})} + \gamma_{\downarrow}' 	, \nn \\
		\gamma_{\uparrow,4} =& \abss\alpha \gamma_{\uparrow-}^{(-\omega_q+\omega_r)}  + \big({\abss\alpha} + 1\big) \gamma_{\uparrow+}^{(-\omega_q-\omega_r)} \nn\\
			&+ \gamma_{\uparrow}^{(\omega_{q})} + \gamma_{\uparrow}^{(-\omega_{q})} + \gamma_{\uparrow}' , \nn \\
		\gamma_{\varphi,4} =& \abss\alpha \gamma_{\varphi-}^{(\omega_{r})} + \big( {\abss\alpha} + 1 \big) \gamma_{\varphi+}^{(-\omega_{r})} \nn\\
			&+\gamma_{\varphi}^{(0)} + \gamma_{\varphi}^{(\omega_{q})} + \gamma_{\varphi}^{(-\omega_{q})} + \gamma_{\varphi}' .
		\label{eq:gamma}
	\end{align}
	The expressions for $\gamma_{\uparrow/\downarrow}^{(\pm \omega_{q})}$, $\gamma'_{\uparrow/\downarrow}$, 
	$\gamma_{\varphi}^{(0)}$, $\gamma_{\varphi}^{(\pm \omega_{q})}$, and $\gamma'_{\varphi}$ 
	are somewhat lengthy and are given in full in Appendix~\ref{App:SteadyRates}, where we also discuss their relationship to the $\Gamma_j$'s in \eqn{eq:4thAll}.

	If in the previous calculation the effective resonator temperature $T_{r}$ is large, i.e., when $\kappa_{+} \gtrsim \kappa_{-}$, the expectation value of $\langle\tilde a\hc \tilde a\rangle$ acquires a non-zero value $\abss{\tilde\alpha}$. 
	If necessary, we can simply take this into account by noting that tracing over the resonator degrees of freedom now involves additional factors of $\abss{\tilde\alpha}$.
	As an example Eq.~\eqref{eq:corrDissQubit} becomes 
	\begin{align}
		\text{Tr}_{r}\l\{\diss{D\hc\sigma_{-}a\hc D}\bar\rho\r\} = \big( {\abss{\alpha}} + \abss{\tilde\alpha} + 1 \big) \diss{\sigma_{-}}\bar\rho \,,
	\end{align}
	where $\tilde\alpha$ is determined from the effective thermal state at temperature $T_{r}$.

\section{Application to gain and loss in microwave-driven double quantum-dots\label{sec:Results}}

We now apply this theory to a specific experimental situation, where an open semiconductor double quantum-dot is coupled to a microwave resonator and subject to a transport bias. 
We are interested in calculating the steady-state resonator field depending on qubit parameters in an effort to replicate the experimental findings of Ref.~\onlinecite{Liu:PRL:2014}.  In particular,  we calculate the  gain of the DQD-resonator system, by comparing the internal circulating microwave energy within the driven DQD-resonator with the same quantity for a driven resonator alone (i.e.\ without the DQD).

From Eq.~\eqref{eq:AlphaEigen} we find the resonator steady-state field amplitude is described by
\begin{align}
	 \alpha = - { \epsilon_{d} }/({ 2\,\delta\omega_{r}' - i  \kappa' })\,,
	 \label{Eq:AlphaEigen2}
\end{align} 
where $\delta\omega_r' = \delta\omega_{r} +2\bar\chi \mean{\sigma_{z}}$ is the DQD-renormalised detuning. 
We calculate $\alpha$ by simultaneously solving for the qubit steady-state,~\eqn{eq:qDotME} and for the resonator field,~\eqn{Eq:AlphaEigen2}.  
The power gain is then given by
\begin{align}
	G= \abss{{\alpha}/{\alpha_{0}}}={|2 \,\delta\omega_{r} - i \kappa|^2}/|{2 \,\delta\omega'_{r} - i \kappa'}|^2 ,
	\nn
\end{align}
where $\alpha_{0} = -  { \epsilon_{d} }/({2 \,\delta\omega_{r} - i \kappa})$ is the steady-state resonator field that would be produced in the absence of the DQD coupling, (i.e.\ setting $g=0$).  
We set \mbox{$\delta\omega_{r}=0$}, corresponding to resonant cavity driving, which is the experimental situation  reported in Ref.~\onlinecite{Liu:PRL:2014}.

\begin{figure}[!t]
	\begin{center}
		\includegraphics[width=\columnwidth]{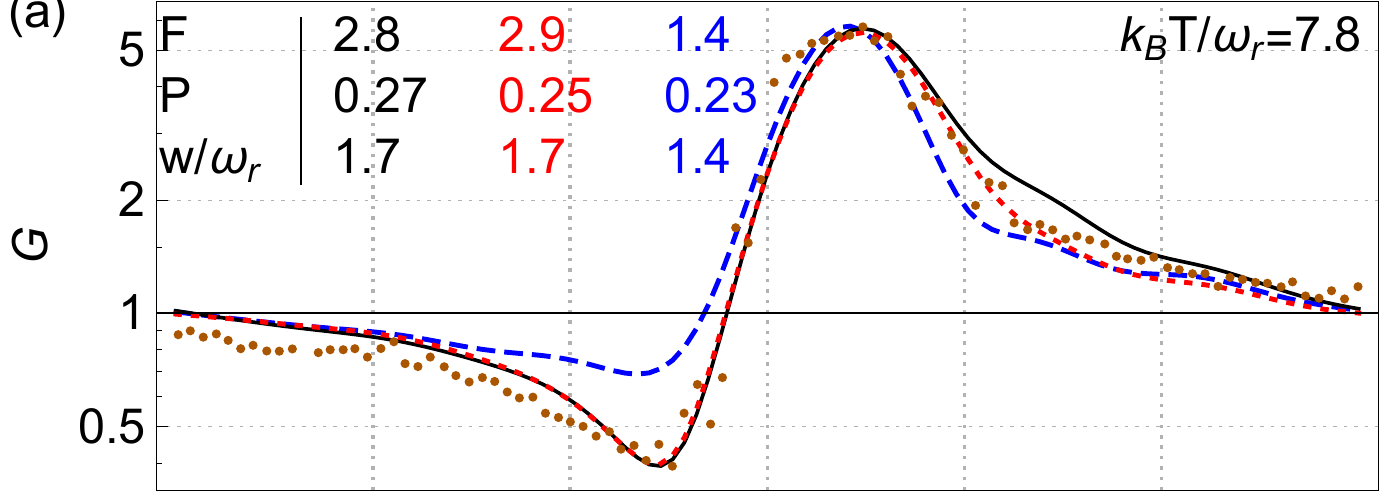}
		\includegraphics[width=\columnwidth]{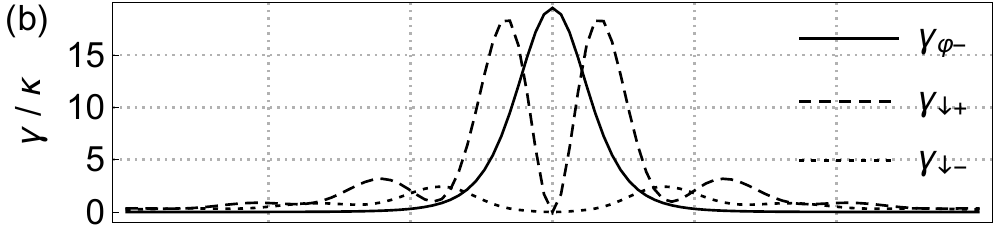}
		\includegraphics[width=\columnwidth]{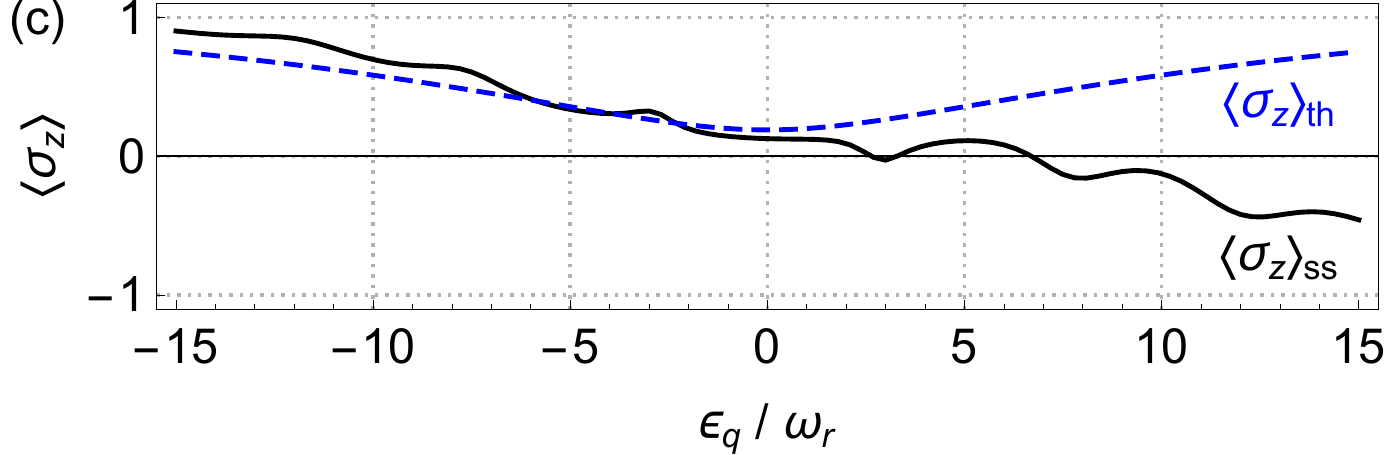}
		\caption{(Color online) {\textbf{(a)}} Logarithmic plot of the microwave power  gain, $G$, versus DQD bias, $\epsilon_{q}$, for  \mbox{$k_{B}T/\omega_{r}=7.8$ (corresponding to $T=3$~K)}.  
			Points are experimental data extracted from Ref.~\onlinecite{Liu:PRL:2014}. 
			The blue-dashed theory curve is generated using terms in the first two lines of~\eqn{eq:4thME}, corresponding to the polaron transformation used in Ref.~\onlinecite{Gullans:PRL:2015}. 
			The solid-black theory curve is generated using the six terms in~\eqn{eq:4thME}. The dotted red curve includes all additional terms in Eq.~\eqref{eq:4thAll}. 
			\textbf{(b)} Correlated rates, $\gamma_{\downarrow+}$ (dashed), $\gamma_{\downarrow -}$ (dotted), and $\gamma_{\varphi-}$ (solid) from~\eqn{eq:CorrelatedRates} corresponding to the black curve in panel (a). 
			\textbf{(c)} Qubit steady state population $\mean{\sigma_{z}}_{ss}$ (solid, black), compared to thermal population of the DQD, \mbox{$\langle\sigma_z\rangle_\textrm{th}$} (dashed, blue). 
			Common parameters for all panels: $ \omega_{d}/\omega_{r}= 1$, $g/\omega_{r}=0.0125$, $\Delta_{q}/\omega_{r} = 3$, $\kappa/\omega_{r}=52 \times10^{-6}$, $\Gamma/\omega_{r}=0.34$, as in Ref.~\onlinecite{Gullans:PRL:2015}. 
			} 
		\label{fig:gain}
	\end{center}
\end{figure}

In order to make a quantitative comparison between theory and experiment, we assume the bath spectral function is \mbox{$C(\omega) = J(\omega) \l( n_{\text{th}} + \theta(\omega) \r)$} with \mbox{$J(\omega) = J_{\text{1D}}(\omega) + J_{\text{3D}}(\omega)$}, 
where the spectral densities for the first phonon mode in the quantum wire and the  bulk  substrate phonons are given by~\cite{Stace:PRL:2005, Gullans:PRL:2015}
\begin{align}
	\frac{J_{\text{1D}}(\omega)}{\omega_{r}} &= \mathsf{F} \frac{c_{n}}{\omega\, d} \big(1-\cos(\omega \, d /  c_n)\big) \ee^{-\omega^{2} a^{2} / 2 c_{n}^{2}},\nn\\
	\frac{J_{\text{3D}}(\omega)}{\omega_{r}} &= \mathsf{P} \frac{\omega}{\omega_{r}} \big(1-\text{sinc}({\omega\, d/c_{s}})\big) \ee^{-\omega^{2} a^{2}  /2c_{s}^{2}},
	\label{eq:SpecFunc}
\end{align}
with the speed of sound in the quantum wire \mbox{$c_{n}= 4000$~m/s} and in the substrate $c_{s} = 5 000$~m/s~\cite{Fate:JAP:1975}, an inter-dot spacing $d = 120$~nm and nanowire radius $a = 25$~nm.  We have taken \mbox{$k_{B}T/\omega_{r}=7.8$}, from \cite{Gullans:PRL:2015}.  
We discuss details of fitting parameters $\mathsf{F}$, $\mathsf{P}$, and a gaussian smoothing parameter $\mathsf{w}$ below.

\subsection*{Theoretical and Experimental Gain-Loss Profiles}

In Fig.~\ref{fig:gain}(a) we plot the  gain $G$ in the microwave resonator due to the DQD coupling for three different theoretical treatments, along with experimental gain data extracted from Ref.~\onlinecite{Liu:PRL:2014}, shown as points.   

The three  theoretical curves shown in \fig{fig:gain}(a)  compare the dependence of the gain on various terms in the fourth-order correlated dissipators.   
The dashed blue curve includes terms in the first two lines of \eqn{eq:4thME} but not the third line (i.e.\ the fourth-order theory restricted to $ \gamma_{\varphi \pm}^{(\mp\omega_r)}=0$), which is equivalent to the polaronic theory in Ref.~\onlinecite{Gullans:PRL:2015}.  
The solid black curve further includes the six terms in \eqn{eq:4thME}, corresponding to DQD-mediated photon-to-phonon interconversion.  
The dotted red curve finally includes all 21 fourth-order rates that appear in the derivation of the full master equation, Eq.~\eqref{eq:4thAll}. 

In \fig{fig:gain}(a), there is a clear qualitative difference between the polaronic theory (blue, dashed) and the experimental data in the regime \mbox{$\epsilon_q/\omega_r\lesssim0$}: the theory is unable to explain the depth of loss (sub-unity gain).  
In contrast, the additional dephasing-mediated processes in the last line of \eqn{eq:4thME}  give rise to enhanced losses beyond the polaronic terms, and are sufficient to quantitatively account for the entire range of gain and loss observed in the experimental data (black).  
We have also shown the results of the full fourth-order theory (red dotted) from Eq.~\eqref{eq:4thAll}, which also consistently captures the peak of the gain for $\epsilon_q>0$, and the depth of the trough for $\epsilon_q<0$. 
In the parameter regime described here, 
there are only small differences between the latter two curves, including a rescaling of $\mathsf{F}$ and $\mathsf{P}$, indicating that the six terms in~\eqn{eq:4thME} quantitatively account for the extremes of gain and loss.

\fig{fig:gain}(b) plots the rates in \eqn{eq:CorrelatedRates}, and shows clearly that the dephasing-assisted loss rate (solid), $\gamma_{\varphi-}^{(\omega_r)}$, is significant compared to the other correlated decay processes, $\gamma_{\downarrow \pm}^{(\omega_q\mp\omega_r)}$ (dashed and dotted).  
This is the main reason for the qualitative difference between the theory curves in \fig{fig:gain}(a).  
Since the full fourth-order theory accounts for the experimental data, whilst the polaron-only theory does not, we conclude that dephasing-assisted loss is a substantial contribution to the system dynamics.

\fig{fig:gain}(c) shows the steady-state DQD population imbalance, $\langle\sigma_z\rangle_{ss}$ (black), compared to the thermal equilibrium value, $\langle\sigma_z\rangle_\textrm{th}=\Tr\{\sigma_z e^{-\beta H_S}\}/\mathcal{Z}$ (blue).  
In conventional gain/loss models, positive values of the difference  $\langle\sigma_z\rangle_\textrm{th}-\langle\sigma_z\rangle_{ss}$ (i.e.\ population inversion) drive gain, while negative differences drive loss.  
This is manifest in $\kappa'_{\pm}$, where enhancement of $P_e$ leads to an increase of $\kappa'_{+}$, a corresponding decrease in $\kappa'$, and thus an overall increase of $\alpha$.  
In contrast, the dephasing-assisted loss contribution to \eqn{eq:kappa} is independent of the state of the DQD in the limit that $P_{\O}\ll 1$, which is the case here. 

\subsection*{Note on Parameter Fitting}

As in Ref.~\onlinecite{Gullans:PRL:2015}, we treat the dimensionless spectral strengths $\mathsf{F}$ and $\mathsf{P}$ as free parameters.  
We follow the same fitting approach, in which we choose parameter values so that the theory curves satisfactorily replicate the strong gain peak evident for $\epsilon_q>0$.  The specific values are shown as insets in Fig.~\ref{fig:gain}(a). 

Likewise, we also convolve the bare theory gain curves with a normalised Gaussian smoothing kernel~$\propto e^{-\epsilon_q^{2}/ 2\mathsf w^{2}}$ (corresponding to a full-width-at-half-maximum of $\sqrt{8\ln{2}}\, \mathsf w$) \cite{Gullans:PRL:2015}, 
to account for low-frequency voltage noise in the gates defining the inter-dot bias~\cite{Petersson:N:2012}.

In effect, all theory curves have 3 fitting parameters,  $\mathsf{F}$, $\mathsf{P}$, and  $\mathsf{w}$.  Very roughly, $\mathsf{F}$ controls the gain peak height, $\mathsf{P}$ controls the tail behaviour of the gain at large $\epsilon_{q}$, and  $\mathsf{w}$ controls the gain peak width.  
Thus, once we find a parameter set for a given theory curve that adequately accounts for the gain profile for $\epsilon_{q}>0$, the dependence for  $\epsilon_{q}<0$ is determined. 
The overall scale for $\mathsf{F}$ and $\mathsf{P}$  is ultimately constrained by the value of the resonator linewidth $\kappa$: it is the relative contributions from the correlated decay rates compared to $\kappa$ that  set overall gain and loss through the effective resonator line-width $\kappa'$, c.f.~\eqn{eq:KappaFull}.

To make contact with the microscopic electron-phonon coupling, Appendix~\ref{App:Phonons} gives estimates of the values of $\mathsf{F}$ and $\mathsf{P}$ calculated using material coupling constants, 
densities and speeds of sound, assuming simplified geometries for phonon modes of the 1D InAs wire and bulk SiN substrate ~\cite{Stace:PRL:2005}. We find $\mathsf{F_{piezo}}=0.85$ and $\mathsf{P_{piezo}}=0.16$, which are in reasonable order-of-magnitude agreement with values shown inset in \fig{fig:gain}(a).

The high  phonon temperature assumed in \fig{fig:gain}(a) ($T= 3$K) may arise from ohmic heating from current flowing through the DQD, 
as estimated in Ref.~\onlinecite{Liu:PRL:2014}. Increasing the phonon bath temperature in the theory leads to an increase in the loss rate, with commensurate improvement in the agreement with the experimental data for $\epsilon_q/\omega_r < -3$, as shown in Appendix~\ref{App:HighTemp}.

\section{  Discussion and Conclusions \label{Sec:Discussion} }
	
\subsection*{Effective Rates}

\begin{figure*}[t]
	\begin{center}
		\begin{tabular}{c|cc|cc}
		 & $\kappa_{-,4}/g^{2}\:@\:P_{e}=0$ & $\kappa_{+,4}/g^{2}\:@\:P_{e}=0$ & $\kappa_{-,4}/g^{2}\:@\:P_{e}=1$ & $\kappa_{+,4}/g^{2}\:@\:P_{e}=1$\\
		\hline
		\begin{turn}{90}$T=0$\end{turn}& 	
			\includegraphics[width=.45\columnwidth]{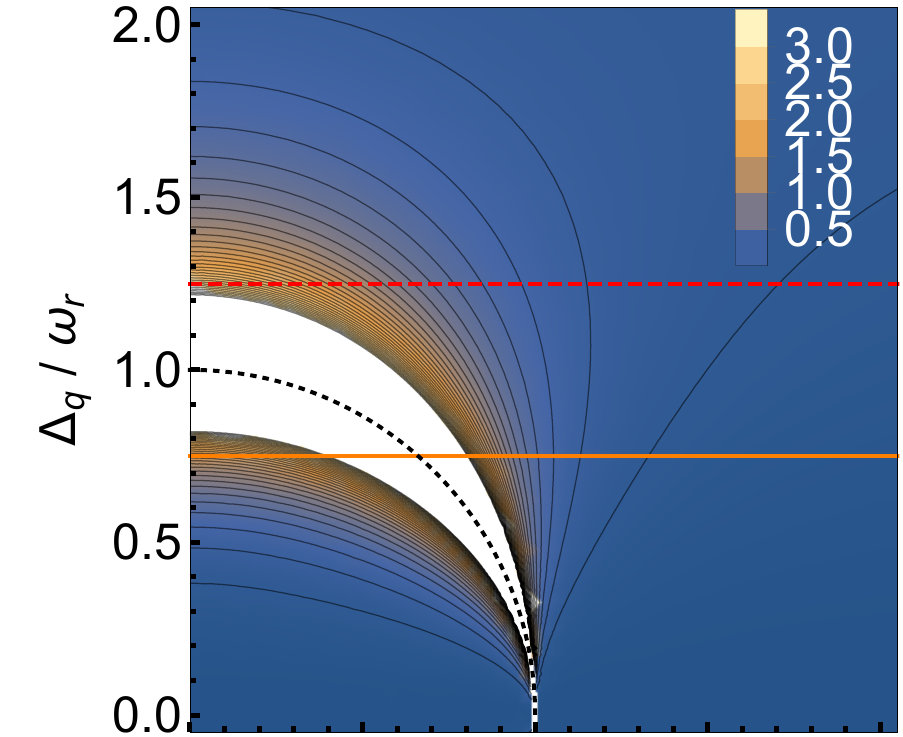}  &
			\includegraphics[width=.45\columnwidth]{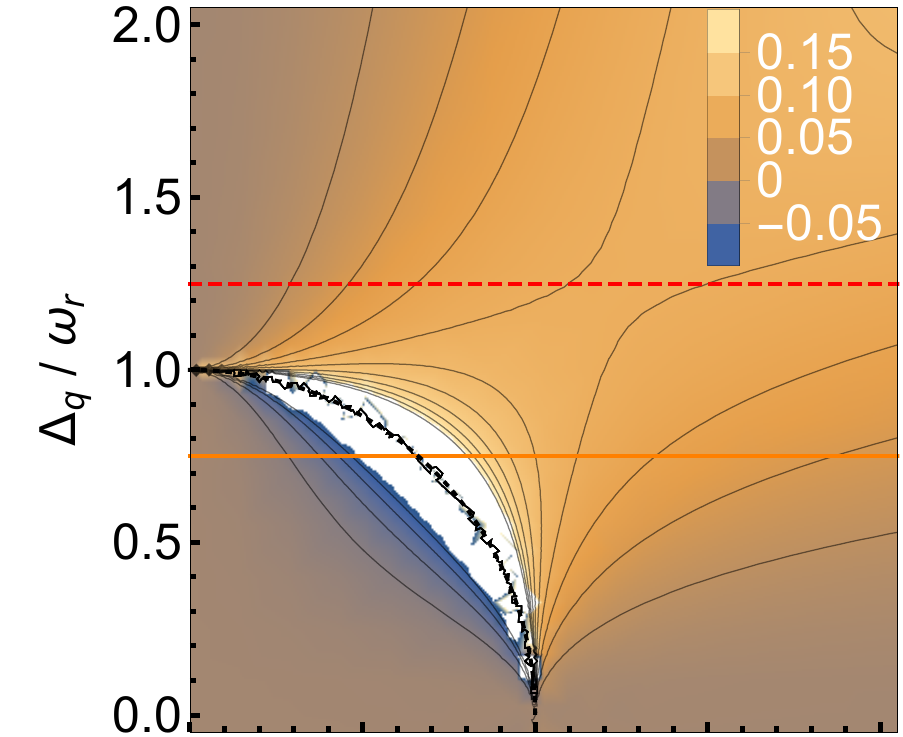} & 
			\includegraphics[width=.45\columnwidth]{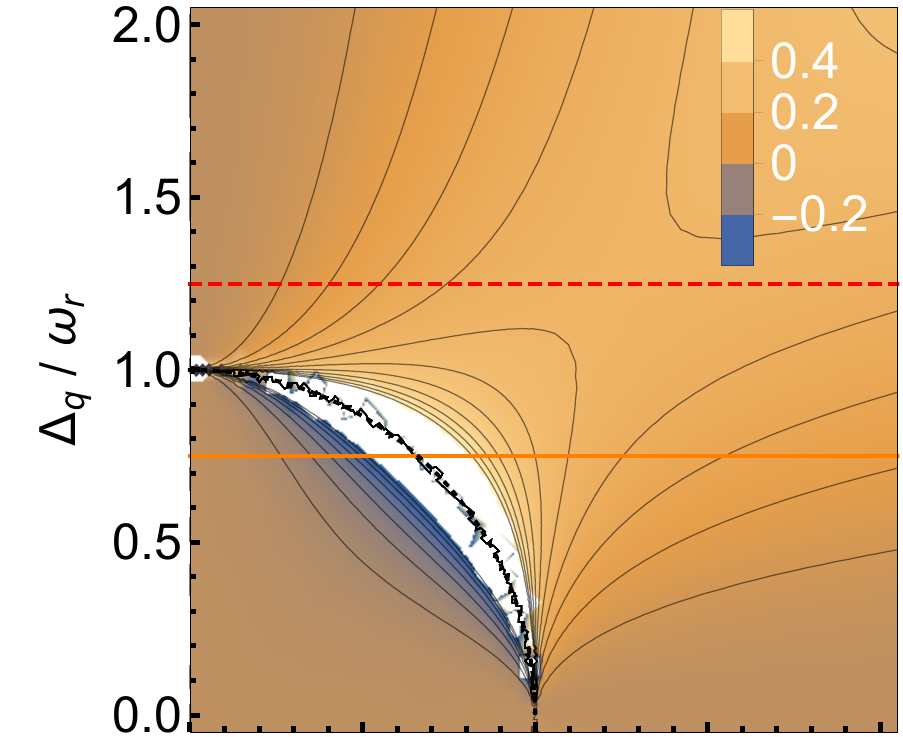} & 	
			\includegraphics[width=.45\columnwidth]{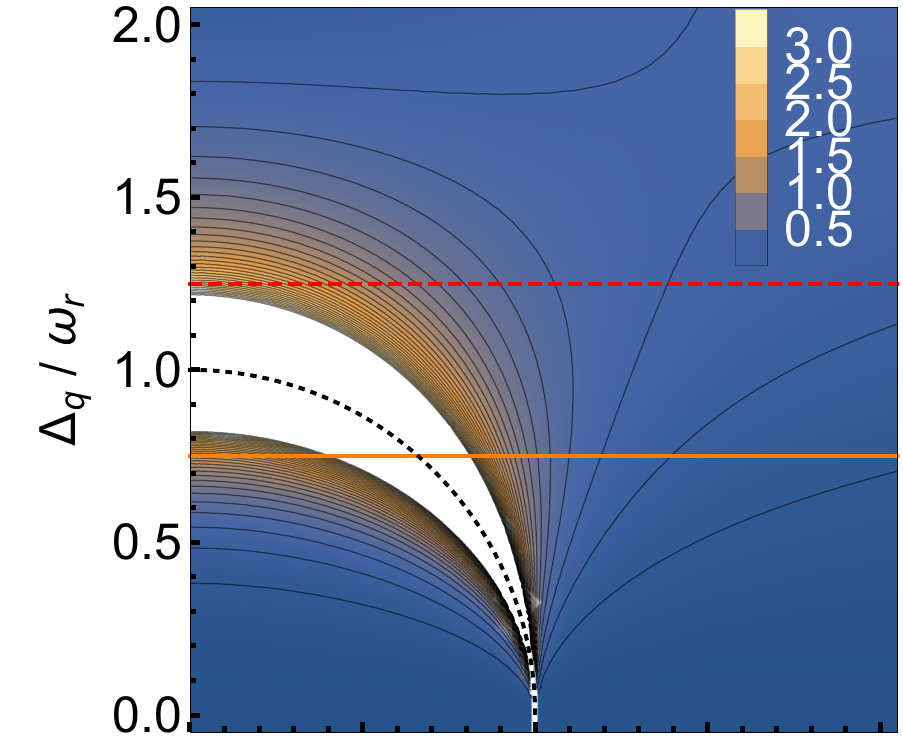}\\
			 & 
			\includegraphics[width=.45\columnwidth]{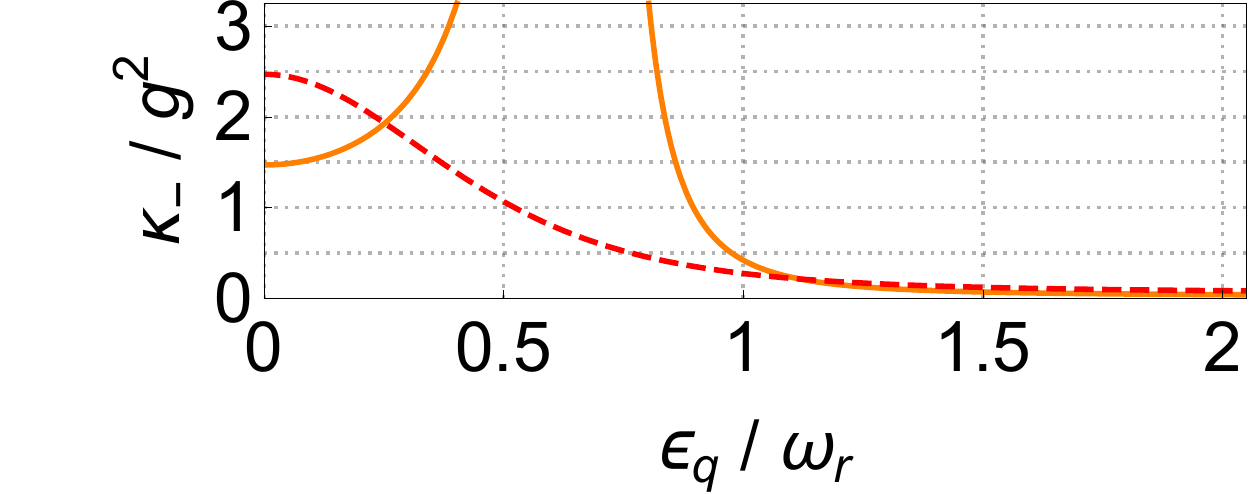} & 
			\includegraphics[width=.45\columnwidth]{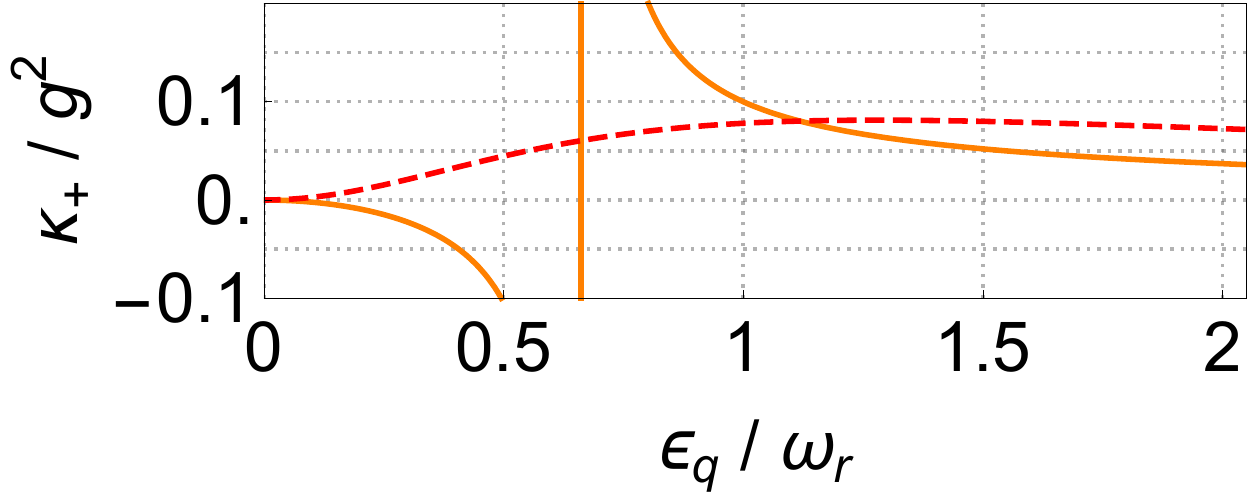} & 
			\includegraphics[width=.45\columnwidth]{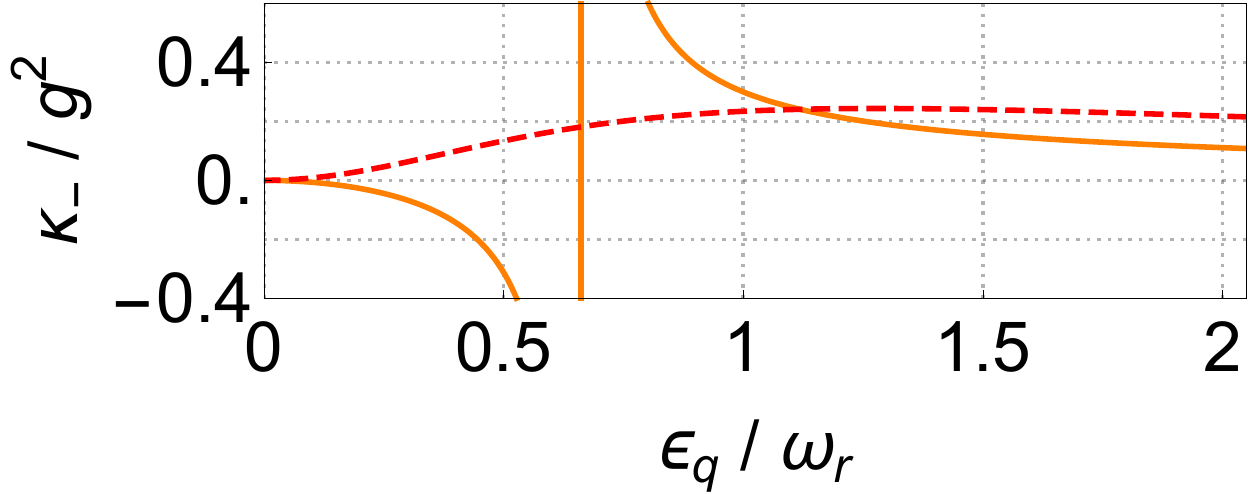} & 
			\includegraphics[width=.45\columnwidth]{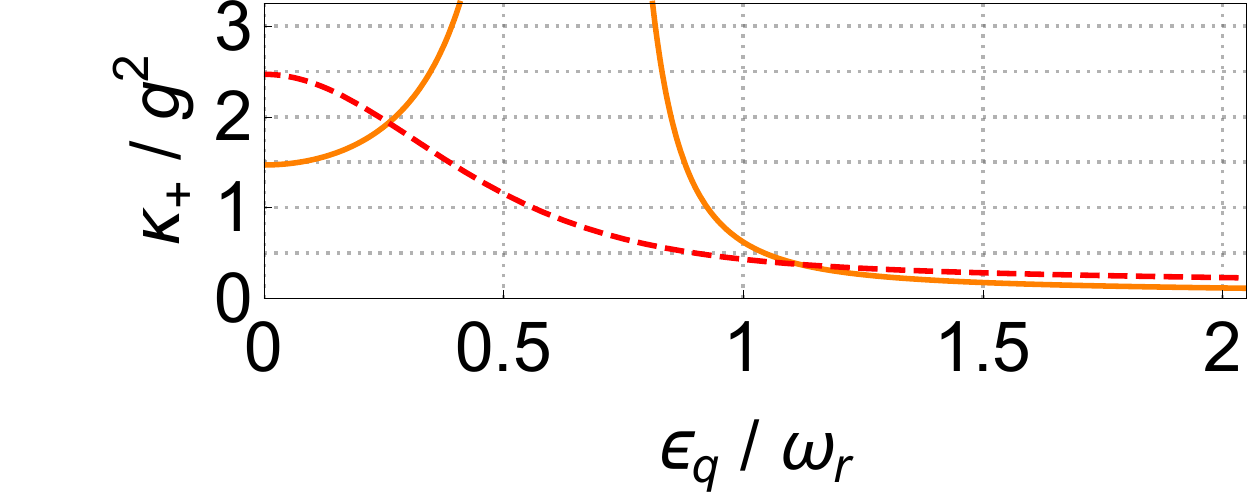} \\
			\hline
		\begin{turn}{90}$T=0.2 \omega_{r}$\end{turn}&
			\includegraphics[width=.45\columnwidth]{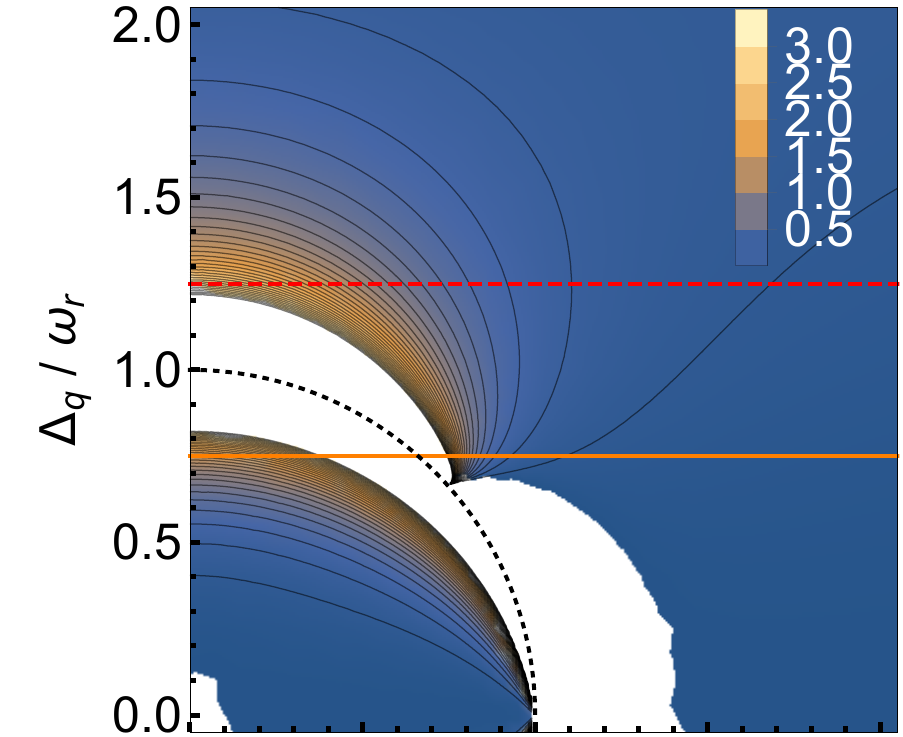} &
			\includegraphics[width=.45\columnwidth]{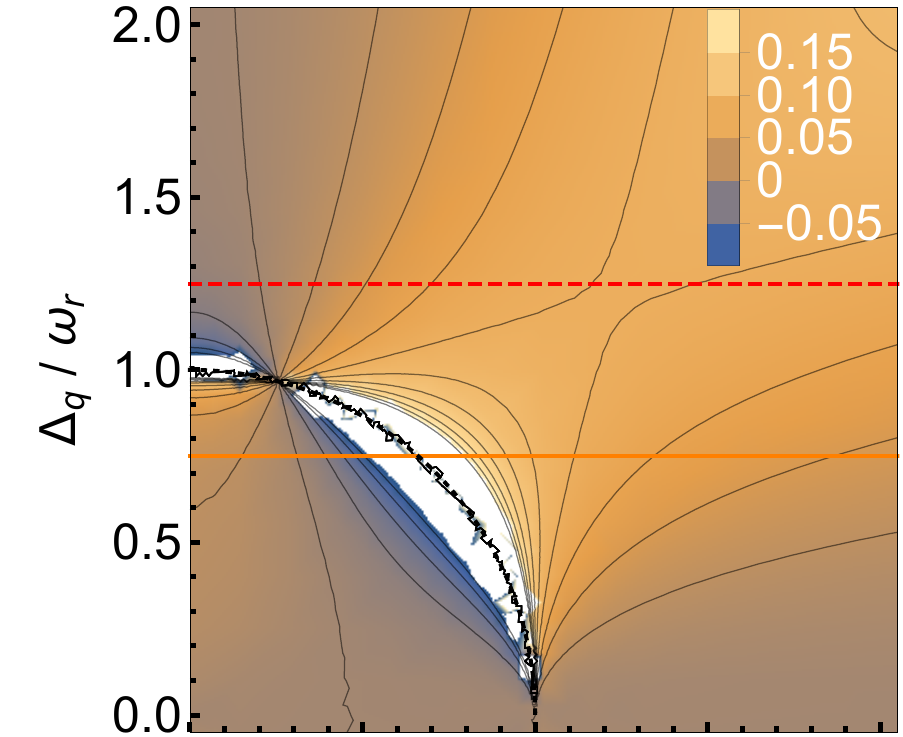} &
			\includegraphics[width=.45\columnwidth]{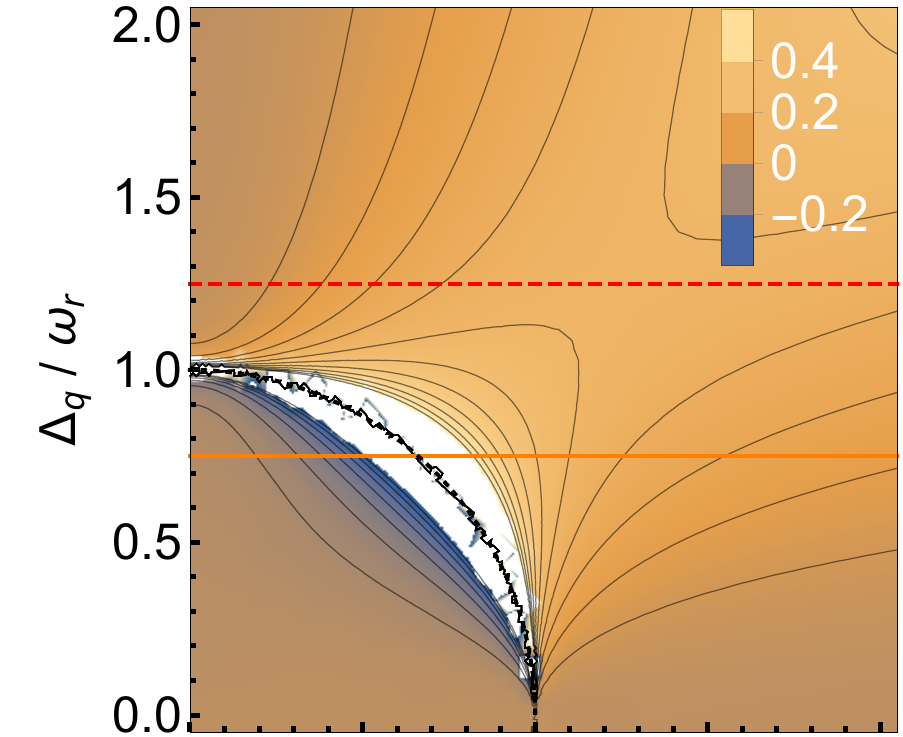} &
			\includegraphics[width=.45\columnwidth]{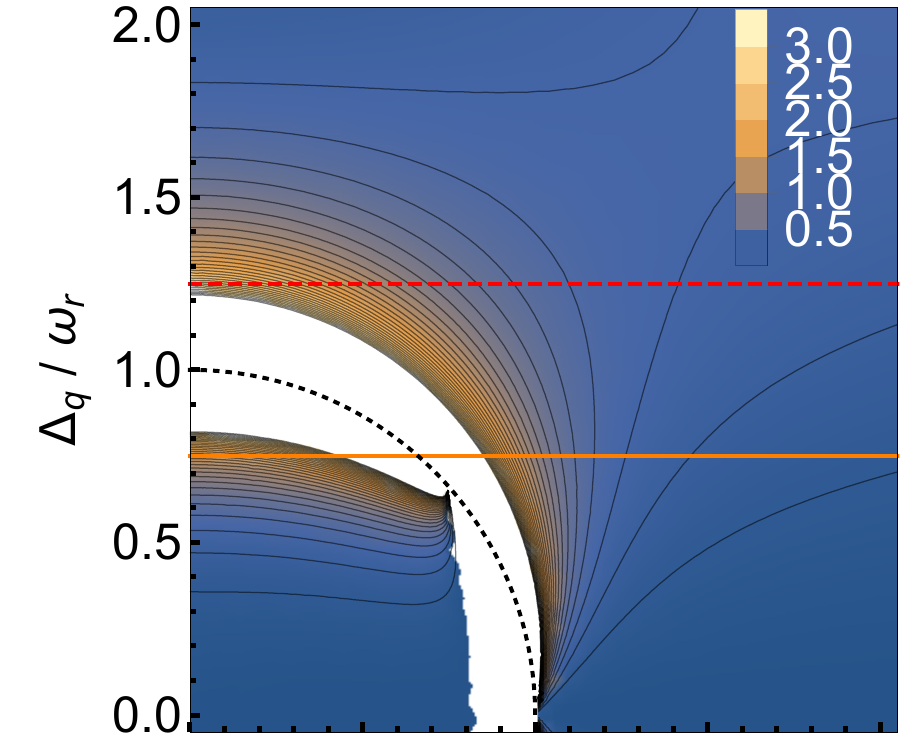}\\
			&
			\includegraphics[width=.45\columnwidth]{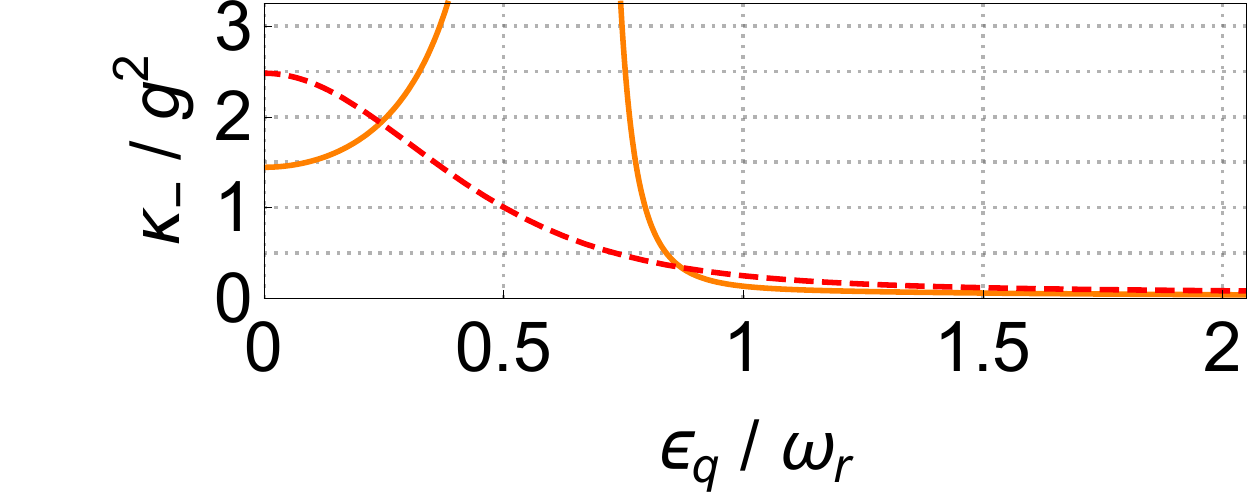} &
			\includegraphics[width=.45\columnwidth]{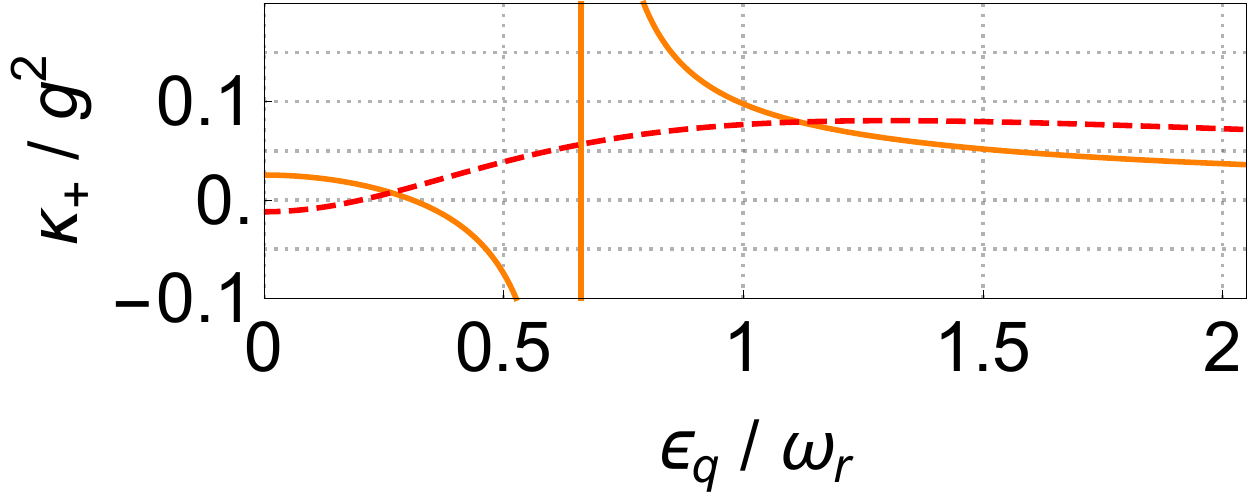} &
			\includegraphics[width=.45\columnwidth]{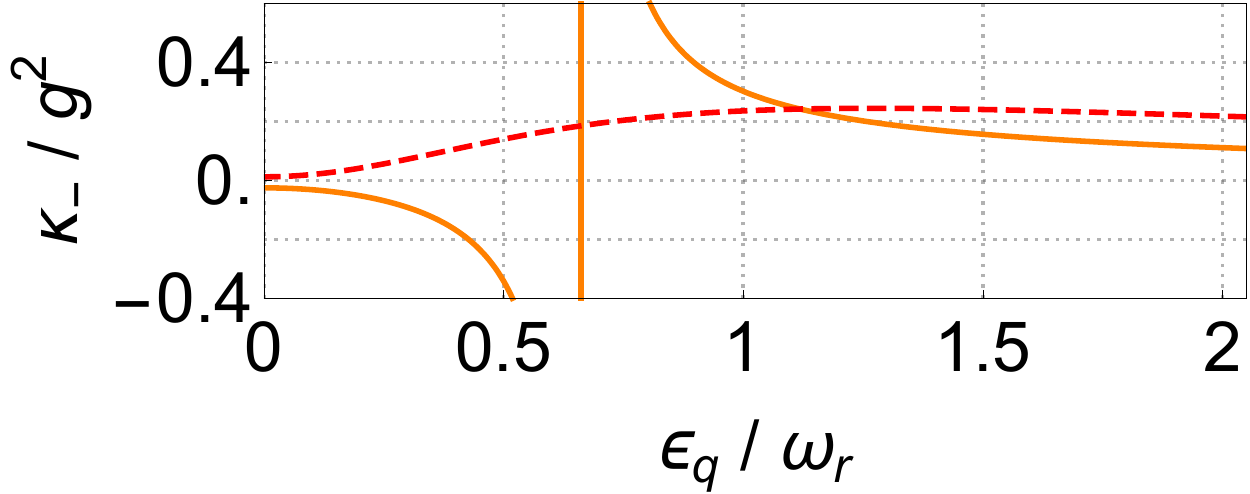}&
			\includegraphics[width=.45\columnwidth]{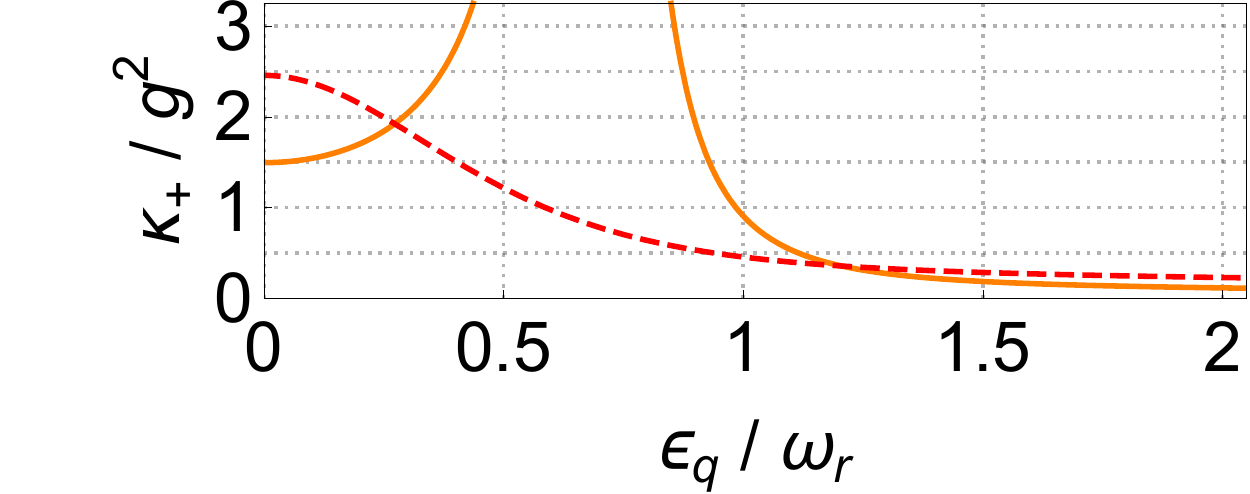}\\
		\end{tabular} 
		\caption{(Color online) Plots of the effective dissipative resonator rates $\kappa_{\pm,4}$ from~\eqn{eq:kappa} for an ohmic bath spectral function $C(\omega) = \omega (n_{\text{th}}(\omega)+\theta(\omega))$ 
			with the bosonic thermal occupation factor $n_{\text{th}} (\omega)=1/(\ee^{\beta\omega}-1)$, $\beta=1/k_{B}T$.
			We show $\kappa_{\pm,4}$ for the qubit in its groundstate, $P_{e}= 0$ (left), and in its excited state, $P_{e} = 1$ (right), for the two situations of zero temperature, $T=0$ (top row), and non-zero temperature, $T=0.2 \omega_{r}$ (bottom row).
			Horizontal lines in the contour plots indicate the values of $\Delta_{q}$ used in the line plots below.
			The thin dotted line in the contour plots indicates the resonance condition $\omega_{q} = \omega_{r}$.
		}
		\label{fig:kappa}
	\end{center}
\end{figure*}
	
	The large number of rates that appear in~\eqn{eq:4thAll} is too great to analyse individually in detail, however the mean-field approximation embodied in~\eqn{eqn:separation} 
	helps to distil significant combinations of the rates in~\eqn{eq:4thAll} into effective decay rates on the resonator,~\eqn{eq:kappa}, and the qubit,~\eqn{eq:gamma}, separately.  
	
	Figure~\ref{fig:kappa} (and Figs.~\ref{fig:gamma} and~\ref{fig:gamma0} in the appendix) show plots of the effective rates appearing in the qubit and resonator steady-state equations,~\eqn{eq:resME} and~\eqn{eq:qDotME}, in the region $\omega_{q} \sim \omega_{r}$. 
	Many of the rates show divergences at resonance $\omega_{q}= \omega_{r}$ or become negative in the vicinity of very small detuning.
	This behaviour is not unexpected, since our perturbative series relies on a small qubit-resonator coupling compared to the qubit and resonator detuning, $g<|\omega_{q} - \omega_{r}|$.
	Proximity to resonance between qubit and resonator will therefore make it necessary to take higher orders in perturbation series into account to accurately describe the system dynamics.
	
	The effective resonator decay rates $\kappa_{\pm,4}$ are shown in~\fig{fig:kappa} in various limits of $P_e$ and for an ohmic bath spectral function, \mbox{$C(\omega) = \omega (n_{\text{th}} + \theta(\omega))$}, both for zero and non-zero bath temperature $T$.
	For a given choice of qubit parameters, and temperature, the `true' value of the effective rates $\kappa_{\pm,4}$ is a convex combination of the limits plotted.  
	For $\omega_q>\omega_r$, these effective resonator rates are all positive, so that the Lindblad operators generate well-defined completely positive (CP) maps on the resonator. 
	
	As discussed before, the displaced resonator mode $\tilde a$ is un-driven such that its steady-state will be a thermal state at an effective temperature $T_{r}$ defined by the ratio of its relaxation and excitation rates $\kappa_{+} / \kappa_{-} = e^{-\omega_{r} / (k_{B} T_{r})}$.
	Due to the effective coupling of resonator and qubit steady-states through the fourth-order dissipative rates, this effective temperature will be influenced by both the original resonator temperature as well as the temperature of the qubit environment.
	The effective resonator temperature depends specifically on the qubit steady-state population $P_{e}$ as seen from Eq.~\eqref{eq:kappa} and Fig.~\ref{fig:kappa}. 
	In the case of population inversion in the qubit, the effective resonator excitation rate $\kappa_{+,4}$ can dominate over $\kappa_{-,4}$.
	
	In appendix~\ref{sec:EffectiveRates} we show plots of the correlated qubit relaxation, excitation and dephasing rates from Eq.~\eqref{eq:gamma}. Their behaviour is similar to that of the effective resonator rates shown in Fig.~\ref{fig:kappa}.

\subsection*{Negative Lindblad superoperators}
	For $\omega_q\lesssim\omega_r$, there are regimes in which the $\kappa_{\pm,4}$ can become negative. The situation is similar for the effective qubit rates $\gamma_{\downarrow/\uparrow,4}$ and $\gamma_{\varphi,4}$, as shown in appendix~\ref{sec:EffectiveRates}.
	More generally,~\eqn{eq:4thAll} has a number of Lindblad superoperators with explicitly negative coefficients.  
	This can be seen e.g. in the prefactors of $\diss{\sigma_{z} a\hc a}\bar\rho$ and $\diss{\sigma_{z} + \sigma_{z}a\hc a}\bar \rho $, which have a relative minus sign relating them, so that one coefficient will always be negative.  
	This mathematical phenomenon arises naturally  at fourth-order perturbation theory, and has previously been 
	connected to the description of explicitly non-Markovian dynamics~\cite{Breuer:PRL:2009, Breuer:EPL:2009, Laine:PRA:2010, Ribeiro:PRB:2015} and its unravelling into a quantum jump description~\cite{Laine:JPB:2012}.  	
	In contrast, the second order rates in~\eqn{eq:QDRates0} are always explicitly positive.	
	
	The appearance of negative coefficients to Lindblad superoperators is controversial, since there is no longer a guarantee that the resulting map is completely positive.  
		
	There are a number of possible resolutions to this mathematical problem.  
	Firstly, the dominant fourth-order terms (i.e.\ those in \eqn{eq:4thME}) are indeed explicitly positive, so that they do correspond to well-behaved Lindblad superoperators.   
	That is, the most significant physics is represented by a CP map.
	
	Secondly, the main objective of this paper is to find steady states of the propagator, so that we do not necessarily require complete positivity.  Rather, we simply require that the steady state is a positive operator, as is indeed the case in the situation described here, illustrated by the red dotted curve in Fig.~\ref{fig:gain}.  
	However, this argument is unsatisfactory in general, since the dissipators can be interpreted in a dynamical equation, just as \mbox{$\mathcal L_{2}\bar \rho$} in \eqn{eqn:L2} appears in the dissipative dynamics of the quantum optical master equation.
	
	Thirdly, the explicitly negative signs in~\eqn{eq:4thAll} are associated with somewhat anomalous dissipators such as \mbox{$\diss{\sigma_{z} + \sigma_{z}a\hc a}\bar \rho $}, which on their own contain products of four resonator creation and annihilation operators 
	(e.g.\ the expansion of this Lindblad dissipator includes the operator product \mbox{$\sigma_{z}a\hc a \bar \rho \sigma_{z}a\hc a$}).  These dissipators all arise from the diagonalisation procedure described in Appendix~\ref{app:nondiag}. 
	In fact, such dissipators always appear in combination with another dissipator (in this case \mbox{$\diss{\sigma_{z} a\hc a}\bar\rho$}), such that quartic products of resonator operators cancel.  
	 At higher orders of perturbation theory, we expect to see  dissipators that are truly quartic in the resonator operators appearing, and these would renormalise the negative coefficients seen at the order we evaluate.  
	
	 This suggests that the apparent negative signs appearing in~\eqn{eq:4thAll} are a consequence of the truncation of an infinite power series; including higher-order terms should renormalise the dissipative rate at each order.  
	  As such, we speculate that resumming a large class of higher order Keldysh diagrams will lead to well defined Lindblad superoperators with positive prefactors.
	  
	 This can be understood as an analogue to the series-expansion of a unitary operator, $U=e^{-i H}$ generated by a Hamiltonian $H$. While the zeroth and infinite order expansion are both explicitly unitary (corresponding to the physical requirement that probability is conserved), 
	 this is not  guaranteed to hold for any finite order (e.g.\ the approximation $ U\approx1-i H$ is not generically a unitary operator), so that  conservation of probability is strictly violated.  Nevertheless, linear-response theory explicitly makes these kinds of approximations, to good effect.
	  By analogy, it is not surprising that the fourth-order Keldysh-Lindblad master equation we have derived includes terms that are non-CP, which are nevertheless useful for describing the system.
	 
\subsection*{Keldysh-Lindblad synthesis}

	Equation~\eqref{eq:4thAll} is the principal theory result of this paper.  	
	It represents the synthesis of the  Keldysh and Lindblad formalisms, which both are widely used in mesoscopic physics, but with relatively little overlap in the community of practitioners who deploy them; 
	the Lindblad formalism has been carried over from the quantum optics literature, whilst the Keldysh formalism has its roots in the non-equilibrium condensed matter literature.  
	
	Equation~\eq{eq:4thAll} contains a large class of  Lindblad superoperators  that go beyond the standard dissipators in the quantum-optical master equation.  
	The first six terms in~\eqn{eq:4thAll} are dominant in a typical experimental situation, and these have a natural interpretation as qubit mediated energy exchange between the resonator and the bath.  
	The first four terms have previously been identified using a polaron transformation~\cite{Gullans:PRL:2015}.  
	The  terms $\gamma_{\varphi +} \diss{\sigma_{z} a\hc}\bar\rho + \gamma_{\varphi - }\diss{\sigma_{z}a} \bar\rho$, which correspond to a dephasing assisted energy exchange, represent the most significant new contribution presented here.

\subsection*{Conclusions}
	
	We conclude that the new dephasing-mediated gain and loss Lindblad superoperators in~\eqn{eq:4thME} account for substantial additional loss observed in recent experiments.  
	These terms were derived using the Keldysh diagrammatic techniques, and arise at the same order of perturbation theory as other terms previously derived using a polaron transformation.  
	Including all additional dissipators derived at the same order in perturbation theory does not qualitatively change the results apart from a rescaling of numerical parameters. 
	
	Synthesising Lindblad and Keldysh techniques to derive higher-order dissipative terms is thus a powerful approach to a consistent, quantitative understanding of quantum phenomena in mesoscopic systems.  
	Lifting the simplifying mean-field approximation to study the effects of correlations between the DQD and resonator will be the subject of future work.
	 
	 This paper opens a variety of avenues to explore. 
	 We have not evaluated the fourth-order dispersive terms, nor the fourth order pure bath (multi-phonon) terms.  
	 In some regimes, two-phonon emission may be significant, particularly if phononic engineering is used to suppress the spectral density at dominant single-phonon frequencies (e.g.\ by designing phononic band-gaps).  
	 In such a situation, two (or more) phonon emission rates can be calculated within the Keldysh formalism, and collected into Lindblad superoperators.  

	In future work we intend to analyse the dynamics and steady-state of the fully correlated equation,~\eqn{eq:4thAll}, without the simplifying mean-field approximations,~\eqn{eqn:separation}.
	Also, the master equation~\eqn{eq:4thAll} naturally lends itself to an input-output treatment allowing the calculation of qubit and resonator output correlation functions.

\acknowledgements{We thank J.~H.~Cole, A.~Doherty, M.~Marthaler, J.~Petta, A.~Shnirman, J.~Taylor, J.~Vaccaro, and H.~Wiseman for discussions.}

\bibliography{../QDotPhonon}
\clearpage

\appendix
\makeatletter

\section{Keldysh technique}\label{App:K}
\renewcommand{\thefigure}{\ref{App:K}\arabic{figure}}
\renewcommand{\thetable}{\ref{App:K}\arabic{table}}
\setcounter{figure}{0}
\setcounter{table}{0}

\subsection{Derivation of Keldysh master equation\label{App:Keldysh}}

	We describe a system interacting with an environment using Keldysh diagrams. The basic Hamiltonian has the form
	\begin{align}
		\H = \H_{0} + \H_{I} ,
	\end{align}
	where $\H_{0}= \H_{S} + \H_{B}$ describes the unperturbed system $\H_{S}$ and the environment $\H_{B}$, and $\H_{I}$ contains all the interactions terms and is considered small.
	In the following we will employ an interaction picture with respect to $\H_{I}$, where operators now evolve in time with respect to $\H_{0}$ while states evolve according to $\H_{I}$. 
	
	In the Schr\"odinger picture, the time evolution of the systems density matrix elements is 
	\begin{align}
		\rho_{\s'\s}^{(S)}(t) 
		 =& \text{Tr}\{ \rho^{(S)}(t) {\ket{\s}}{\bra{\s'}} \}\nn\\
			=& \text{Tr}\big\{ U_{0} U_{0}\hc \rho^{(S)}(t) U_{0} U_{0}\hc {\ket{\s}}{\bra{\s'}} \big\} \nn\\
			=& \text{Tr}\big\{ \rho^{(I)}(t) U_{0}\hc {\ket{\s}}{\bra{\s'}} U_{0} \big\} \nn\\
			=& \text{Tr}\big\{ U_{I} \rho^{(I)}(t_{0}) U_{I}\hc U_{0}\hc {\ket{\s}}{\bra{\s'}} U_{0} \big\} \nn\\
			=& \sum_{\bar \s' \bar \s} \rho_{\bar \s' \bar \s}(t_{0})\bra{\bar \s} \text{Tr}_{B}\big\{ \rho_{B}  U_{I}\hc P_{\s'\s}(t) U_{I}  \big\}\ket{\bar \s'}\nn\\
			\equiv&\sum_{\bar \s \bar \s'} \rho_{\bar \s' \bar \s}(t_{0}) \Pi_{\bar \s' \bar \s \rightarrow \s' \s}(t_{0}, t),	\label{eq:rhot}
	\end{align}
	where we have traced out the bath degrees of freedom, defined the projector unto system states $P_{\s\s'} = {\ket{\s}}{\bra{\s'}}$ with
	\begin{align}
		P_{\s\s'}(t) = U_{0}(t,t_{0})\hc {\ket{\s}}{\bra{\s'}} U_{0}(t,t_{0})\,,
	\end{align}
	where $U_{0}\equiv U_{0}(t,t_{0}) = \ee^{-\ii \H_{0} (t-t_{0})}$,
	and we have defined the forward and backwards time-evolution operators in the interaction picture by
	\begin{align}
		U_I\equiv U_{I}(t,t_{0}) &= T \exp\left\{ -\ii \int_{t_{0}}^{t} dt_{1}\: \H_{I}(t_{1}) \right\} , \nn\\
		U_I^\dagger\equiv  U_{I}(t_{0},t) &= \bar T \exp\left\{ \ii \int_{t_{0}}^{t} dt_{1}\: \H_{I}(t_{1}) \right\}, 
		\label{eq:TimeEvolve}
	\end{align}
	where $T$ indicates time-ordering (later times left) and $\bar T$ anti-time-ordering (later times right). 
	The interaction term $\H_{I}(t)$ is itself in the interaction picture, i.e., evolves in time according to the unperturbed Hamiltonian $\H_{0}$ as
	\begin{align}
		\H_{I}(t) = U_{0}(t_{0}, t) \H_{I} U_{0}(t,t_{0}) \,.
	\end{align}
	We have also assumed in \eqn{eq:rhot} that the system and bath factorise at the initial time, i.e.\
	\begin{align}
		\rho(t_{0}) = \rho_{B}(t_{0}) \otimes \sum_{\bar \s \bar \s'} \rho_{\bar \s' \bar \s} (t_{0}) \ket{\bar \s'}\bra{\bar \s}\,.
	\end{align}
	\eqn{eq:rhot} implicitly defines the density matrix propagator $\Pi(t_{0},t)$.
	 
	We now expand the time-evolution operators $U_I^\dagger$ and $U_I$ on both sides of $P_{\s\s'}(t)$ in \eqn{eq:rhot}  in a perturbative expansion in powers of the coupling coefficients $g$ and $\beta_j$ that appear in $H_I$. 
	At a given order, $m$, of perturbation theory in this expansion, the propagator $\Pi$ acts on  the density matrix by application of $m$ instances of $H_I$, which appear on the left and right of $\rho$ in every possible configuration.  
	
	\begin{figure*}[!t!]
		\begin{center}
			\includegraphics[width=.95\textwidth]{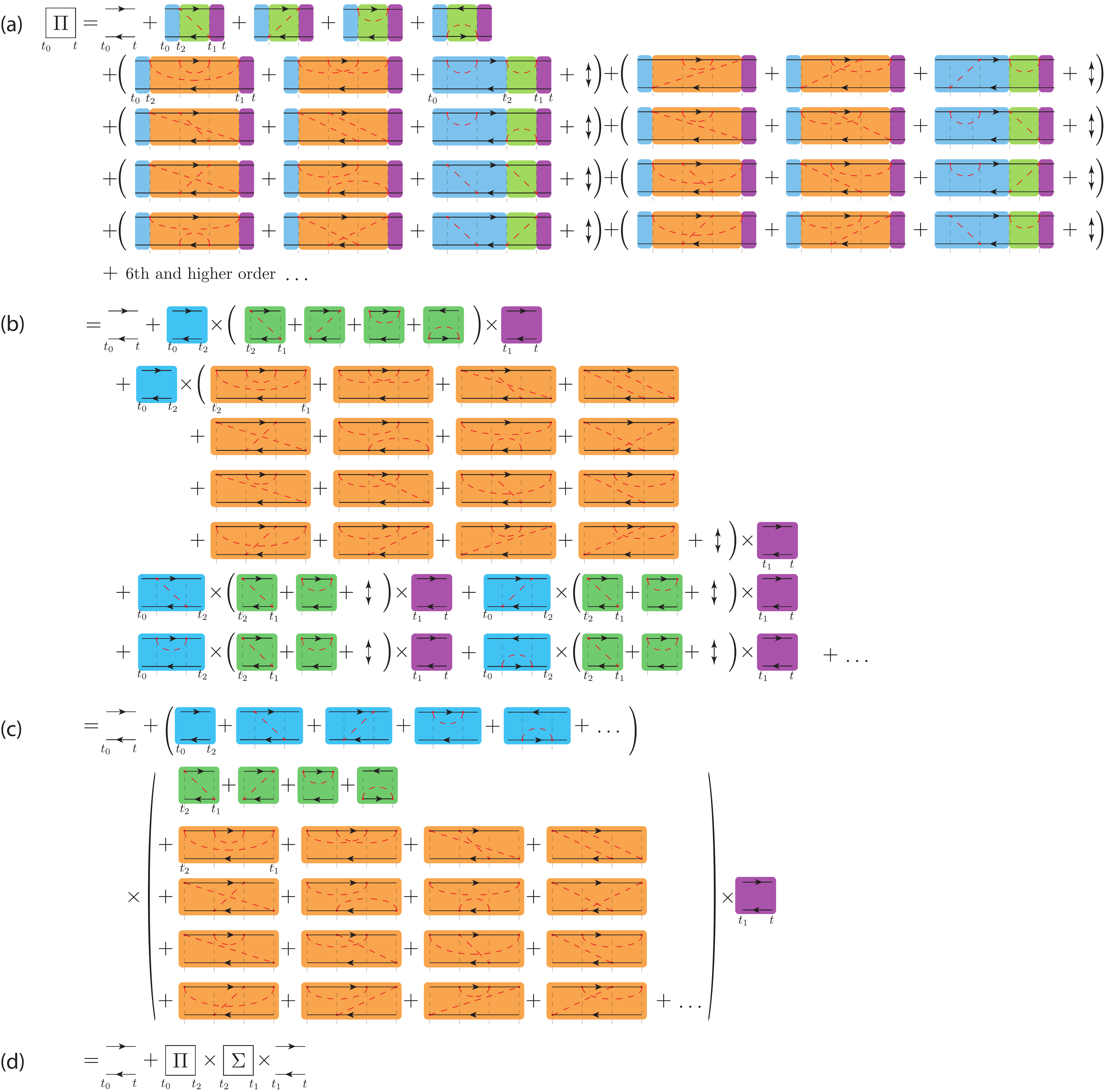}
			\caption{
			(Colour online) (a) Illustration of the density matrix propagator $\Pi$ in diagrammatic perturbation theory for evolution from the initial time $t_0$ to the final time $t$.  Other times depicted are dummy integration variables, satisfying $t_0<...<t_2<t_1<t$. 
			Lines with arrows correspond to periods of free evolution. Dots denote the interaction Hamiltonian, $H_I(t_j)$ acting at  time $t_j$.  $\updownarrow$ denotes swapping of upper and lower vertices. Background colours are used to help track terms in the following panels.   
			Only terms at even orders in perturbation theory are shown, and dashed lines denote contractions of bath operators into two-time correlation functions.
			In our convention, $\Pi$ acts on the density matrix from the right, so that $\rho(t)=\rho(t_0)\Pi_{t_0\rightarrow t}$, as in \eqn{eq:rhoDyson}. 
			(b) Diagrammatic reordering of the  $\Pi$. Here we factor each diagram into products of irreducible diagrams, which are those that cannot be divided by vertical lines without cutting any contractions.
			(c) Diagrammatic expression of the Dyson equation. 
			The first term in the product (blue diagrams) is the propagator, $\Pi_{t_0\rightarrow t_2}$, and the second term (green and orange diagrams) is the self energy, $\Sigma_{t_2\rightarrow t_1}$, composed of irreducible self-energy diagrams, followed by a period of free evolution. 
			(d) Dyson equation expressed algebraically, which implicitly defines the self energy $\Sigma$ in terms of irreducible diagrams.
			}
			\label{fig:DysonEquationAll}
		\end{center}
	\end{figure*}
	
	This expansion has a graphical representation in the form of Keldysh diagrams, shown in Fig.~\ref{fig:DysonEquationAll}.  
	The action of $H_I$ in \eqn{eq:rhot} acting on the left (which come from $U_I^\dagger$) are drawn as vertices on the lower branch of the diagram; those acting on the right (which come from $U_I$) are drawn as vertices on the upper branch. 
	These diagrams are operator-valued, schematically acting on $\rho(t_0)$ from the right. 
	Diagrams up to fourth-order are shown in Fig.~\ref{fig:DysonEquationAll}a.
	Here we have already made use of Wick's theorem to express multi-operator bath correlation functions as sums of two-time correlation functions, 
	indicated by red dashed lines in the diagrams of Fig.~\ref{fig:DysonEquationAll}.
	
	\fig{fig:DysonEquationAll}(b) and (c) show a diagrammatic refactoring of the expansion, to derive the Dyson equation, which relates the propagator, $\Pi$, to the self-energy $\Sigma$, and the free evolution $\Pi^0$ through
	\begin{equation}
		\Pi(t_0,t)=\Pi^0(t_0,t)+\int dt_2 dt_1\Pi(t_0,t_2)\Sigma(t_2,t_1)\Pi^0(t_1,t).
	\end{equation}
	Diagrammatically, the self-energy $\Sigma$ is given by the sum over all irreducible diagrams, shown in~\fig{fig:DysonEquationAll}(c). 
	
	When we substitute the the Dyson equation (shown in Fig.~\ref{fig:DysonEquationAll}d) into the propagator in \eqn{eq:rhot}, we find the time-dependent evolution of the density matrix is given by
	\begin{widetext}
	\begin{align}
		\rho_{\s'\s}(t) =& \sum_{\bar \s \bar \s'} \rho_{\bar \s' \bar \s}(t_{0})\: \Pi^{0}_{\bar \s' \bar \s \rightarrow \s' \s}(t_{0}, t) 
		+ \sum_{\bar \s \bar \s'}\sum_{\s_{1}\s'_{1}}\sum_{\s_{2}\s'_{2}} \int_{t_{0}}^{t}dt_{1}\: \int_{t_{0}}^{t_{1}} dt_{2}\: \rho_{\bar\s' \bar\s} (t_{0}) \Pi_{\bar \s' \bar \s\rightarrow \s'_{1}\s_{1}}(t_{0},t_{2}) 
				\Sigma_{\s'_{1}\s_{1}\rightarrow \s'_{2}\s_{2}}(t_{2},t_{1}) \Pi^{0}_{\s'_{2}\s_{2}\rightarrow \s'\s}(t_{1},t) \nn\\
			=\,&\rho_{ \s' \s}(t_{0})\: \ee^{-\ii (E_{\s'} - E_{\s}) (t-t_{0})} 
			+ \sum_{\bar \s \bar \s'} \int_{t_{0}}^{t}dt_{1}\: \int_{t_{0}}^{t_{1}} dt_{2}\: \rho_{\bar \s' \bar \s} (t_{2}) \Sigma_{\bar \s' \bar \s\rightarrow \s'\s}(t_{2},t_{1}) \ee^{-\ii (E_{\s'} - E_{\s}) (t-t_{1})} ,
		\label{eq:rhoDyson}
	\end{align}
	\end{widetext}
	where in the second line the free propagator has been explicitly written as
	\begin{align}
		\Pi^{0}_{\bar\s' \bar\s \rightarrow\s'\s}(t,t_{0}) = \delta_{\bar\s,\s}\delta_{\bar\s', \s'} \ee^{-\ii (E_{\s'} - E_{\s})(t-t_{0})}.
	\end{align}
	
	Taking the time-derivate of \eqn{eq:rhoDyson} finally leads to the Keldysh master equation
	\begin{align}
		\partial_{t} \rho_{\s'\s}(t) = \,& -\ii (E_{\s'} - E_{\s}) \rho_{\s'\s}(t) \nn\\
			&{}+ \sum_{\bar\s \bar\s'} \int_{t_{0}}^{t} dt_{1}\: \rho_{\bar\s' \bar\s}(t_{1}) \Sigma_{\bar\s' \bar\s \rightarrow\s'\s} (t_{1},t),
		\label{eq:MasterKeldysh}
	\end{align}
	where $E_{\s}$ is the energy of system state $\ket\s$.
	From this equation we see that the self-energy superoperators $\Sigma_{\bar\s' \bar\s \rightarrow\s'\s}$ generate the residual dynamics arising from the interaction hamiltonian, and are responsible for both dispersive and dissipative effects at higher order.
	Note that the self-energy in \eqn{eq:MasterKeldysh} is still formally exact, since it contains all orders of perturbation theory. 

	Noting that $\rho(t) = \sum_{\s'\s}\rho_{\s'\s}(t) \ket{\s'}\bra{\s}$, we  write the master equation for the full density matrix 
	\begin{align}
		\dot\rho^{(S)}(t) = -\ii [{H_{0}},{\rho^{(S)}(t)}] + \int_{t_{0}}^{t} dt_{1}\: \rho^{(S)}(t_{1}) \Sigma(t_{1},t) ,
		\label{eq:MasterKeldysh2}
	\end{align}
	where the self-energy superoperator $\Sigma(t_{1}, t)$ acts on the density matrix from the right. This convention ensures that the Keldysh diagrams we draw later are correctly understood as acting with earlier times on the left.  In the interaction picture, we have 
	\begin{align}
		\dot\rho^{(I)}(t) =  \int_{t_{0}}^{t} dt_{1}\: \rho^{(I)}(t_{1}) \Sigma(t_{1},t).
		\label{eq:MasterKeldyshIPa}
	\end{align}
	In what follows, we will work in the interaction picture, and drop the notation ${}^{(I)}$.
	
	The detailed relationship between the last term in each of \eqn{eq:MasterKeldysh} and \eqn{eq:MasterKeldysh2} is established by defining $\Sigma_{\bar\s' \bar\s \rightarrow\s'\s} (t_{1},t) \equiv\sum_j  \bra{\bar\s}\op L_{j} {\ket\s}{\bra{\s'}}\op R_{j} {\ket{\bar\s'}}$ 
	for some set of time-dependent operators $\op L_{j}$, $\op R_{j}$. 
	Then
	\begin{align}
			\sum_{\s'\s}& \sum_{\bar\s' \bar\s}  \rho_{\bar\s' \bar\s}(t_{1}) \sum_{j}\bra{\bar\s}\op L_{j} {\ket\s}{\bra{\s'}}\op R_{j} {\ket{\bar\s'}}{\ket{\s'}}{\bra{\s}}, \nn\\
			&= \sum_{\s'\s} \sum_{\bar\s' \bar\s} \sum_{j} {\bra{\s'}}\op R_{j} {\ket{\bar\s'}}\rho_{\bar\s' \bar\s}(t_{1}) \bra{\bar\s}\op L_{j} {\ket\s}{\ket{\s'}}{\bra{\s}}, \nn\\
			&= \sum_{\s'\s} \sum_{j} {\ket{\s'}}{\bra{\s'}}\op R_{j} \rho(t_{1})\op L_{j} {\ket\s}{\bra{\s}} ,\nn\\
			&= \sum_{j}  \op R_{j} \rho(t_{1})\op L_{j}, \nn\\
			&\equiv \rho(t_{1})\Sigma(t_{1},t).
	\end{align}
	We note that in order to preserve hermiticity of $\rho$,  the self-energy operator is self-adjoint.
 
	In what follows, we evaluate the self-energy operator up to fourth order, and collect terms in order to express the master equation in Lindblad form.

\subsection{Note on environmental spectral functions\label{App:SpectralFunction}}

	As in most treatments of open quantum systems, the bath spectral function plays a critical role in capturing the open-systems dynamics. 
	In anticipation of later derivations, here we define the bath spectral function through the Fourier transformation of the two-time correlation function of the bath coupling operators as
	\begin{align}
		C(\omega) = \frac12 \int_{-\infty}^{\infty} dt\: \ee^{\ii \omega (t_{1} - t_{2})} \mean{\hat X(t_{1}) \hat X(t_{2})} ,
	\end{align}
	such that in all integral expression we can replace
	\begin{align}
		\mean{\op X(t_{1}) \op X(t_{2})} = \frac1\pi \int_{-\infty}^{\infty}d\omega\: \ee^{-\ii \omega(t_{1} - t_{2})}\: C(\omega) .
		\label{eq:BathSpectral}
	\end{align}
	Here, $\hat X(t)$ is the bath coupling operator in the interaction picture, which in our case, for bosonic phonon modes, takes the form
	\begin{align}
		\hat X(t) = \sum_{j} \beta_{j} \big( b_{j}\: \ee^{-\ii\omega_{j }t} + b_{j}\hc\: \ee^{\ii \omega_{j} t} \big) .
	\end{align}
	Defining the coupling operator spectral density $J(\omega)$ in the continuum limit of the bath modes with spectral density $\nu(\omega)$ as
	\begin{align}
		\sum_{j} \beta_{j}^{2} \rightarrow \frac1\pi \int_{0}^{\infty}d\omega\: \beta^{2}(\omega) \nu(\omega) =  \frac1\pi \int_{0}^{\infty}d\omega\: J(\omega) ,
	\end{align}
	and assuming the bath modes to be in thermal equilibrium with a bose distribution, $\langle{b_j\hc b_j}\rangle = n_{\text{th}}(\omega_j) = \frac{1}{\ee^{\beta\omega_j} - 1}$, we can explicitly evaluate the spectral function as
	\begin{align}
		C(\omega) &= \frac12 J(|\omega|) \left( \coth{\frac{\abs\omega}{2T}} + \text{sign}(\omega) \right), \nn\\
			&=  \frac12 J(\omega) \l( n_{\text{th}} (\omega) + \theta(\omega) \r),
	\end{align}
	which follows the detailed balance relation $C(\omega) / C(-\omega) = \ee^{\beta \omega}$ with the thermal factor $\beta = 1/k_{B}T$.

\subsection{Laplace space integration\label{App:Laplace}}

	To evaluate the self-energy, we transform the master equation into Laplace space. 
	In Laplace space, the time-integration is particularly simple, as all necessary time-integrals take the form of convolutions. 
	Additionally, we will focus our evaluation on finding the steady-state density matrix of the system. 
	We make use of the technique developed in Ref.~\onlinecite{Stace:PRL:2013}, which enables one in general to find dynamical steady-states, i.e. steady-states with a residual, periodic time-dependence.
	Here, we will only be interested in the time-independent steady-state. 
	In the following we provide a short overview of the technique, more details can be found in Ref.~\onlinecite{Stace:PRL:2013}.

	We first transform the Keldysh master equation into Laplace space. Since the time-integrals can be written as multi-product convolutions, their solution in Laplace space is particularly simple. 
	Defining 
	\begin{align}
		\text{LT} \l\{ \ee^{-\ii \omega t} \r\} &= \frac{1}{s+\ii \omega} ,\nn\\
		\text{LT} \l\{ f(t) \ee^{-\ii \omega t} \r\} &= f_{s+\ii \omega} ,
	\end{align}
	where $f_{s}$ is the Laplace transformed function $f(t)$, the Keldysh master-equation, \eqn{eq:MasterKeldyshIPa}, in Laplace space becomes 
	\begin{align}
		s\rho_{s} -\rho(0) &= \int d\omega \sum_{j} \xi_{j} \op o_{j,1} \rho_{s+\ii \omega_{j}}  \op o_{j,2} \prod_{k=1}^{N_{j}-1} \frac{1}{s+\ii \omega_{k}} ,\nn\\
			&= \sum_{\omega_j} \rho_{s+i\omega_j} \bar\Sigma_s
		\label{eq:masterKeldyshLaplace}
	\end{align}
	where $\rho_{s}$ is the Laplace transformed density matrix and the index $j$ enumerates the different diagrams in the self-energy.  
	The operators $\op o_{j}$ depend on the specific diagram and the index $k$ is dependent on the order $N_{j}$ of perturbation theory of the $j$-th term.
	The coefficients $\xi_{j}$ contain bath spectral functions $C(\omega)$ each of which comes with an integral over the Fourier frequency $\omega$, c.f. App.~\ref{App:SpectralFunction}.
	Finally, the frequencies $\omega_{k}$ are linear combinations of system frequencies from the time-evolution of the system operators, and Fourier frequencies from the definition of the bath spectral function.  
	Examples of this expression for specific diagrams are evaluated in \eqn{eqnexample1} and \eqn{eqn:degen}.
		
	To solve \eqn{eq:MasterKeldyshIPa} for the dynamical steady state of the system, the following ansatz has been shown to be useful \cite{Stace:PRL:2013}
	\begin{align}
		\bar\rho(t) = \bar\rho_{0} + \sum_{j} \bar\rho_{j} \ee^{-\ii \omega_{j} t} ,
	\end{align}
	where $\bar\rho_{0}$ is the time-independent part of the steady-state and the $\bar\rho_{j}$ components of the steady-state show residual oscillations at frequencies $\omega_{j}$.
	In Laplace space, this becomes
	\begin{align}
		\rho_{s} &=\int_{t_0}^\infty dt\, e^{-s t} \rho(t),\nn\\
		&= \frac{\bar\rho_{0}}{s} + \sum_{j} \frac{\bar\rho_{j}}{s+\ii \omega_{j}} .
		\label{eq:RhoAnsatz}
	\end{align}
	Inserting the ansatz into the Keldysh master equation and restricting ourselves to the time-independent steady-state $\bar \rho_{0}$ leads to a transcendental equation for $\bar\rho_{0}$.
	Comparing residuals on both sides of the equation finally leads to the time-averaged steady-state equation of the form
	\begin{align}
		0 = \lim_{s\rightarrow 0^+} \int d\omega \sum_{j} \xi_{j} \op o_{j,1} \bar\rho_{0}  \op o_{j,2} \prod_{k=1}^{N_{j}-1} \frac{1}{s+\ii \omega_{k}} ,
	\end{align} 
	where the limit $\lim_{s\rightarrow 0+}$ corresponds to calculating the residuals of the right hand-terms.
	Here we have assumed that the zero pole on the rhs of Eq.~\eqref{eq:masterKeldyshLaplace} originates with the density matrix $\rho_{s}$, and not from a term in the self-energy $\Sigma_{s}$, as is the case in all diagrams considered below.
	More details can be found in Ref.~\onlinecite{Stace:PRL:2013}.

	When evaluating the residue of the zero frequency pole of the master equation, the following limit will become important
	\begin{align}
		\lim_{s \rightarrow 0^+} \frac{1}{s-\ii \omega} =& \lim_{s\rightarrow 0^+} \frac{s+\ii \omega}{s^{2} + \omega^{2}} \nn\\ =&\, \pi\, \delta(\omega) + \frac{\ii}{\omega} ,\label{eqn:sslimit}
	\end{align}
	where the $\delta$-function allows us to restrict the frequency integral over the bath spectral functions $C(\omega)$ and the imaginary parts may contribute as frequency shifts to the effective Hamiltonian.
	
	\subsection{Example  second order Keldysh diagrams\label{App:2ndK}}
	
	\subsubsection{Second order dissipative Keldysh diagrams\label{App:2ndKdiss}}

	\begin{figure}[t]
		\begin{center}
			\includegraphics[width=.6\columnwidth]{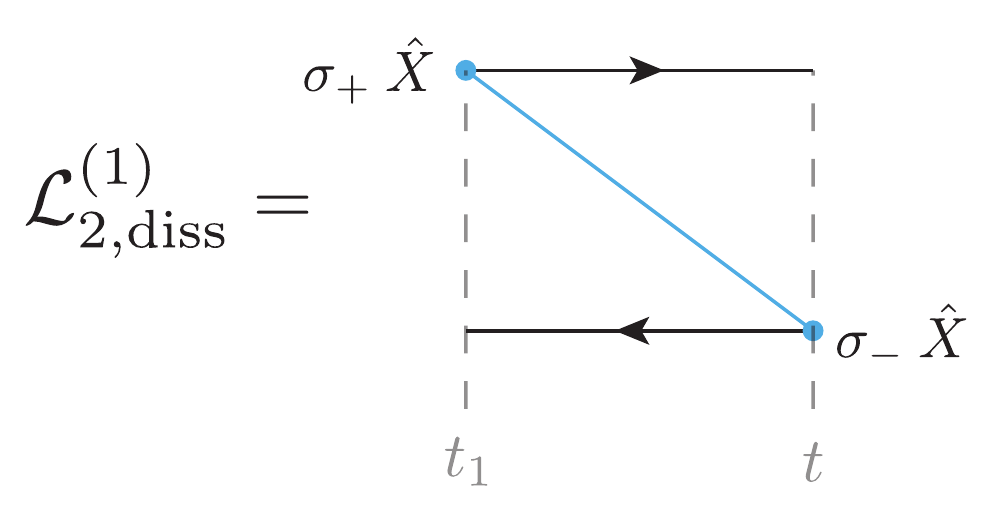}
			\caption{(Color online) One specific self-energy diagram in 2nd order, with each of the vertex operators specified. 
				Here both vertices correspond to coupling to bath operators $\op X$, and their contraction is indicated by the blue solid line. 
				This diagram contributes to the the standard qubit decay process $\sim \diss{\sigma_{-}} \rho$ as detailed in the text.
			}
			\label{fig:2ndExample}
		\end{center}
	\end{figure}

	To give a practical example of the general steps, we will show three specific examples. First we consider a simple dissipative diagram in second order shown in Fig.~\ref{fig:2ndExample}. 
	Here we use the convention that when collecting the terms, we trace the lines in the direction of the arrows, starting from the bottom right. Also, remember that the density matrix at the earliest dummy integration time (here $t_{1}$) multiplies this diagram from the left. 
	Evaluating this diagram leads to the the master equation term
	\begin{align}
		\L_{2,\text{diss}}^{(1)} \rho(t_1) \equiv \text{Tr}_{\text B}\l\{ \int_{t_{0}}^{t} dt_{1}\: \sigma_{-}(t) \op X(t) \, \rho(t_{1})  \, \sigma_{+}(t_{1}) \op X(t_{1}) \r\} ,
	\end{align}
	where for the sake of readability we neglected the prefactors $\sim \sin{\theta}$.

	We now assume separability of the bath and system, $\rho(t')=\rho_q(t')\otimes \rho_B$ over the interval $t_0<t'<t_1$ (n.b.\ the first vertex at the dummy time variable $t_1$), 
	and that the bath is in its thermal steady state throughout this time interval. 
	This gives
	\begin{align}
		\L_{2,\text{diss}}^{(1)} \rho(t_1) = \int_{t_{0}}^{t} dt_{1}\: \sigma_{-}(t)  \rho_q(t_{1}) \sigma_{+}(t_{1})\mean{\op X(t_{1}) \op X(t)} ,
	\end{align}
	which, after inserting the definition of the bath spectral function, \eqn{eq:BathSpectral} becomes
	\begin{align}
		\L_{2,\text{diss}}^{(1)} \rho(t_1) = \frac1\pi \int d\omega\: C(\omega) \int_{t_{0}}^{t} dt_{1}\: \sigma_{-} \rho_q(t_{1}) \sigma_{+} \ee^{\ii (\omega - \omega_{q}) (t-t_{1})} .
	\end{align}
	Here we have expressed the result in the form of temporal convolution, 
	\begin{align*}
		\l( f \star g \r) = \int dt_{1}\: f(t_{1}) g(t-t_{1})
	\end{align*}
	with $f(t) = \rho_q(t)$ and $g(t) = \exp{\l( \ii (\omega-\omega_{q}) t \r)}$.
	Transforming into Laplace space allows us to evaluate the time integral and we get
	\begin{align}
		\L_{2,\text{diss}}^{(1)}  \bar\rho_s&=\int_{t_0}^\infty dt \,e^{-s t}   \L_{2,\text{diss}}^{(1)} \rho(t_1)\nn\\
			&= \frac1\pi \int d\omega\: C(\omega) \frac{1}{s-\ii (\omega-\omega_{q})} \sigma_{-} \bar\rho_{s} \sigma_{+} ,
		\label{eqnexample1}
	\end{align} 
	with the Laplace transformed density matrix $\rho_{s}$.
	Inserting the Ansatz, \eqn{eq:RhoAnsatz}, and taking the steady-state limit using \eqn{eqn:sslimit}  leaves us with 
	\begin{align}
		\L_{2,\text{diss}}^{(1)} \bar\rho_0 &= \int d\omega\: C(\omega) \delta(\omega-\omega_{q}) \sigma_{-} \bar\rho_0 \sigma_{+} \nn\\
			&= C(\omega_{q})  \sigma_{-} \bar\rho_0 \sigma_{+} +i(\text{Dispersive part}),
	\end{align}
	which is one of the terms appearing in the dissipator describing qubit relaxation $\gamma_{\downarrow,0} \diss{\sigma_{-}}\bar\rho$, cf. \eqn{eq:QDRates0}.
	Here  the imaginary term in  limit $s\rightarrow 0^+$ will contribute only as a small renormalization of Hamiltonian parameters, which we neglect.
		
	\subsubsection{Second order dispersive Keldysh diagrams\label{App:2ndKdisp}}
		
\begin{figure}[t]
		\begin{center}
			\includegraphics[width=.6\columnwidth]{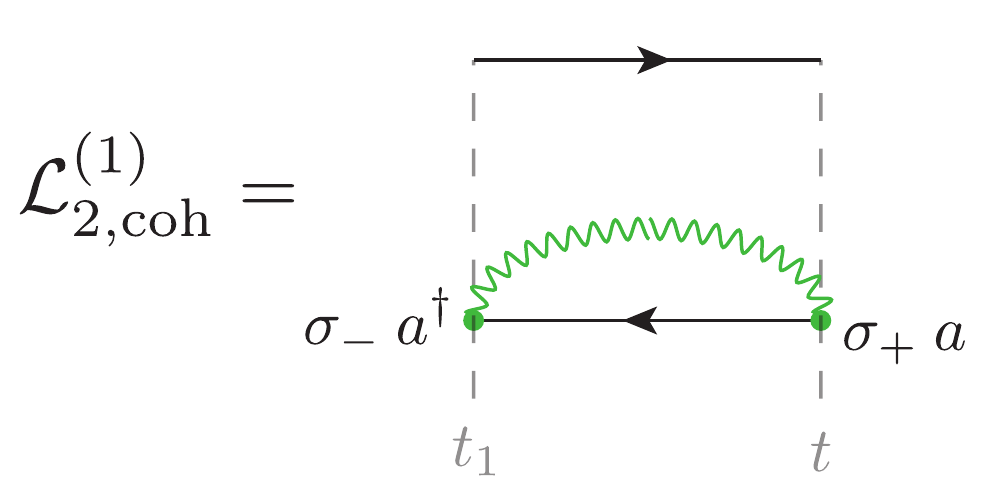}
			\caption{(Color online) Another example of a self-energy diagram in 2nd order, with each of the vertex operators specified. 
				Here both vertices correspond to coherent coupling between the qubit and the resonator, as indicated by the wavy green line. 
			}
			\label{fig:2ndExample2}
		\end{center}
	\end{figure}

	We now show another example in second order, illustrated by the diagram in Fig.~\ref{fig:2ndExample2}.
	Here, both vertices represent operators from the interaction between the qubit and the resonator. 
	The contribution to the Keldysh master equation takes the form
	\begin{align}
		\L_{2,\text{coh}}^{(1)}  \rho(t_1)= \int dt_{1}\: \sigma_{+} a \,\, \sigma_{-} a\hc\,\, \rho(t_{1})  \ee^{\ii (\omega_{q}- \omega_{r})(t-t_{1})} ,
	\end{align}
	where prefactors are ignored. The integral can again be evaluated in Laplace space to arrive at
	\begin{align}
		\L_{2,\text{coh}}^{(1)} \bar\rho_s = \frac{1}{s-\ii (\omega_{q} - \omega_{r})}\sigma_{+} a\,\, \sigma_{-} a\hc\,\, \bar\rho_{s}   .
	\end{align}
	Since there is no frequency integral now, the only relevant term in the limit $s\rightarrow 0^+$ is the imaginary contribution and we get the final result
	\begin{align}
		\L_{2,\text{coh}}^{(1)} \bar\rho_0 =\frac{\ii}{2(\omega_{q} - \omega_{r})} (\mathds 1 - \sigma_{z}) (a\hc a + \mathds 1) \bar\rho_0 ,
	\end{align}
	which forms part of the effective dispersive Hamiltonian $\H_{2}$, \eqn{eq:HDispersive}.

\subsection{4th order diagrams\label{App:4thDiag}}
	
	\begin{figure}[!t]
		\begin{center}
			\includegraphics[width=.9\columnwidth]{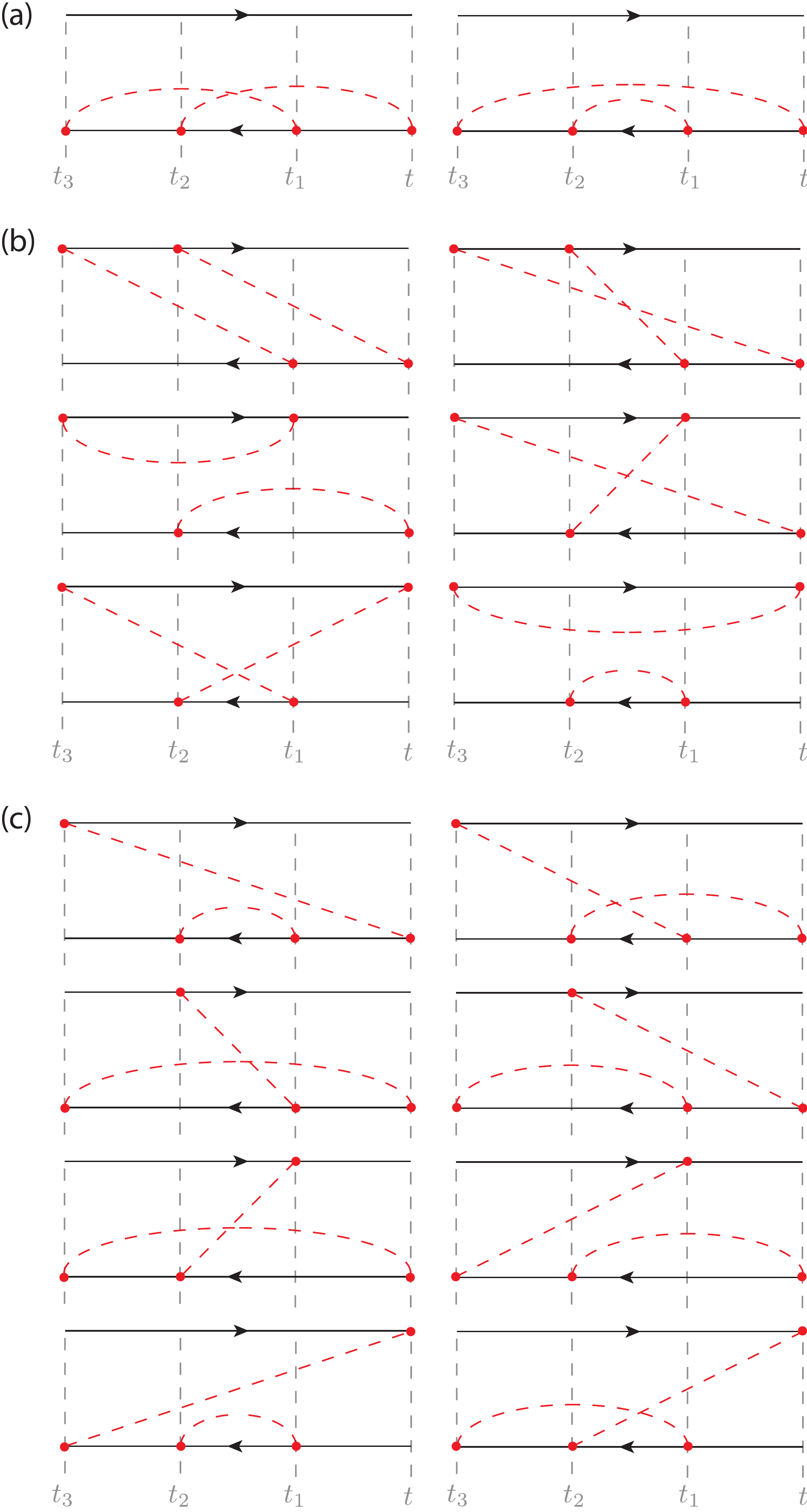}
			\caption{(Color online) (a) Irreducible Keldysh self-energy diagrams at 4th order representing terms where all four interaction vertices appear to the left of the density matrix. 
				 (b) Irreducible Keldysh self-energy diagrams at 4th order representing terms where two interaction vertices appear to the right of the density matrix and two vertices appear to the left. 
 				(c) Irreducible Keldysh self-energy diagrams at 4th order representing terms where three interaction vertices appear to the left of the density matrix and one vertex appears to the right. 
				The remaining irreducible diagrams at 4th order are formed by swapping indices from the upper to the lower line and vice-versa.
			}
			\label{fig:Keldysh4All}
		\end{center}
	\end{figure}

	At 4th order in the interaction, a total of 32 irreducible diagrams contribute to the Keldysh self-energy. 
	Half of these diagrams are schematically depicted in Figs.~\ref{fig:Keldysh4All}, from which the other 16 can be obtained by swapping all vertices between the upper and the lower line.
	
	\subsubsection{Fourth order dissipative Keldysh diagrams\label{App:4thKdiss}}
					
	\begin{figure}[t]
		\begin{center}
			\includegraphics[width=.8\columnwidth]{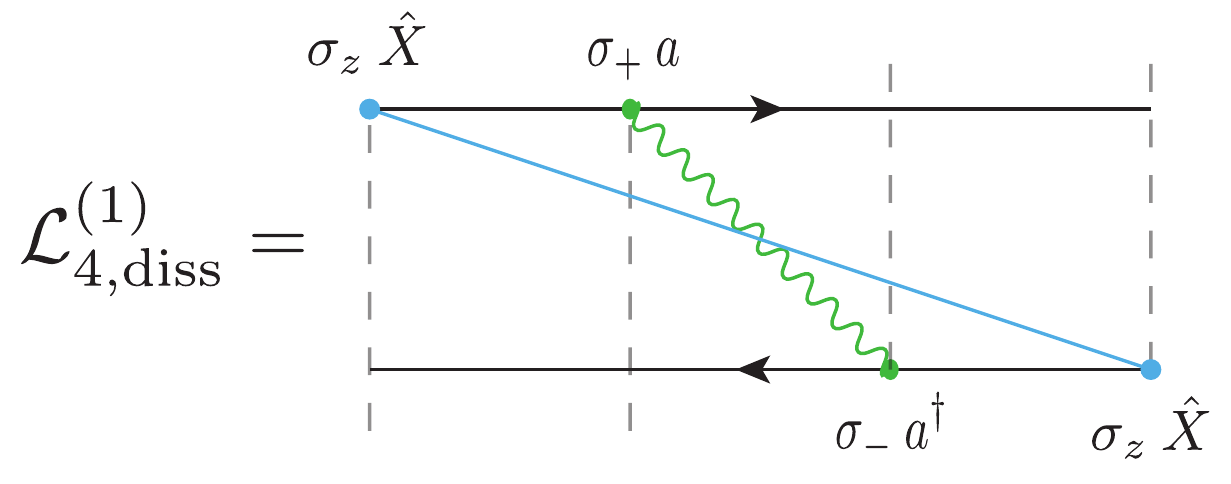}
			\caption{(Color online) Particular realisation of a self-energy diagram in 4th order, with each of the vertex operators specified. 
				Here one pair of operators represents coupling to the dissipative environment and is connected by a blue solid line. 
				The second pair of vertex operators describes coupling between the qubit and the resonator, and is connected by a wavy green line.
				This diagram contributes to the correlated decay process $\sim \diss{\sigma_{-} a\hc} \rho$.
			}
			\label{fig:4thExample1}
		\end{center}
	\end{figure}
		
	We consider one particular dissipative diagram in fourth order, Fig.~\ref{fig:4thExample1}.
	This diagram acts on the state at $t_3$, so that the corresponding term in the Keldysh master equation is
	\begin{align}
		\L_{4,\text{diss}}^{(1)} \rho(t_3) &= \int  dt_{1}dt_{2}dt_{3}\: \Bigl\{ \mean{X(t_{3}) X(t)}  \nn\\
			&\times\ee^{\ii \omega_{r} (t_{1}-t_{2})} \ee^{-\ii \omega_{q} (t_{1}-t_{2})} \sigma_{z} a\hc \sigma_{-} \: \rho(t_{3})\: \sigma_{z} a \sigma_{+}  \Bigr\} ,
	\end{align}
	where we again left off all prefactors for the vertex operators and already assumed the Born initial condition of separable system and bath density operators for $t_0<t'<t_3$.
	Inserting the definition of the bath spectral functions gives
	\begin{align}
		\L_{4,\text{diss}}^{(1)} \rho(t_3) =& -\frac1\pi \int d\omega\: C(\omega) \int dt_{1}dt_{2}dt_{3}\: \sigma_{-} a\hc \rho(t_{3}) a\sigma_{+}   \nn\\
			&\quad\quad \times \ee^{\ii \omega (t-t_{1})} \ee^{\ii (\omega-\omega_{q} + \omega_{r}) (t_{1} - t_{2})} \ee^{\ii \omega (t_{2}- t_{3})}.
	\end{align}
	This is a multi-product convolution over the integration variables. 
	In Laplace space the time integrals can then be evaluated trivially and we get
	\begin{align}
		\L_{4,\text{diss}}^{(1)} \bar\rho_s =& -\frac1\pi \int d\omega\: C(\omega) \sigma_{-} a\hc \bar \rho_{s} a \sigma_{+}  \nn\\
			&\quad\quad \times\frac{1}{(s-\ii \omega)^{2}} \frac{1}{s-\ii (\omega-\omega_{q}+\omega_{r})} .\label{eqn:degen}
	\end{align}
	We now make use of a partial fraction decomposition of the integral kernel, with the added subtlety that one of the poles in \eqn{eqn:degen} is degenerate. 
	We note that
	\begin{align}
		&\frac{1}{(s-\ii \omega)^{2} (s-\ii \omega_{1})} =\nn\\
		& \quad -\frac{1}{(s-\ii \omega_{1})(\omega_{1} - \omega)^{2}} - \frac{\partial}{\partial \omega}\frac{1}{(s-\ii \omega)(\omega - \omega_{1})},
	\end{align}
	where we used shorthand $\omega_{1} = \omega-\omega_{q}+\omega_{r}$.
	The second term deserves special attention since in the steady-state limit, $s\rightarrow 0^+$, it leads to
	\begin{align}
		\int d\omega C(\omega) \frac{\partial}{\partial \omega} &\frac{\delta(\omega)}{(\omega - \omega_{1})} = 
			 -\frac{C'(\omega)}{(\omega - \omega_{1})}  - \frac{C(\omega)}{(\omega - \omega_{1})^{2}} \Biggr|_{\omega= 0},
	\end{align}
	i.e., a degenerate pole leads to a derivative of the bath spectral function, as  has been observed before~\cite{Shnirman:2016}.
	Note that above we neglected the write the imaginary components appearing in the limiting procedure, since we will neglect those small Hamiltonian terms in the end.
	For the complete diagram Fig.~\ref{fig:4thExample1}, we thus get
	\begin{align}
		\L_{4,\text{diss}}^{(1)} \bar\rho =& \sigma_{-} a\hc \bar\rho a\sigma_{+} \nn\\
			& \times \Big\{ \frac{C(\omega_{q} - \omega_{r})}{(\omega_{q} - \omega_{r})^{2}} - \frac{C(0)}{(\omega_{q}- \omega_{r})^{2}} - \frac{C'(0)}{(\omega_{q} - \omega_{r})} \Biggr\} ,
		\label{eqn:fourexample}
	\end{align}
	which contributes to the process of correlated qubit decay and photon generation. Specifically the first term is a relevant contribution to the primary rate expression $\gamma_{\downarrow +}^{(\omega_q-\omega_r)}$, \eqn{eq:CorrelatedRates}. 
	We note in passing that the terms proportional to $C'(0)$ and $C(0)$ as well as any imaginary parts in \eqn{eqn:fourexample} ultimately cancel with terms arising from other, similar diagrams.
	
\subsection{Dissipators from master equation\label{App:Diss}}

	This section describes how to obtain the Lindblad dissipators from the Keldysh steady-state equation. After evaluating the self-energy  to a given order  can  write the Keldysh master equation in the form
	\begin{align}
		\dot {\bar\rho} = \op \sigma A_{\text{m}} \bar\rho \op \sigma\hc -\frac12 \left\{ \op \sigma\hc A_{\text{s}} \op \sigma \bar\rho + \bar\rho \op \sigma\hc A_{\text{e}} \op \sigma \right\}
		\label{eq:MEDiag}
	\end{align}
	where $\op \sigma$ is a row-vector of operators and the objects $A$ are coefficients matrices in the space of these operators.  For example, the scalar factor appearing in \eqn{eqn:fourexample} would contribute to $A_{\text{m}}$.
	
	Hermiticity of the  master equation requires that the matrices $A$ be symmetric.  Trace-conservation of $\rho$ also constrains the $A$'s, so that $A_{\text{s}} = A_{\text{e}}$.  
	These conditions imply that the RHS \eqn{eq:MEDiag} can always be expressed as a sum of hermitian commutators, and lindblad dissipators $\diss$, with real coefficients, $\gamma_j$.  
	They do not automatically constrain the $\gamma_j$ to be positive, which has implications for complete positivity of the evolution.  We comment on this in the main text.
	
	The vector $\op \sigma$ contains all operators that can be made up of combinations of terms in the interaction Hamiltonian $\H_{\text{I}}$ at the appropriate order.
	Specifically in the case of the Rabi model and with the interaction Hamiltonian given by \eqn{eq:HI} up to fourth order in perturbation theory, we have
	\begin{align}
		\op \sigma =\bigl( & \mathds{1}, \sigma_{z}, a\hc a, \sigma_{z} a\hc a, a\hc, \sigma_{z} a\hc, a, \sigma_{z} a, \nn\\
			& \sigma_{-}, \sigma_{+}, \sigma_{-}a\hc a, \sigma_{+} a\hc a, \sigma_{-} a\hc, \sigma_{+} a, \sigma_{-}a, \sigma_{+} a\hc \bigr) .
		\label{eq:opBasis}
	\end{align}
	As is obvious from our choice of parametrization of the matrices $A$, any diagonal term that appears in both $A_{\text{m}}$ as well as $A_{\text{s,e}}$, directly corresponds to a dissipator in the master equation. 
	The dissipative operator is then given by the eigenvector belonging to that particular diagonal element.
	As a simple example, collecting all terms in the RHS of \eqn{eq:MEDiag}  that are proportional to the  phonon spectral function at the difference frequency between the qubit and the resonator, $C(\omega_{q} -\omega_{r})$, we find 
	the matrices $A$ only have a single non-zero entry:
	\begin{align}
		&A_{\text{m,s,e}}^{(\omega_q-\omega_r)} \nn\\
			&= \text{diag}\l( 0,0,0,0,0,0,0,0,0,0,0,0,\gamma_{\downarrow+}^{(\omega_q-\omega_r)},0,0,0 \r) ,
	\end{align}
	in the basis defined by \eqn{eq:opBasis}.  The non-zero entry on the diagonal corresponds to the operator $\sigma_{-} a\hc$ in \eqn{eq:opBasis}, so we  immediately identify this part as the Lindblad dissipator 
	\begin{align}
		\op \sigma A_{\text m}^{(\omega_q-\omega_r)} \bar\rho \op\sigma\hc &-\frac12 \left\{ \op \sigma\hc A_{\text{s}}^{(\omega_q-\omega_r)} \op \sigma \bar\rho + \bar\rho \op \sigma\hc A_{\text{e}}^{(\omega_q-\omega_r)} \op \sigma \right\} \nn\\
			=&\gamma_{\downarrow+}^{(\omega_q-\omega_r)} \diss{\sigma_{-} a\hc}\bar\rho .\label{eqn:Awqwr}
	\end{align}
	Similarly simple diagonal forms can be found for all terms probing the phonon spectral function at frequencies $\pm (\omega_{q} - \omega_{r})$, $\pm (\omega_{q} + \omega_{r})$ and $\pm \omega_{r}$, leading to the 
	six correlated dissipators in \eqn{eq:4thME}.
	
	\subsubsection{Non-diagonal forms for $A$}\label{app:nondiag}

	In general the matrices $A$ in the RHS of \eqn{eq:MEDiag}  are not purely diagonal, and
	 some manipulation is required to bring them into a recognisable form in terms of sums of dissipators, $\diss$.
	Specifically in our case it is the cyclic property of the pauli matrices which makes this a non-trivial challenge.
	
	To diagonalise the matrices $A$, we make use of a particular property of composite dissipators, namely that for any two operators $\op o_{1}$ and $\op o_{2}$, we can write the dissipator of the sum of these operators as
	\begin{align}
		\diss{\op o_{1} + \op o_{2}}\rho =& \diss{\op o_{1}}\rho + \diss{\op o_{2}}\rho + \op o_{1} \rho \op o_{2}\hc + \op o_{2} \rho \op o_{1}\hc \nn\\
			&\hspace*{-10mm} - \frac12 \l( \op o_{1}\hc \op o_{2} \rho + \op o_{2}\hc \op o_{1} \rho + \rho \op o_{1}\hc \op o_{2} + \rho \op o_{2}\hc \op o_{1} \r) .
		\label{eq:DissSum}
	\end{align}
	The last six terms in this expression correspond to symmetric off-diagonal terms in the matrices $A$ in \eqn{eq:MEDiag}, whereas the first two terms are purely diagonal. 
	Thus, combinations of off-diagonal elements in the $A$'s can  be directly related to combinations of  dissipators using \eqn{eq:DissSum}:
	\begin{align}
	&\op o_{1} \rho \op o_{2}\hc + \op o_{2} \rho \op o_{1}\hc- \frac12 \l( \op o_{1}\hc \op o_{2} \rho + \op o_{2}\hc \op o_{1} \rho + \rho \op o_{1}\hc \op o_{2} + \rho \op o_{2}\hc \op o_{1} \r)\nn\\
	&= \diss{\op o_{1} + \op o_{2}}\rho- \diss{\op o_{1}}\rho - \diss{\op o_{2}}\rho
	\end{align}
	Since the matrices $A$ are symmetric, we can thus ascribe dissipators to all off-diagonal elements. 
	All remaining terms are then already diagonal and correspond to dissipators of single operators of our basis vector $\op \sigma$.
	
	In practise, we then perform this diagonalization procedure on the matrix $A_{\text m}$ according to the replacement rule
		\begin{equation}
	\op o_{1} \rho \op o_{2}\hc + \op o_{2} \rho \op o_{1}\hc\rightarrow \diss{\op o_{1} + \op o_{2}}\rho- \diss{\op o_{1}}\rho - \diss{\op o_{2}}\rho
	\end{equation}
	 and verify that the resulting dissipator-like terms are sufficient to construct the matrices $A_{\text e}$ and $A_{\text s}$. 
	If additional terms are required to construct these matrices, those terms will appear as contributions to the Hamiltonian in the steady-state equation, $\sim -\ii \comm{\H}{\bar\rho}$, as opposed to dissipators.
	
\subsubsection{Dissipators proportional to $C(0)$}	
	As an illustration we now consider an example where the matrices $A$ are not diagonal to start with.
	In particular, collecting  all terms proportional to the phonon spectral function at zero frequency, $C(0)$, we find a contribution to the RHS of \eqn{eq:MEDiag} with the $A_m$ matrix taking the form
	\begin{align}
		A_{\text m}^{(0)} = \l( 
			\begin{array}{ccc|ccc|ccc}
				&&&&&&&& \\
				& A_{\text{m},1}^{(0)} & & & 0 & & & 0 &\\
				&&&&&&&& \\
				\hline
				&&&&&&&& \\
				& 0 &&&A_{\text m,2}^{(0)} &&& 0 & \\
				&&&&&&&& \\
				\hline
				&&&&&&&& \\
				& 0 &&& 0 &&& 0_{8\times8} & \\
				&&&&&&&& 
			\end{array} \r)_{16\times16} ,
		\label{eq:A0}
	\end{align}
	with the $4\times4$ sub-matrices
	\begin{align}
		A_{\text{m},1}^{(0)} = \l(  
			\begin{array}{cccc}
				c_{1} & c_{2} & c_{1} & 0 \\
				c_{2} & c_{3} & 0 & c_{3} \\
				c_{1} & 0 & 0 & 0 \\
				0 & c_{3} & 0 & 0
			\end{array} \r) \: C(0) , 
	\end{align}
	and 
	\begin{align}
		A_{\text m,2}^{(0)} = \text{diag}\l( c_{4}, -c_{4}, c_{4}, -c_{4} \r) \: C(0) ,
	\end{align}
	where the matrix coefficients are listed in table~\ref{tab:coeffs}.
	As $A_{\text m,2}^{(0)}$ is already diagonal, we only have to diagonalise $A_{\text{m},1}^{(0)}$ to determine all contributions $\propto C(0)$.
	Since $\diss{\mathds 1} \rho = 0$ and also $\diss{\mathds 1 + \op o}\rho = \diss{\op o} \rho$ for any hermitian operator $\op o$, 
	any off-diagonal elements involving the unit matrix $\mathds 1$ cancel in the diagonalisation procedure as $\diss{\mathds 1 + \op o}\bar\rho - \diss{\op o}\bar\rho = 0$.
	The only relevant contributions from $A_{\text{m},1}^{(0)}$ then come from the block containing the coefficient $c_{3}$ and we find the result
	\begin{align}
		\op \sigma A_{\text m}^{(0)} \bar\rho \op\sigma\hc &-\frac12 \left\{ \op \sigma\hc A_{\text{s}}^{(0)} \op \sigma \bar\rho + \bar\rho \op \sigma\hc A_{\text{e}}^{(0)} \op \sigma \right\} \nn\\
			&\hspace{-15mm}=c_{3}\: C(0) \l( \diss{\sigma_{z} + \sigma_{z}a\hc a}\bar\rho - \diss{\sigma_{z} a\hc a}\bar\rho \r) \nn\\
			&\hspace{-12mm}+c_{4}\: C(0) \l( \diss{a\hc}\bar\rho - \diss{\sigma_{z} a\hc}\bar\rho + \diss{a}\bar\rho - \diss{\sigma_{z} a}\bar\rho \r),
		\label{eq:C0Diss}
	\end{align}
	which contributes to the dissipative superoperator at fourth order, $\L_{4,\text{corr}} \bar\rho$.
	
As an aside, we note that some of the individual dissipators in \eqn{eq:C0Diss} contain four photon creation and annihilation operators, such as $\diss{\sigma_{z} a\hc a}\rho$, 
the total expression including all terms only contains a maximum of two photon operators in each term, as is consistent with our choice of self-energy diagrams.
	
	\begin{table}[t]
		\begin{center}
			\begin{tabular}{ccl}
				$c_{1}$ & $=$ & $ -g^{2}\cos^{2}\theta \frac{\cos^{2}\theta\: \omega_{q}^{2}(\omega_{q}^{2}-3\omega_{r}^{2}) + \omega_{r}^{2}( \omega_{q}^{2}+ \omega_{r}^{2} )}{8\omega_{r}^{2}(\omega_{q}^{2}-\omega_{r}^{2})^{2}},$ \\
				$c_{2}$ & $=$ & $ \phantom{-}g^{2} \cos^{2}\theta \frac{3 \sin^{2}\theta\: \omega_{q}^{3}}{8 \omega_{r} (\omega_{q}^{2}- \omega_{r}^{2})^{2}},$\\
				$c_{3}$ & $=$ & $ \phantom{-}g^{2}\cos^{2}\theta\frac{\cos^{2}\theta\: (\omega_{q}^{4}+ 2 \omega_{q}^{2} \omega_{r}^{2} - \omega_{r}^{4}) -2\omega_{r}^{2}(2\omega_{q}^{2} - \omega_{r}^{2})}{4\omega_{r}^{2}(\omega_{q}^{2}- \omega_{r}^{2})^{2}},$\\
				$c_{4}$ & $=$ & $ -g^{2} \cos^{2}\theta \frac{\cos^{2}\theta\:\omega_{q}^{2} - \omega_{r}^{2} }{4 (\omega_{q}^{2} - \omega_{r}^{2})},$\\
				$c_{5}$ & $=$ & $ -g^{2} \sin^{2}\theta \frac{\omega_{r}^{2} (\omega_{q}^{2}+ \omega_{r}^{2})+ \cos^{2}\theta\omega_{q}^{2} (\omega_{q}^{2} - 3\omega_{r}^{2}) }{16\omega_{r}^{2}(\omega_{q}^{2}-\omega_{r}^{2})^{2}} ,$\\
				$c_{6}$ & $=$ & $ -g^{2} \sin^{2}\theta \frac{ (\omega_{q} - \omega_{r} ) \omega_{r}^{2} + \cos^{2}\theta\: \omega_{q}^{2} ( \omega_{q} + 3\omega_{r} ) }{32 \omega_{r}^{2} (\omega_{q}^{2}- \omega_{r}^{2}) (\omega_{q}+\omega_{r}) },$\\
				$c_{7}$ & $=$ & $ \phantom{-}g^{2} \sin^{2}\theta \frac{\cos^{2}\theta (\omega_{q}^{2}-\omega_{r}^{2}) + \sin^{2}\theta\: \omega_{q}\omega_{r}}{8 (\omega_{q}^{2}-\omega_{r}^{2})^{2}},$\\
				$c_{8}$ & $=$ & $ \phantom{-}g^{2}\st \frac{\cos^{2}\theta}{8(\omega_{q}^{2}-\omega_{r}^{2})},$\\
				$c_{9}$ & $=$ & $ \phantom{-}g^{2}\sin^{2}\theta \frac{\cos^{2}\theta\: \omega_{q}^{2}-\omega_{r}^{2}}{8 \omega_{r}( \omega_{q}^{2}-\omega_{r}^{2} )( \omega_{q}-\omega_{r})},$\\
				$c_{10}$ & $=$ & $ \phantom{-}g^{2}\st \frac{\cos^{2}\theta\: \omega_{q}^{2}}{16\omega_{r}^{2} (\omega_{q}^{2}-\omega_{r}^{2})},$\\
				$c_{11}$ & $=$ & $ \phantom{-}g^{2}\sin^{2}\theta \frac{\cos^{2}\theta\: \omega_{q}^{2}(\omega_{q}-\omega_{r}) - \sin^{2}\theta\:\omega_{q}\omega_{r}^{2}}{8\omega_{r}^{2}(\omega_{q}^{2}-\omega_{r}^{2})(\omega_{q}-\omega_{r}) },$\\
				$c_{12}$ & $=$ & $ \phantom{-}g^{2}\sin^{2}\theta \frac{\omega_{r}^{2}-\cos^{2}\theta \omega_{q}^{2}}{8\omega_{r} (\omega_{q}^{2}-\omega_{r}^{2}) (\omega_{q}+\omega_{r})},$\\
				$c_{13}$ & $=$ & $ \phantom{-}g^{2}\sin^{2}\theta \frac{\ct\: \omega_{q}^{2}(\omega_{q}+\omega_{r}) -\st\: \omega_{q}\omega_{r}^{2} }{8\omega_{r}^{2}(\omega_{q}^{2}-\omega_{r}^{2})(\omega_{q}+\omega_{r})},$\\
				$c_{14}$ & $=$ & $ -g^{2}\st \frac{\ct}{8(\omega_{q}\omega_{r} - \omega_{r}^{2})},$\\
				$c_{15}$ & $=$ & $ -g^{2}\sin^{2}\theta \frac{\ct\: \omega_{q} + \ct\: (\omega_{q}^{2}-\omega_{r}^{2}) + 2\omega_{r}^{2}}{8\omega_{r}^{2}(\omega_{q}^{2}-\omega_{r}^{2})},$\\
				$c_{16}$ & $=$ & $ -g^{2}\sin^{2}\theta \frac{\ct\: (2\omega_{q}^{2} + \omega_{r}^{2}) + 2\st\:\omega_{r}^{2}}{8 \omega_{r}^{2} (\omega_{q}^{2}-\omega_{r}^{2}) } ,$\\
				$c_{17}$ & $=$ & $ \phantom{-}g^{2}\sin^{2}\theta \frac{\omega_{r}-\ct\: \omega_{q}}{8\omega_{r}(\omega_{q}-\omega_{r})^{2}},$\\
				$c_{18}$ & $=$ & $ \phantom{-}g^{2}\sin^{2}\theta \frac{\omega_{r}^{2} -\ct\: \omega_{q} (2\omega_{q}-\omega_{r})}{8\omega_{r}^{2}(\omega_{q}-\omega_{r})^{2}},$\\
				$c_{19}$ & $=$ & $ \phantom{-}g^{2}\sin^{2}\theta \frac{\omega_{r}+\ct\: \omega_{q}}{8\omega_{r}(\omega_{q}+\omega_{r})^{2}} ,$\\
				$c_{20}$ & $=$ & $ \phantom{-}g^{2}\sin^{2}\theta \frac{\omega_{r}^{2} -\ct\: \omega_{q}(2\omega_{q}+\omega_{r})}{8\omega_{r}^{2}(\omega_{q}+\omega_{r})^{2}},$\\
				$c_{21}$ & $=$ & $ \phantom{-}g^{2}\sin^{2}\theta \frac{\omega_{r}^{2}(\omega_{q}+ \omega_{r}) + \ct\: \omega_{q}^{2} (\omega_{q}-3\omega_{r})}{32 \omega_{r}^{2} (\omega_{q}^{2}-\omega_{r}^{2})(\omega_{q}-\omega_{r})},$\\
				$c_{22}$ & $=$ & $ \phantom{-}g^{2}\sin^{2}\theta \frac{\ct\: (\omega_{q}^{2}-\omega_{r}^{2}) -\st\: \omega_{q}\omega_{r}}{8(\omega_{q}^{2}-\omega_{r}^{2})^{2}},$\\
				$c_{23}$ & $=$ & $ -g^{2}\sin^{2}\theta \frac{\ct\: (2\omega_{q}^{2}-\omega_{q}\omega_{r}+\omega_{r}^{2}) + 2\st\: \omega_{r}^{2}}{8\omega_{r}^{2}(\omega_{q}^{2}-\omega_{r}^{2})},$\\
				$c_{24}$ & $=$ & $ \phantom{-}g^{2}\sin^{2}\theta \frac{\ct}{8\omega_{r}(\omega_{q}+\omega_{r})},$\\
				$c_{25}$ & $=$ & $ -g^{2}\sin^{2}\theta \frac{\omega_{r}^{2} -\ct\: \omega_{q}^{2}}{16\omega_{r}(\omega_{q}^{2}-\omega_{r}^{2})} ,$\\
				$c_{26}$ & $=$ & $ \phantom{-}g^{2}\sin^{2}\theta \frac{\omega_{r} - \ct\: \omega_{q}}{32\omega_{r}(\omega_{q}-\omega_{r})},$\\
				$c_{27}$ & $=$ & $ -g^{2}\sin^{2}\theta \frac{\st\: \omega_{q}}{16(\omega_{q}^{2}-\omega_{r}^{2})},$\\
				$c_{28}$ & $=$ & $ \phantom{-}g^{2}\sin^{2}\theta \frac{\omega_{r} + \ct\: \omega_{q}}{32 \omega_{r}(\omega_{q}+ \omega_{r})},$\\
			\end{tabular}
			\caption{Table of matrix coefficients obtained when evaluating the Keldysh self-energy at 4th order. 
				The coefficients $c_{1}$, $c_{2}$, $c_{6}$, $c_{21}$, $c_{25}$, $c_{26}$, and $c_{28}$ do not appear in any of the rates summarized in Table~\ref{tab:AllRates}, since 
				their contributions ultimately cancel out in the diagonalisation procedure.
			}
			\label{tab:coeffs}
		\end{center}
	\end{table}	
	
\subsubsection{General form for dissipators proportional to $C(\omega)$}	

	The preceding subsection illustrated the diagonalisation procedure for terms in the master equation that are proportional to $C(0)$.  
	There are other terms proportional to  $C(-\omega_q), C(\omega_q), C'(-\omega_q)$ and $C'(\omega_q)$.  Generically, each of these terms take the form 
	\begin{equation}
		\op \sigma A_{\text m}^{(\omega)} \bar\rho \op\sigma\hc -\frac12 \left\{ \op \sigma\hc A_{\text{s}}^{(\omega)} \op \sigma \bar\rho + \bar\rho \op \sigma\hc A_{\text{e}}^{(\omega)} \op \sigma \right\} \label{eqn:genA}
	\end{equation}
	with
		\begin{align}
			A_{\text m}^{(\omega)} = \l( 
				\begin{array}{ccc|ccc|ccc|ccc}
					&&&&&&&& &&& \\
					& A_{\text{m},1}^{(\omega)} & & & 0 & & & 0 & && 0 &\\
					&&&&&&&& &&&\\
					\hline
					&&&&&&&& &&&\\
					& 0 &&&A_{\text m,2}^{(\omega)} &&& 0 & && 0 &\\
					&&&&&&&& &&&\\
					\hline
					&&&&&&&& &&&\\
					& 0 &&& 0 &&& A_{\text{m},3}^{(\omega)} & && 0 &\\
					&&&&&&&& &&&\\
					\hline
					&&&&&&&& &&&\\
					& 0 &&& 0 &&& 0 & && A_{\text{m},4}^{(\omega)} &\\
					&&&&&&&& &&&
				\end{array} \r) ,
			\label{eq:Awq}
		\end{align}
	being block diagonal in the operator basis~\eqn{eq:opBasis},
	and we find $A_{\text s}^{(\omega)} = A_{\text e}^{(\omega)}$, consistent with the form required to ensure that~\eqn{eqn:genA} is of Lindblad form.
	Here the blocks  $A_{\text m,j}^{(\omega)}$ are $4\times4$ sub-matrices. 
		
	For completeness, in the following subsections we explicitly provide the 	forms for these block sub-matrices, and explicitly write out the results of the diagonalisation procedure in terms of dissipators, $\diss$.
	
\subsubsection{Dissipators proportional to $C(-\omega_q)$}	

	For the terms proportional to the spectral function at the negative qubit frequency, $C(-\omega_{q})$, we find
	\begin{align}
		A_{\text{m},1}^{(-\omega_{q})} &= \l(  
			\begin{array}{cccc}
				c_{5} & c_{6} & c_{5} & 0 \\
				c_{6} & c_{7} & c_{5} & c_{8} \\
				c_{5} & c_{5} & 0 & 0 \\
				0 & c_{8} & 0 & 0
			\end{array} \r) \: C(-\omega_{q}) , \nn\\
		A_{\text{m},2}^{(-\omega_{q})} &= \l(  
			\begin{array}{cccc}
				c_{9} & c_{10} & 0 & 0 \\
				c_{10} & c_{11} & 0 & 0 \\
				0 & 0 & c_{12} & c_{10} \\
				0 & 0 & c_{10} & c_{13}
			\end{array} \r) \: C(-\omega_{q}) , \nn\\
		A_{\text{m},3}^{(-\omega_{q})} &= \l(  
			\begin{array}{cccc}
				c_{14} & 0 & -c_{8} & 0 \\
				0 & c_{15} & 0 & c_{16} \\
				-c_{8} & 0 & 0 & 0 \\
				0 & c_{16} & 0 & 0
			\end{array} \r) \: C(-\omega_{q}) , \nn\\
		A_{\text{m},4}^{(-\omega_{q})} &=  \text{diag}\l( c_{17}, c_{18}, c_{19}, c_{20} \r) \: C(-\omega_{q}) ,
	\end{align}
	with the coefficients specified in Table~\ref{tab:coeffs}.
	Performing the above specified diagonalisation procedure leads to the expression
	\begin{align}
		\op \sigma A_{\text m}^{(-\omega_{q})} \bar\rho \op\sigma\hc &-\frac12 \left\{ \op \sigma\hc A_{\text{s}}^{(-\omega_{q})} \op \sigma \bar\rho + \bar\rho \op \sigma\hc A_{\text{e}}^{(-\omega_{q})} \op \sigma \right\} \nn\\
			&\hspace*{-15mm}= C(-\omega_{q}) \Bigl[ c_{17} \diss{\sigma_{-} a\hc} + c_{18}\diss{\sigma_{+} a}  \nn\\
			&\hspace*{-10mm}+ c_{19}\diss{\sigma_{-}a} + c_{20} \diss{\sigma_{+}a\hc} \nn\\
			&\hspace*{-10mm}+ (c_{11} - c_{10}) \diss{\sigma_{z} a\hc} + (c_{13} - c_{10})\diss{\sigma_{z}a} \nn\\
			&\hspace*{-10mm}+ (c_{12} - c_{10}) \diss{a} +  (c_{9} - c_{10} )\diss{a\hc} \nn\\
			&\hspace*{-10mm}+ (c_{15} - c_{16}) \diss{\sigma_{+}} + (c_{14} + c_{8}) \diss{\sigma_{-}}  \nn\\
			&\hspace*{-10mm}- c_{5} \l( \diss{a\hc a} - \diss{\sigma_{z}+a\hc a} \r) + (c_{22}+ c_{5}- c_{8})\diss{\sigma_{z}} \nn\\
			&\hspace*{-10mm}+ c_{10} \l( \diss{a+ \sigma_{z}a} + \diss{a\hc + \sigma_{z}a\hc} \r) \nn\\
			&\hspace*{-10mm}- c_{8} \l(\diss{\sigma_{z} a\hc a} - \diss{\sigma_{z} + \sigma_{z}a\hc a} \r) \nn\\
			&\hspace*{-10mm}+ c_{8} \l( \diss{\sigma_{-} a\hc a} - \diss{\sigma_{-} + \sigma_{-}a\hc a } \r)\nn\\
			&\hspace*{-10mm}- c_{16} \l(\diss{\sigma_{+} a\hc a} -\diss{\sigma_{+} + \sigma_{+}a\hc a } \r) \Bigr] \bar\rho ,
		\label{eq:DiagOmega}
	\end{align}

\subsubsection{Dissipators proportional to $C(\omega_q)$}
	
	The matrix representing terms proportional to the spectral function at the positive qubit frequency, $C(\omega_{q})$, is represented by the blocks
	\begin{align}
		A_{\text{m},1}^{(\omega_{q})} &= \l(  
			\begin{array}{cccc}
				c_{5} & c_{21} & c_{5} & 0 \\
				c_{21} & c_{22} & -c_{5} & c_{8} \\
				c_{5} & -c_{5} & 0 & 0 \\
				0 & c_{8} & 0 & 0
			\end{array} \r) \: C(\omega_{q}) , \nn\\
		A_{\text{m},2}^{(\omega_{q})} &= \l(  
			\begin{array}{cccc}
				c_{12} & -c_{10} & 0 & 0 \\
				-c_{10} & c_{13} & 0 & 0 \\
				0 & 0 & c_{9} & -c_{10} \\
				0 & 0 & -c_{10} & c_{11}
			\end{array} \r) \: C(\omega_{q}) , \nn\\
		A_{\text{m},3}^{(\omega_{q})} &= \l(  
			\begin{array}{cccc}
				c_{23} & 0 & c_{16} & 0 \\
				0 & c_{24} & 0 & -c_{8} \\
				c_{16} & 0 & 0 & 0 \\
				0 & -c_{8} & 0 & 0
			\end{array} \r) \: C(\omega_{q}) , \nn\\
		A_{\text{m},4}^{(\omega_{q})} &=  \text{diag}\l( c_{18}, c_{17}, c_{20}, c_{19} \r) \: C(\omega_{q}) ,
	\end{align}
	with diagonalisation resulting in a similar expression to Eq~\eqref{eq:DiagOmega},
	\begin{align}
		\op \sigma A_{\text m}^{(\omega_{q})} \bar\rho \op\sigma\hc &-\frac12 \left\{ \op \sigma\hc A_{\text{s}}^{(\omega_{q})} \op \sigma \bar\rho + \bar\rho \op \sigma\hc A_{\text{e}}^{(\omega_{q})} \op \sigma \right\} \nn\\
			&\hspace*{-15mm}= C(\omega_{q}) \Bigl[ c_{18} \diss{\sigma_{-} a\hc} + c_{17}\diss{\sigma_{+} a}  \nn\\
			&\hspace*{-10mm}+ c_{20}\diss{\sigma_{-}a} + c_{19} \diss{\sigma_{+}a\hc} \nn\\
			&\hspace*{-10mm}+ (c_{13}+ c_{10}) \diss{\sigma_{z} a\hc} + (c_{11}+ c_{10})\diss{\sigma_{z}a} \nn\\
			&\hspace*{-10mm}+ (c_{9}+ c_{10}) \diss{a} +  (c_{12}+ c_{10} )\diss{a\hc} \nn\\
			&\hspace*{-10mm}+ (c_{24}+ c_{8}) \diss{\sigma_{+}} + (c_{23}-c_{16}) \diss{\sigma_{-}}  \nn\\
			&\hspace*{-10mm}+ c_{5} \l( \diss{a\hc a} - \diss{\sigma_{z}+a\hc a} \r) + (c_{22}+ c_{5}- c_{8})\diss{\sigma_{z}} \nn\\
			&\hspace*{-10mm}- c_{10} \l( \diss{a+ \sigma_{z}a} + \diss{a\hc + \sigma_{z}a\hc} \r) \nn\\
			&\hspace*{-10mm}- c_{8} \l(\diss{\sigma_{z} a\hc a} - \diss{\sigma_{z} + \sigma_{z}a\hc a} \r) \nn\\
			&\hspace*{-10mm}- c_{16} \l( \diss{\sigma_{-} a\hc a} - \diss{\sigma_{-} + \sigma_{-}a\hc a } \r)\nn\\
			&\hspace*{-10mm}+ c_{8} \l(\diss{\sigma_{+} a\hc a} -\diss{\sigma_{+} + \sigma_{+}a\hc a } \r) \Bigr] \bar\rho ,
	\end{align}
	
\subsubsection{Dissipators proportional to $C'(\omega_q)$}
	
	The final remaining terms are proportional to derivatives of the spectral function. 
	Terms $\propto C'(\omega_{q})$ can be written as
	\begin{align}
		{A'}_{\text{m},1}^{(\omega_{q})} &= \l(  
			\begin{array}{cccc}
				c_{25} & c_{26} & 0 & -c_{27} \\
				c_{26} & c_{27} & 0 & c_{27} \\
				0 & 0 & 0 & 0 \\
				-c_{27} & c_{27} & 0 & 0
			\end{array} \r)\: C'(\omega_{q}) , \nn\\
		{A'}_{\text{m},3}^{(\omega_{q})} &= \l(  
			\begin{array}{cccc}
				-4c_{27} & 0 & -4c_{27} & 0 \\
				0 & 0 & 0 & 0 \\
				-4c_{27} & 0 & 0 & 0 \\
				0 & 0 & 0 & 0
			\end{array} \r)\: C'(\omega_{q}) , 
	\end{align}
	where the other two blocks are zero. Diagonalisation gives
	\begin{align}
		\op \sigma {A'}_{\text m}^{(\omega_{q})} \bar\rho \op\sigma\hc &-\frac12 \left\{ \op \sigma\hc {A'}_{\text{s}}^{(\omega_{q})} \op \sigma \bar\rho + \bar\rho \op \sigma\hc {A'}_{\text{e}}^{(\omega_{q})} \op \sigma \right\} \nn\\
			&\hspace*{-20mm}= c_{27}\:C'(\omega_{q}) \Biggl( \diss{\sigma_{z}+\sigma_{z}a\hc a} - \diss{\sigma_{z}a\hc a} \nn\\
				&+ 4\diss{\sigma_{-}a\hc a} - 4\diss{\sigma_{-}+ \sigma_{-}a\hc a} \Biggr)\bar\rho.
		\label{eq:DiagDerivPlus}
	\end{align}
	
\subsubsection{Dissipators proportional to $C'(-\omega_q)$}	
	The terms $\propto C'(-\omega_{q})$ lead to
	\begin{align}
		{A'}_{\text{m},1}^{(-\omega_{q})} &= \l(  
			\begin{array}{cccc}
				c_{25} & c_{28} & 0 & -c_{27} \\
				c_{28} & -c_{27} & 0 & -c_{27} \\
				0 & 0 & 0 & 0 \\
				-c_{27} & -c_{27} & 0 & 0
			\end{array} \r)\: C'(-\omega_{q}) , \nn\\
		{A'}_{\text{m},3}^{(-\omega_{q})} &= \l(  
			\begin{array}{cccc}
				0 & 0 & 0 & 0\\
				0 & 4c_{27} & 0 & 4c_{27} \\
				0 & 0 & 0 & 0 \\
				0 & 4c_{27} & 0 & 0
			\end{array} \r)\: C'(-\omega_{q}) , 
	\end{align}
	with the other two sub-matrices again zero. 
	Similar to \eqn{eq:DiagDerivPlus}, diagonalisation leads to the Lindblad dissipators
	\begin{align}
		\op \sigma {A'}_{\text m}^{(-\omega_{q})} \bar\rho \op\sigma\hc &-\frac12 \left\{ \op \sigma\hc {A'}_{\text{s}}^{(-\omega_{q})} \op \sigma \bar\rho + \bar\rho \op \sigma\hc {A'}_{\text{e}}^{(-\omega_{q})} \op \sigma \right\} \nn\\
			&\hspace*{-20mm}= c_{27}\:C'(-\omega_{q}) \Big( -\diss{\sigma_{z}+\sigma_{z}a\hc a} + \diss{\sigma_{z}a\hc a} \nn\\
				&- 4\diss{\sigma_{+}a\hc a} + 4\diss{\sigma_{+}+ \sigma_{+}a\hc a} \Big)\bar\rho.
		\label{eq:DiagDerivMinus}
	\end{align}

\section{Fourth order master equation of the Rabi Model}\label{app:fourth}
\renewcommand{\thefigure}{\ref{app:fourth}\arabic{figure}}
\renewcommand{\thetable}{\ref{app:fourth}\arabic{table}}
\setcounter{figure}{0}
\setcounter{table}{0}

\subsection{All fourth order correlated dissipators}\label{App:AllRates}

	\begin{table}[t]
		\begin{center}
		\begin{tabular}{lcl}
			$\Gamma_{\downarrow+}$ & $=$ & $\gamma_{\downarrow+}^{(\omega_q-\omega_r)} + c_{17} C(-\omega_{q}) + c_{18} C(\omega_{q})$ \\
			$\Gamma_{\uparrow-}$ &$=$ & $ \gamma_{\uparrow-}^{(-\omega_q-\omega_r)} + c_{18} C(-\omega_{q}) + c_{17} C(\omega_{q}) $\\
			$\Gamma_{\downarrow-}$ &$=$ & $ \gamma_{\downarrow-}^{(\omega_q+\omega_r)} + c_{19} C(-\omega_{q}) + c_{20} C(\omega_{q})$\\
			$\Gamma_{\uparrow+}$ &$=$ & $\gamma_{\uparrow+}^{(-\omega_q+\omega_r)} + c_{20} C(-\omega_{q}) + c_{19} C(\omega_{q}) $\\
			$\Gamma_{\varphi+}$ &$=$ & $\gamma_{\varphi+}^{(-\omega_{r})} \!\!-\! c_{4} C(0)\! +\! (c_{11}\! - \!c_{10}) C(-\omega_{q}) +(c_{13}+ c_{10} ) C(\omega_{q}) $\\
			$\Gamma_{\varphi-} $ &$=$ & $\gamma_{\varphi-}^{(\omega_{r})}\!\! - \!c_{4} C(0) + (c_{13}-c_{10}) C(-\omega_{q}) + ( c_{11}+ c_{10} ) C(\omega_{q}) $\\
			$\Gamma_{\updownarrow}$ &$=$ & $ (c_{23} - c_{16}) C_{-}(\omega_{q})$\\
			$\Gamma_{\varphi,4}$ &$=$ & $(c_{7} - c_{5}- c_{8}) C_{-}(\omega_{q})$\\
			$\Gamma_{-}$ &$=$ & $c_{4} C(0) + (c_{12}- c_{10}) C(-\omega_{q}) + (c_{9}+ c_{10}) C(\omega_{q})$\\
			$\Gamma_{+}$ &$=$ & $c_{4} C(0) + (c_{9} - c_{10}) C(-\omega_{q}) + (c_{12}+ C_{10}) C(\omega_{q})$\\
			$\Gamma_{n} $ &$=$ & $c_{5} C_{-}(\omega_{q})$\\
			$\Gamma_{\varphi n}$ &$=$ & $- c_{3} C(0) - c_{8} C_{+}(\omega_{q})  - c_{27} C'_{-}(\omega_{q}) $\\
			$\Gamma_{\downarrow n} $ &$=$ & $ c_{8} C(-\omega_{q}) -c_{16} C(\omega_{q}) +4c_{27} C'(\omega_{q}) $\\
			$\Gamma_{\uparrow n}$ &$=$ & $ -c_{16} C(-\omega_{q}) - c_{8} C(\omega_{q}) -4 c_{27} C'(-\omega_{q})  $\\
			$\Gamma_{\pm,\varphi\pm}$ &$=$ & $-c_{10} C_{-}(\omega_{q})$\\
		\end{tabular}
		\caption{Expressions for dissipative rates in the 4th-order contribution to the steady-state equation, \eqn{eq:4thAllapp}.  Terms $\gamma_{m}^{(\omega)}$ are defined in \eqn{eq:4thME}, while $c_j$'s are defined in Table \ref{tab:coeffs}.
		Here we have additionally introduced the symmetrized and anti-symmetrized bath spectral functions $C_{\pm}(\omega) = C(\omega) \pm C(-\omega)$
		and the unsymmetrized spectral function derivative $C_{-}'(\omega) = C'(\omega) - C'(-\omega)$
		}
		\label{tab:AllRates}
		\end{center}
	\end{table}	
	
	Combining dissipators in equations~\eq{eq:4thME}, \eq{eq:C0Diss}, \eq{eq:DiagOmega}, \eq{eq:DiagDerivPlus}, and~\eq{eq:DiagDerivMinus} gives a total of 21 dissipators and their corresponding rates.  
	These constitute the correlated decay processes at fourth order
	\begin{align}
		\mathcal L_{4,\text{corr}} \bar \rho \: & = \Gamma_{\downarrow +} \diss{\sigma_{-}a\hc}\bar\rho + \Gamma_{\downarrow -}\diss{\sigma_{-}a}\bar\rho \nn\\
			&+ \Gamma_{\uparrow +}\diss{\sigma_{+}a\hc}\bar\rho + \Gamma_{\uparrow-}\diss{\sigma_{+}a}\bar\rho \nn\\
			&+ \Gamma_{\varphi +} \diss{\sigma_{z} a\hc}\bar\rho + \Gamma_{\varphi - }\diss{\sigma_{z}a} \bar\rho \nn\\
			&+ \Gamma_{-} \diss{a}\bar\rho + \Gamma_{+} \diss{a\hc}\bar\rho \nn\\
			&+ \Gamma_{\updownarrow} \l( \diss{\sigma_{+}}\bar\rho +\diss{\sigma_{-}}\bar\rho \r) \nn\\
			&+ \Gamma_{n} \l( \diss{a\hc a}\bar\rho - \diss{\sigma_{z}+a\hc a}\bar \rho \r) + \Gamma_{\varphi,4}\diss{\sigma_{z}}\bar\rho \nn\\
			&+ \Gamma_{\pm,\varphi\pm} \l( \diss{a+ \sigma_{z}a}\bar\rho + \diss{a\hc + \sigma_{z}a\hc}\bar\rho \r) \nn\\
			&+ \Gamma_{\varphi n} \l(\diss{\sigma_{z} a\hc a}\bar\rho - \diss{\sigma_{z} + \sigma_{z}a\hc a}\bar \rho \r) \nn\\
			&+ \Gamma_{\downarrow n} \l( \diss{\sigma_{-} a\hc a}\bar\rho - \diss{\sigma_{-} + \sigma_{-}a\hc a }\bar\rho\r)\nn\\
			&+ \Gamma_{\uparrow n}\l(\diss{\sigma_{+} a\hc a}\bar\rho -\diss{\sigma_{+} + \sigma_{+}a\hc a }\bar\rho\r) ,
		\label{eq:4thAllapp}
	\end{align}
	with the complete expressions for all rates summarized in 
	Table~\ref{tab:AllRates}. There we have additionally introduced the symmetrized and anti-symmetrized bath spectral functions $C_{+}(\omega) = C(\omega) + C(-\omega)$ and $C_{-}(\omega) = C(\omega) - C(-\omega)$
	as well as the unsymmetrized spectral function derivative $C_{-}'(\omega) = C'(\omega) - C'(-\omega)$.

\subsection{Complete coupled steady-state equations\label{App:SteadyRates}}

	Here we give the full expression for all rates and factors appearing in the coupled steady-state equations for resonator and qubit. 
	
	We previously found the resonator steady-state equation as
	\begin{align}
		0 = -\ii \big[{\tilde \H_{r}},{\bar\rho_{r}}\big] + \kappa_{-} \diss{a}\bar\rho_{r} + \kappa_{+}\diss{a\hc} \bar\rho_{r}
	\end{align}
	with the renormalized photon loss and generation rates
	\begin{align}
		\kappa_{-} =& \kappa_{-,r} + P_{e} \gamma_{\downarrow -}^{(\omega_q+\omega_r)} + P_{g} \gamma_{\uparrow -}^{(-\omega_q+\omega_r)} + \gamma_{\varphi -}^{(\omega_{r})} \nn\\
			&+ \mean{\sigma_{z}} \kappa_{\varphi}^{(\omega_{q})} + P_{e} \kappa^{(\omega_{q})} - P_{g} \kappa^{(-\omega_{q})} ,\nn\\
		\kappa_{+} =& \kappa_{+,r} + P_{e} \gamma_{\downarrow +}^{(\omega_q-\omega_r)} + P_{g} \gamma_{\uparrow +}^{(-\omega_q-\omega_r)}  + \gamma_{\varphi +}^{(-\omega_{r})} \nn\\
			&+\mean{\sigma_{z}} \kappa_{\varphi}^{\omega_{q}} - P_{e} \kappa^{(\omega_{q})} + P_{g} \kappa^{(-\omega_{q})},
	\end{align}
	where we ordered the contributions according to their bath frequencies. The previously unspecified additional rates are
	\begin{align}
		\kappa_{\varphi}^{(\omega_{q})} =& \frac18 g^{2} \sin^{2}\theta\: \frac{\omega_{q}^{2}\cos^{2}\theta}{\omega_{r}^{2}(\omega_{q}^{2} - \omega_{r}^{2})} C_{+}(\omega_{q} ) ,\nn\\
		\kappa^{(\omega_{q})} =& \frac12 g^{2} \sin^{2}\theta\: \frac{\omega_{q} (\cos^{2}\theta\: \omega_{q}^{2} - \omega_{r}^{2}) }{\omega_{r} (\omega_{q}^{2} - \omega_{r}^{2})^{2}}\: C(\omega_{q}),\nn\\
		\kappa^{(-\omega_{q})} =& \frac12 g^{2} \sin^{2}\theta\: \frac{\omega_{q} (\cos^{2}\theta\: \omega_{q}^{2} - \omega_{r}^{2}) }{\omega_{r} (\omega_{q}^{2} - \omega_{r}^{2})^{2}}\: C(-\omega_{q}),
	\end{align}
	with all other expressions defined previously.
	
	The self-consistent equation for the resonator field amplitude $\alpha$ we write again as
	\begin{align}
		\ii \dot\alpha =   \epsilon_{d}/2 + \alpha (& \delta\omega_{r} + 2 \tilde\chi \sigma_{z} - i \kappa'/2 ) ,
	\end{align}
	with the renormalized resonator linewidth $\kappa'$
	\begin{align}
		\kappa' =& \kappa_{-} - \kappa_{+} ,
	\end{align}
	with all expressions already previously defined.

	The qubit steady-state is determined from 
	\begin{align}
		0 =& -\ii \big[{\tilde \H_{q}},{\bar\rho_{q}}\big] \nn\\
			&+ \gamma_{\downarrow} \diss{\sigma_{-}}\bar \rho_{q} + \gamma_{\uparrow} \diss{\sigma_{+}} \bar\rho_{q} + \gamma_{\varphi}\diss{\sigma_{z}}\bar\rho_{q} ,
	\end{align}
	with the correlated dissipative rates	
	\begin{align}
		\gamma_{\downarrow} =& \gamma_{\downarrow,2} + \abss\alpha \gamma_{\downarrow-}^{(\omega_q+\omega_r)} + \big(\abss\alpha +1 \big) \gamma_{\downarrow+}^{(\omega_q-\omega_r)} \nn\\
			&+ \gamma_{\downarrow}^{(\omega_{q})} + \gamma_{\downarrow}^{(-\omega_{q})} + \gamma_{\downarrow}' , \nn \\
		\gamma_{\uparrow} =& \gamma_{\uparrow,2} + \abss\alpha \gamma_{\uparrow-}^{(-\omega_q+\omega_r)}  + \big(\abss\alpha + 1\big) \gamma_{\uparrow+}^{(-\omega_q-\omega_r)} \nn\\
			&+ \gamma_{\uparrow}^{(\omega_{q})} + \gamma_{\uparrow}^{(-\omega_{q})} + \gamma_{\uparrow}' , \nn \\
		\gamma_{\varphi} =& \gamma_{\varphi,2} + \abss\alpha \gamma_{\varphi-}^{(\omega_{r})} + \big( \abss\alpha + 1 \big) \gamma_{\varphi +}^{(-\omega_{r})} \nn\\
			&+\gamma_{\varphi}^{(0)} + \gamma_{\varphi}^{(\omega_{q})} + \gamma_{\varphi}^{(-\omega_{q})} + \gamma_{\varphi}' .
	\end{align}
	where we again ordered all contributions according to their bath frequency and we already specified the expressions for the rates $\gamma_{\uparrow/\downarrow,0}$, $\gamma_{\downarrow/\uparrow,\pm}^{(\pm\omega_q\pm\omega_r)}$,
	$\gamma_{\varphi,0}$, and $\gamma_{\varphi,\pm}^{(\pm\omega_{r})}$ in \eqn{eq:CorrelatedRates} of the main paper. 
	
	The  qubit relaxation rates are
	\begin{align}
		\gamma_{\downarrow}^{(\omega_{q})} \!=& \big( \!\abss\alpha\! (c_{18} + c_{20} - 2 c_{16}) \!+\! (c_{18} + c_{23} - 2 c_{16}) \big) C(\omega_{q}) ,\nn\\
		\gamma_{\downarrow}^{(-\omega_{q})} \!=& \big(\! \abss\alpha\! (c_{17} \!+\! c_{19}\! +\! 2c_{8}) \!+\! (c_{17} \!-\! c_{23} \!+\! c_{16} \!+ \!c_{8})  \big)  C(-\omega_{q}) ,\nn\\
				\gamma_{\downarrow}' =& \big( 8 \abss\alpha c_{27} + 4 c_{27} \big) C'(\omega_{q}) .
	\end{align}
	Similarly, we also find the correlated rates for qubit excitations as
	\begin{align}
		\gamma_{\uparrow}^{(\omega_{q})} =& \big(\! \abss\alpha \!(c_{17} \!+\! c_{19}\! -\! 2c_{8}) \! +\! (c_{19} \!+\! c_{23} \!-\! c_{16}\! -\! c_{8}) \big) C(\omega_{q}) ,\nn\\
		\gamma_{\uparrow}^{(-\omega_{q})} =& \big( \!\abss\alpha \!(c_{18} + c_{20} - 2 c_{16})\! +\! (c_{20} - c_{23} )  \big) C(-\omega_{q}) ,\nn\\
		\gamma_{\uparrow}' =& - \big( 8 \abss\alpha c_{27} + 4 c_{27} \big) C'(-\omega_{q}) .
	\end{align}
	Finally, the additional contributions to qubit dephasing are given by
	\begin{align}
		\gamma_{\varphi}^{(0)} =& -\big( \abss\alpha (2 c_{3} + 2 c_{4}) + (c_{3}+ c_{4}) \big) C(0) , \nn \\
		\gamma_{\varphi}^{(\omega_{q}\!)} \!=& \big(\! \abss\alpha \!(c_{13}\!+\!c_{11}\!-\!2c_{8}\!) \!+\! (c_{13}\!+\!c_{10}\!-\!2c_{8}\!+\!c_{7}\!-\!c_{5}) \!\big) C(\omega_{q}\!) ,\nn \\
		\gamma_{\varphi}^{(\!-\!\omega_{q}\!)} =& \big(\! \abss\alpha\! (\!c_{11}\!+\! c_{13}\!-\!2c_{8}\!)\! +\! (\!c_{11}\!-\!c_{10}\!-\!c_{7}\!+\!c_{5}\!)\! \big)  C(-\omega_{q}) , \nn\\
		\gamma_{\varphi}' =& -\big( 2\abss\alpha c_{27} + c_{27} \big) C'_{-}(\omega_{q}) \,.
	\end{align}

\subsection{Effective qubit rates\label{sec:EffectiveRates}}

	\begin{figure*}[t]
		\begin{center}
			\begin{tabular}{c|cc|cc}
			 & $\gamma_{\downarrow,4}/g^{2}\:@\:\alpha=0$ & $\gamma_{\uparrow,4}/g^{2}\:@\:\alpha=0$ & $\gamma_{\downarrow,4}/g^{2}\:@\:\abss\alpha\gg1$ & $\gamma_{\uparrow,4}/g^{2}\:@\:\abss\alpha\gg1$\\
			\hline
			\begin{turn}{90}$T=0$ \end{turn}& 	
				\includegraphics[width=.45\columnwidth]{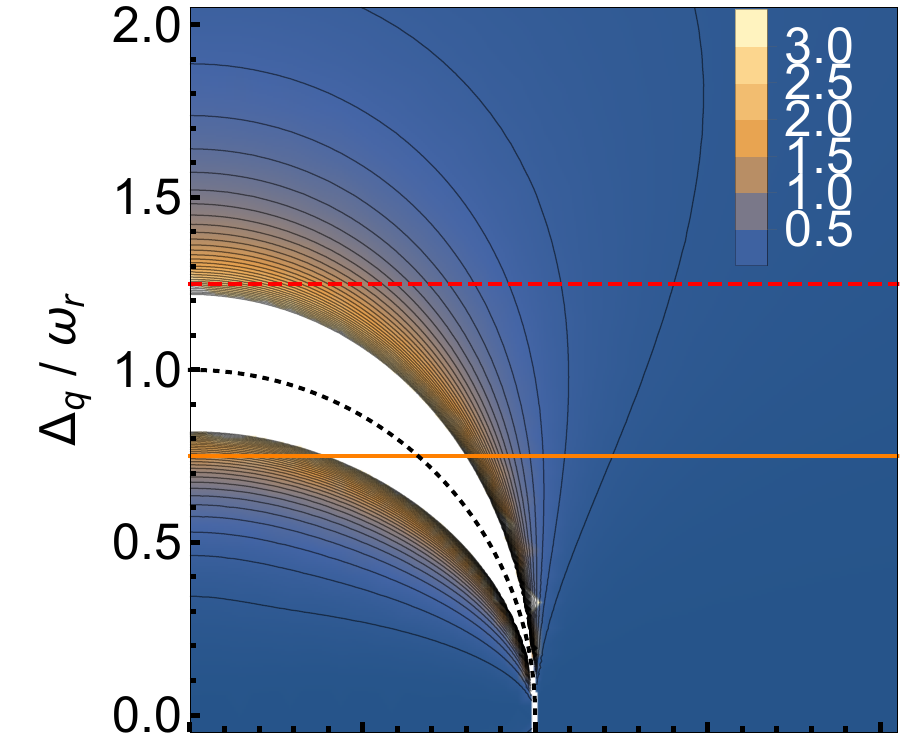}  &
				\includegraphics[width=.45\columnwidth]{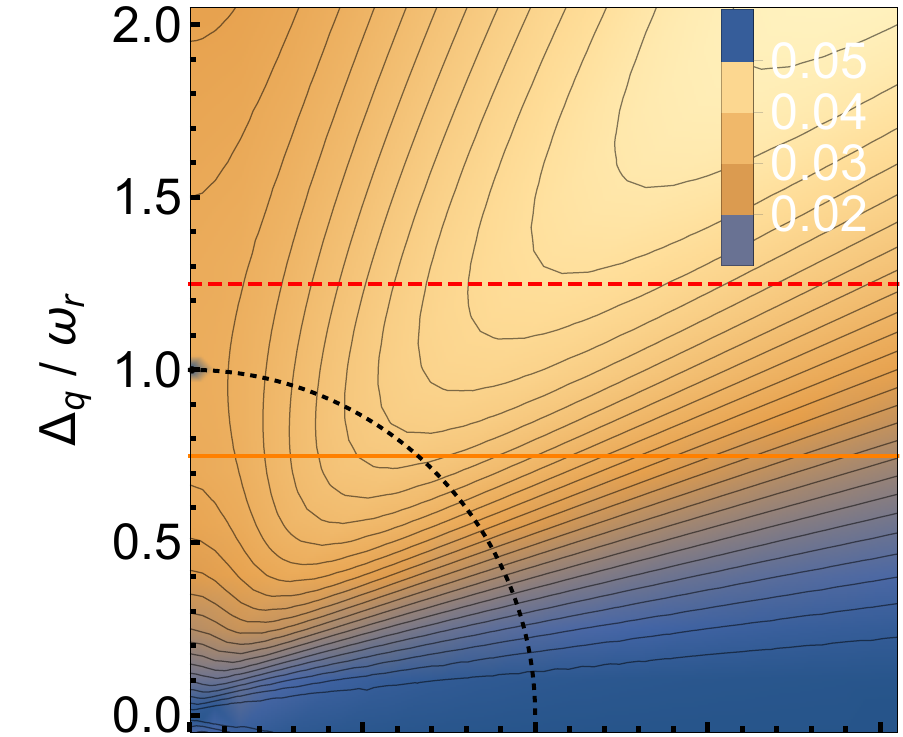} & 
				\includegraphics[width=.45\columnwidth]{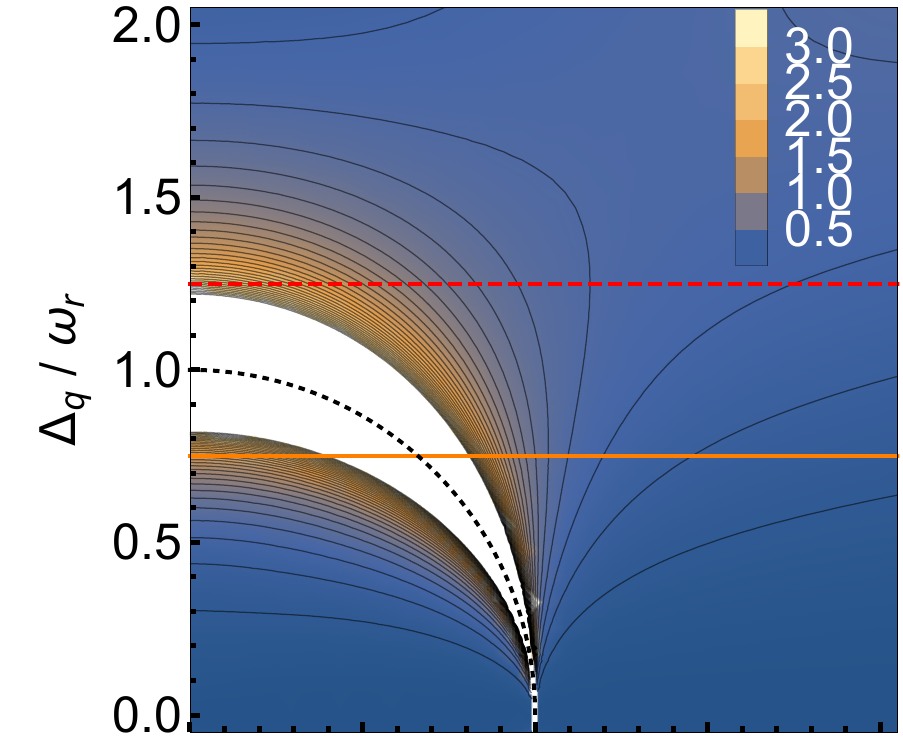} & 	
				\includegraphics[width=.45\columnwidth]{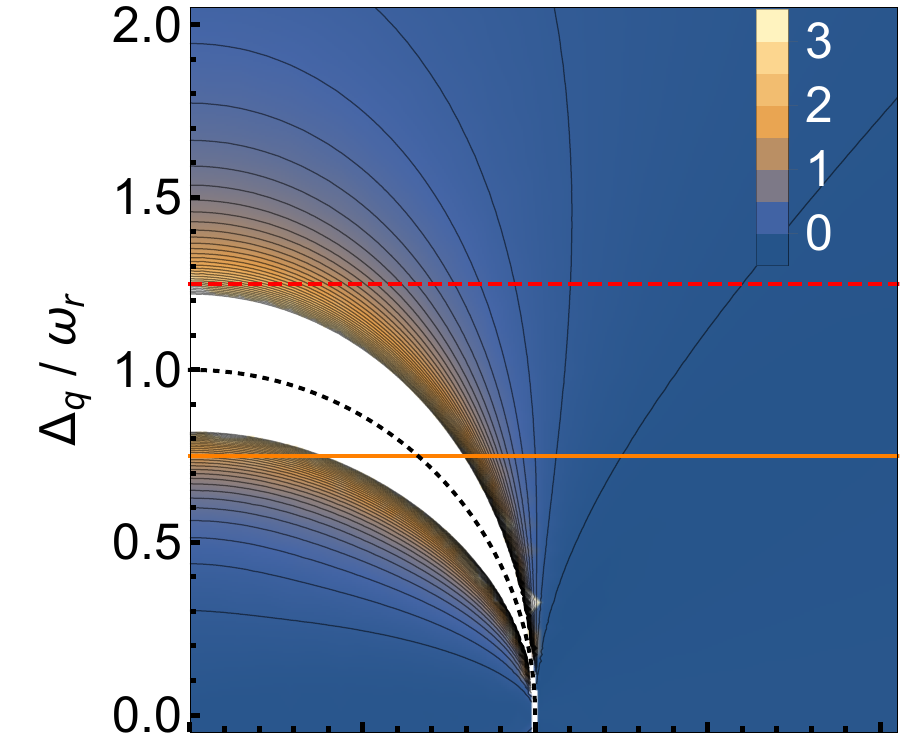}\\
				 & 
				\includegraphics[width=.45\columnwidth]{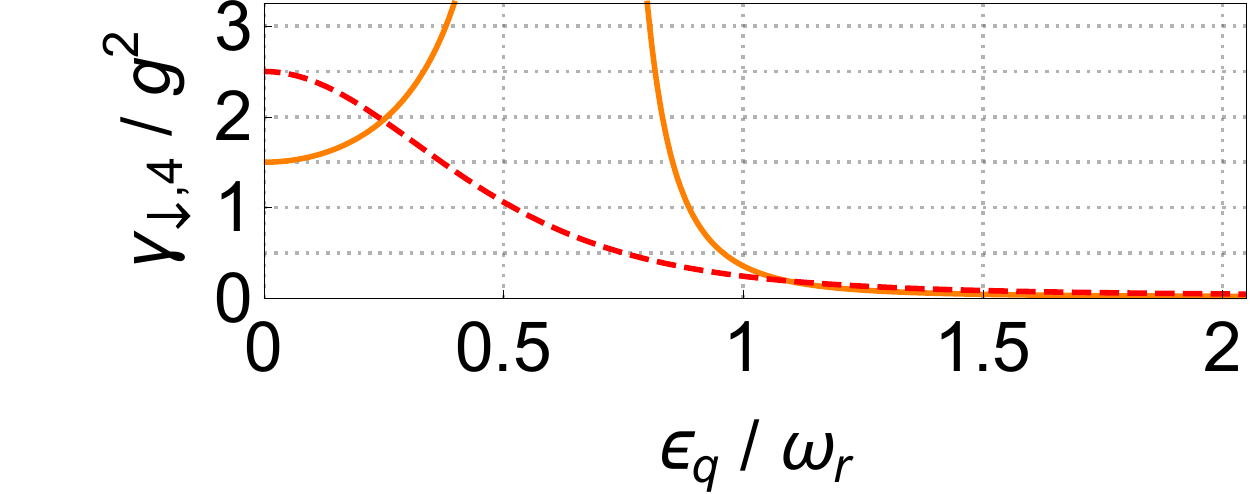} & 
				\includegraphics[width=.45\columnwidth]{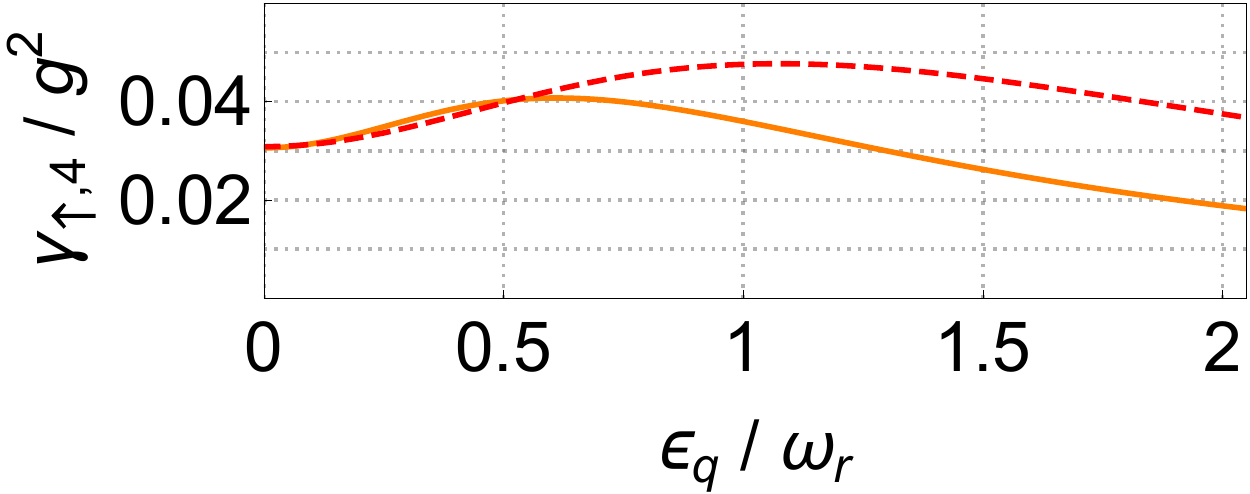} & 
				\includegraphics[width=.45\columnwidth]{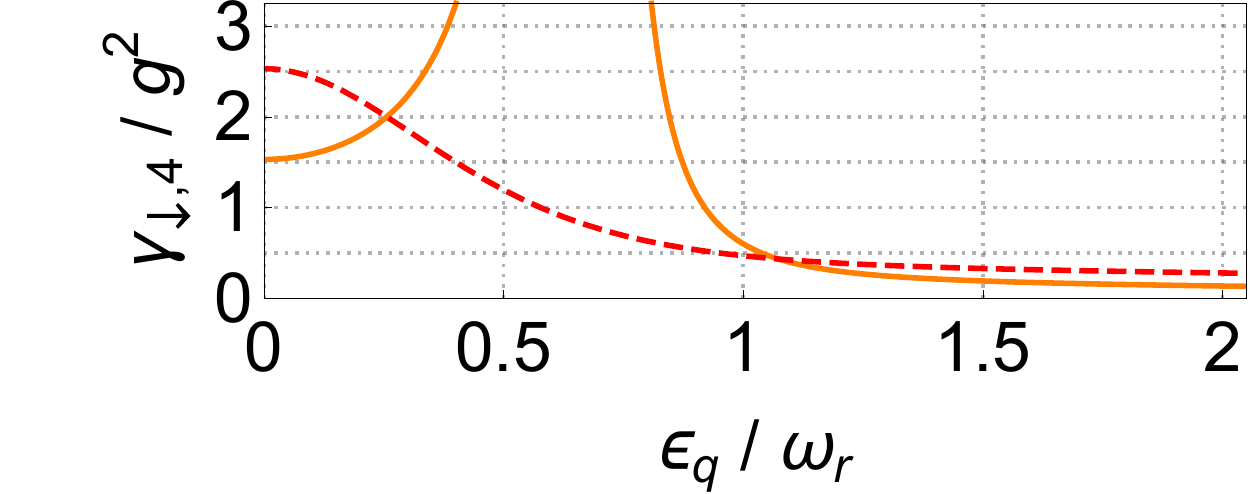} & 
				\includegraphics[width=.45\columnwidth]{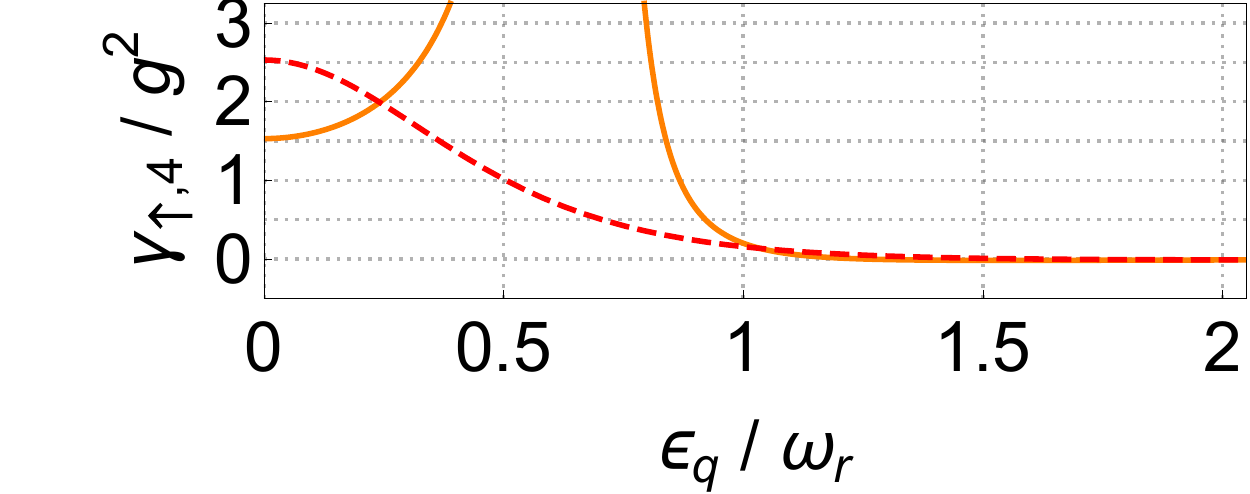} \\
				\hline
			\begin{turn}{90}$T=0.2 \omega_{r}$\end{turn}&
				\includegraphics[width=.45\columnwidth]{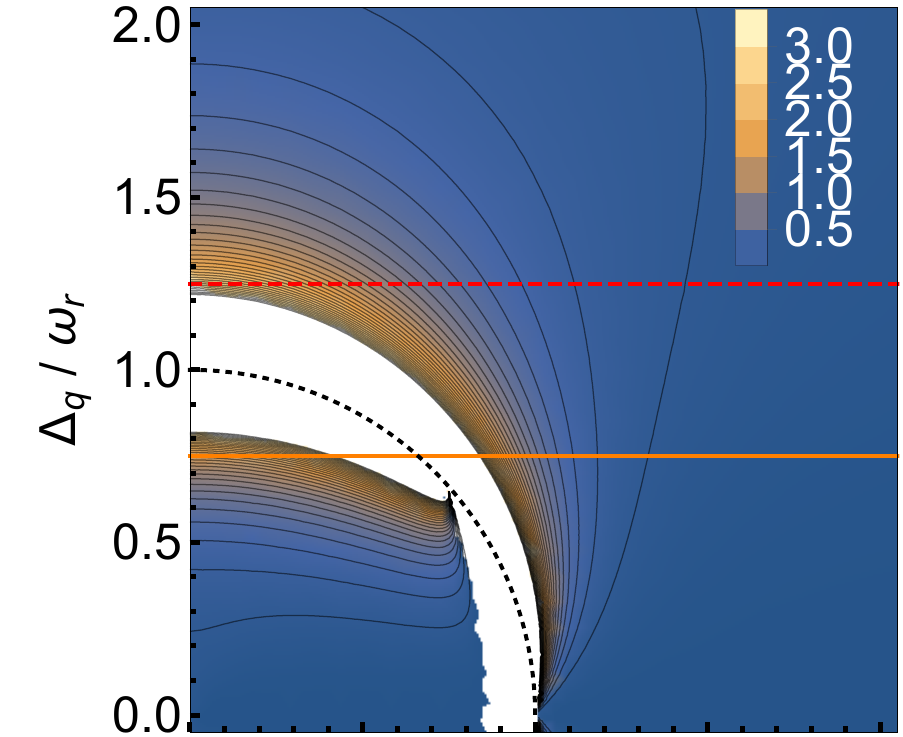} &
				\includegraphics[width=.45\columnwidth]{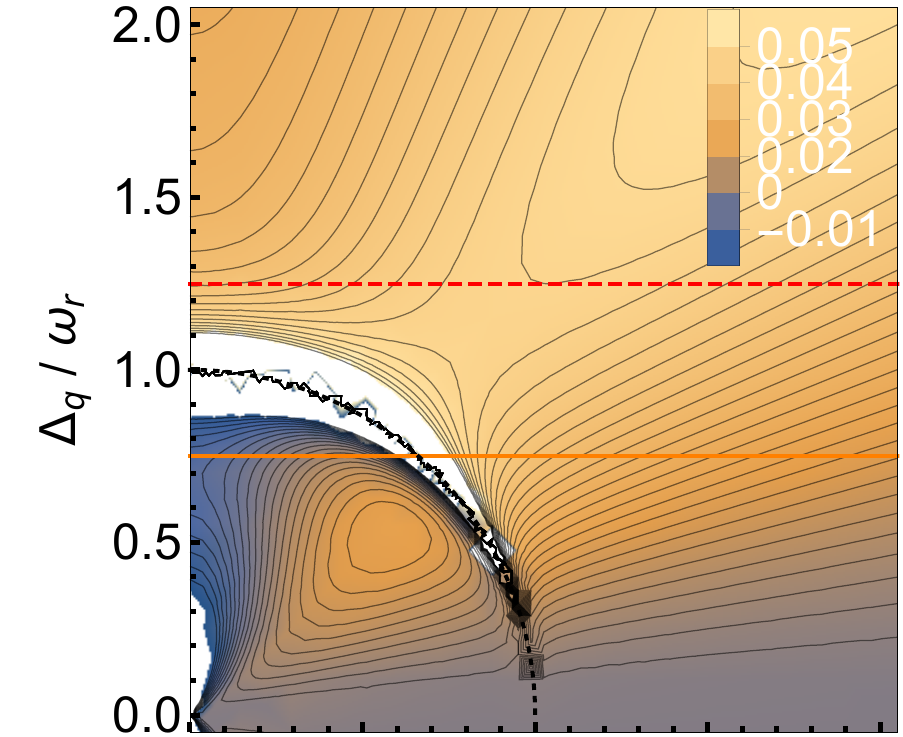} &
				\includegraphics[width=.45\columnwidth]{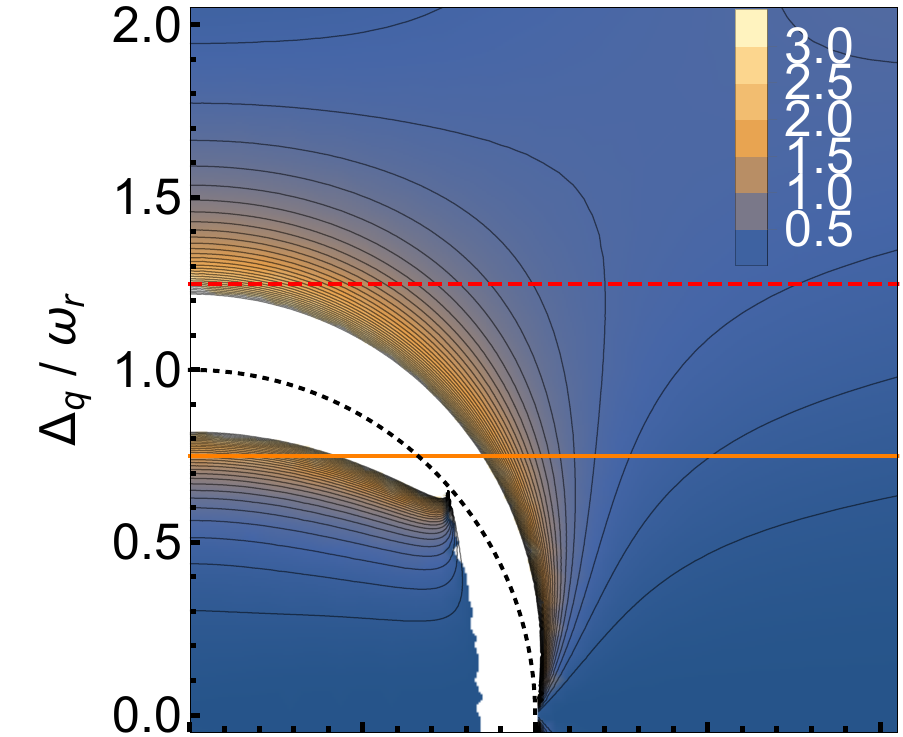} &
				\includegraphics[width=.45\columnwidth]{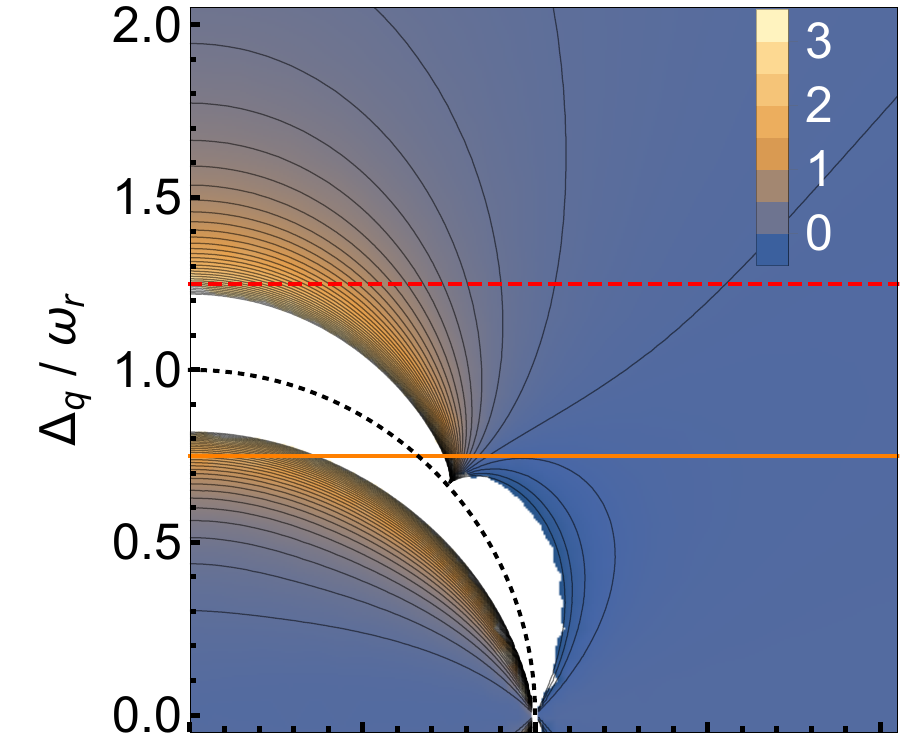}\\
				&
				\includegraphics[width=.45\columnwidth]{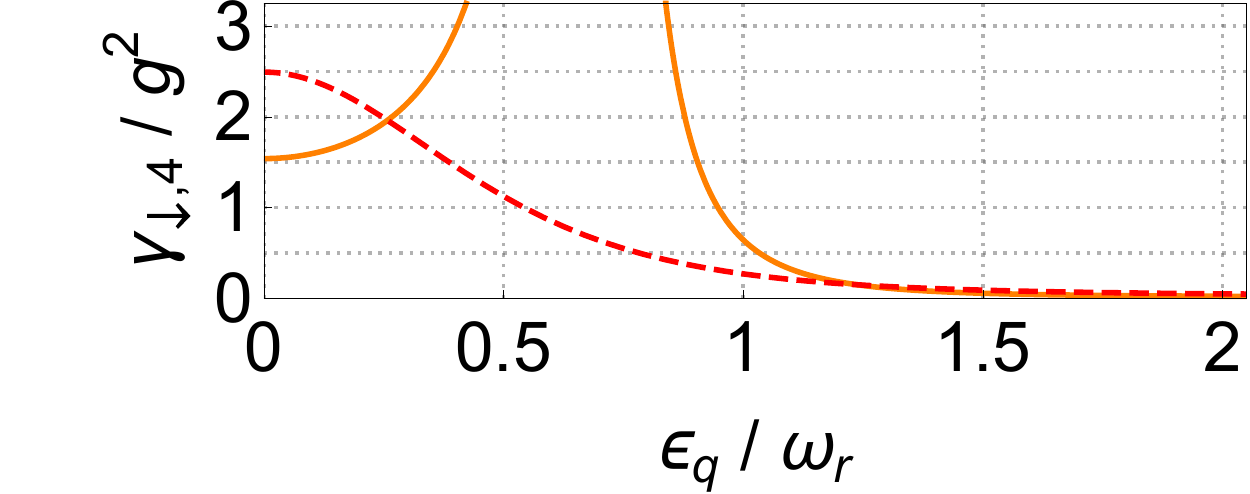} &
				\includegraphics[width=.45\columnwidth]{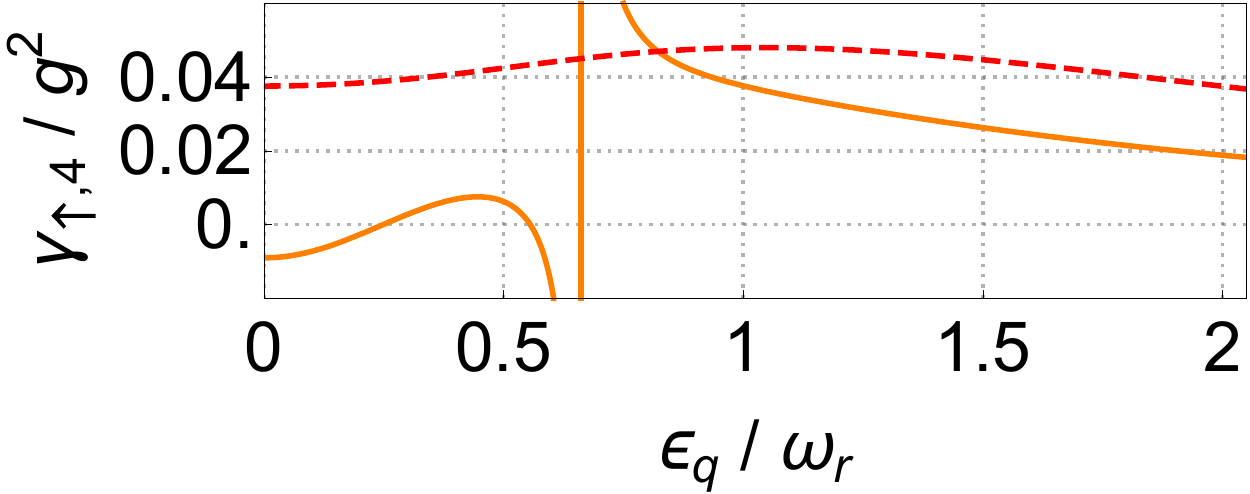} &
				\includegraphics[width=.45\columnwidth]{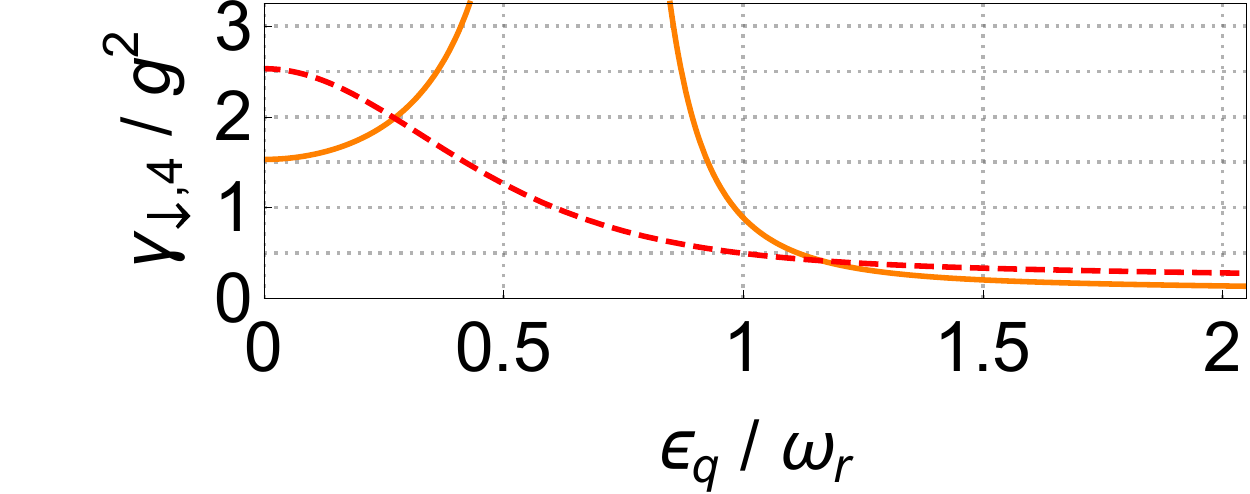}&
				\includegraphics[width=.45\columnwidth]{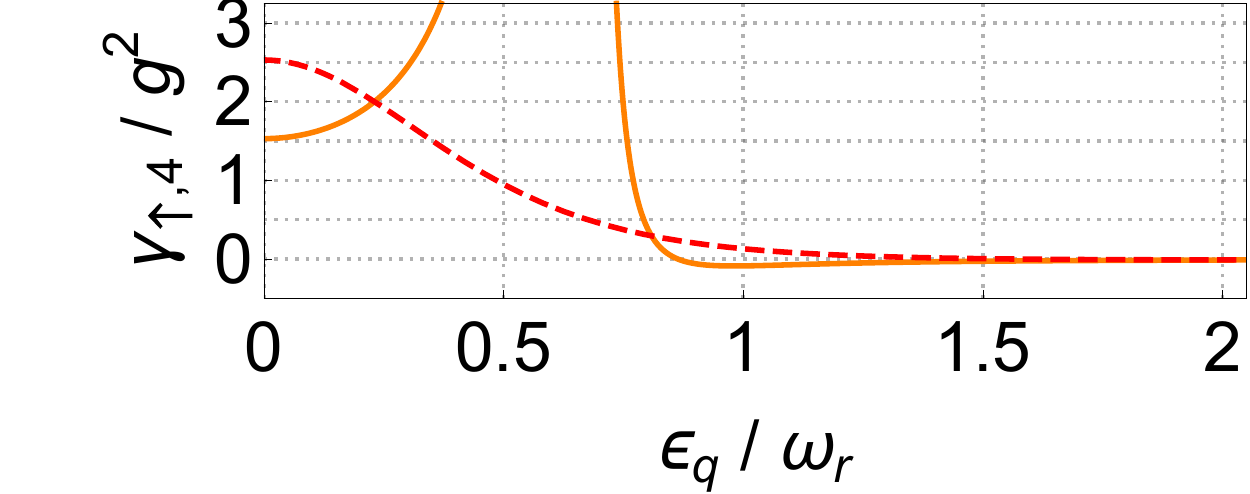}\\
			\end{tabular} 
			\caption{(Color online) Plots of the effective qubit relaxation and excitation rates $\gamma_{\uparrow,4}$ and $\gamma_{\downarrow,4}$ from~\eqn{eq:gamma} for an ohmic bath spectral function $C(\omega) = \omega (n_{\text{th}}(\omega)+\theta(\omega))$ 
				with the bosonic thermal occupation factor $n_{\text{th}} (\omega)=1/(\ee^{\beta\omega}-1)$, $\beta=1/k_{B}T$.
				We show the contributions to the rates for the resonator unpopulated, $\alpha= 0$ (left), and for a large resonator field amplitude, $\alpha \gg 1$ (right), for the two situations of zero temperature, $T=0$ (top), and non-zero temperature, $T=0.2 \omega_{r}$ (bottom).
				Horizontal lines in the contour plots indicate the values of $\Delta_{q}$ used in the line plots below.
				The thin dotted line in the contour plots indicates the resonance condition $\omega_{q} = \omega_{r}$.
			}
			\label{fig:gamma}
		\end{center}
	\end{figure*}
	
	\begin{figure}[t]
		\begin{center}
			\begin{tabular}{c|cc}
			& $\gamma_{\varphi}/g^{2}\:@\:\alpha=0$ & $\gamma_{\varphi}/g^{2}\:@\:\abss\alpha\gg1$\\
			\hline
			\begin{turn}{90}$T=0$\end{turn}& 	
				\includegraphics[width=.45\columnwidth]{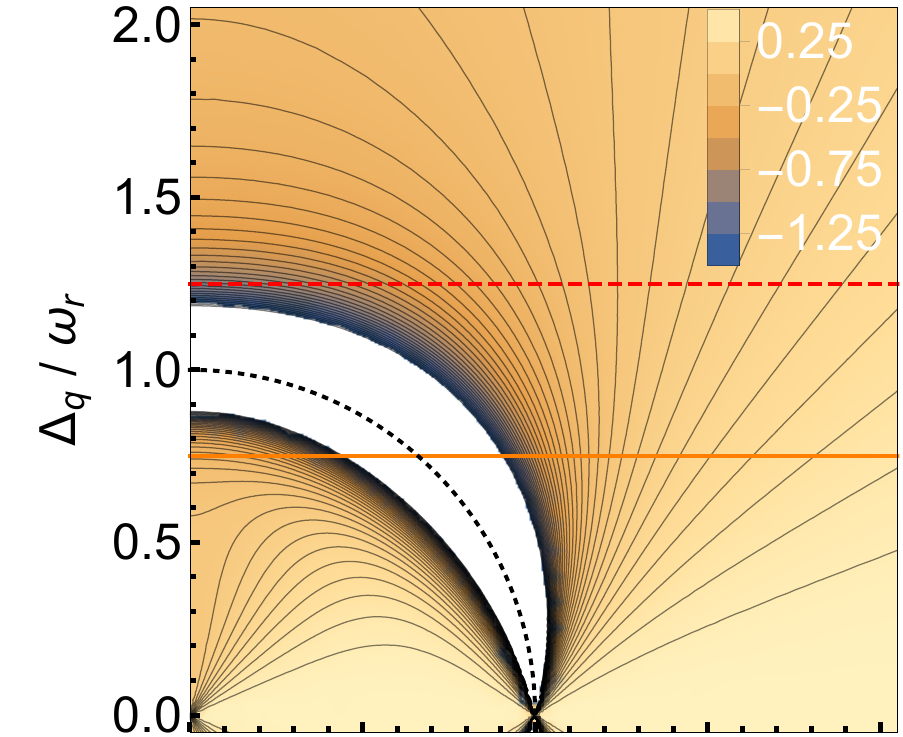}  &
				\includegraphics[width=.45\columnwidth]{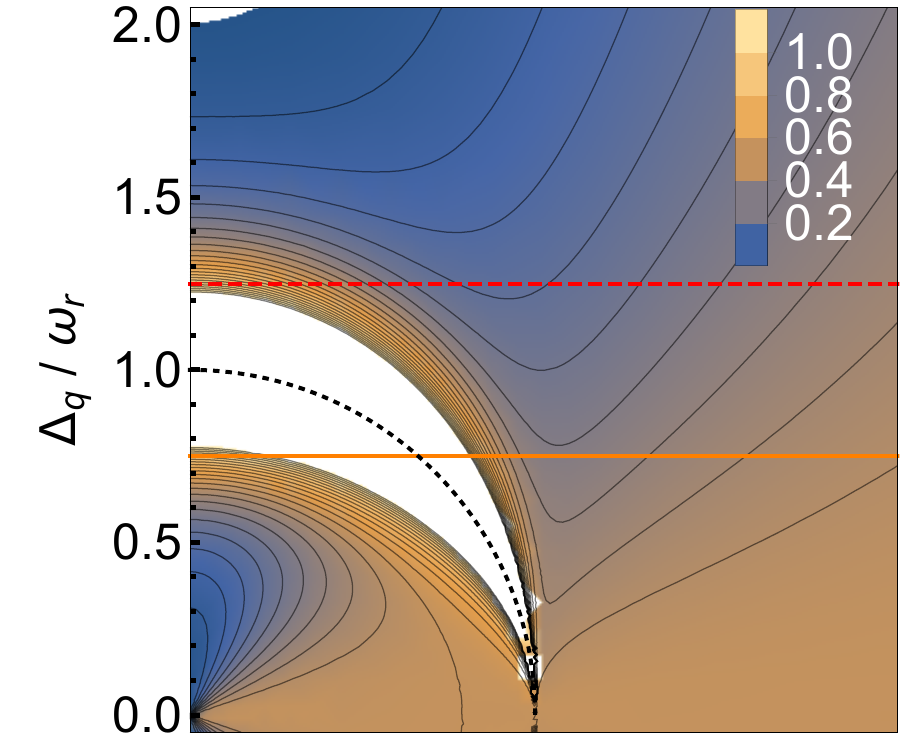} \\	 
			& 	\includegraphics[width=.45\columnwidth]{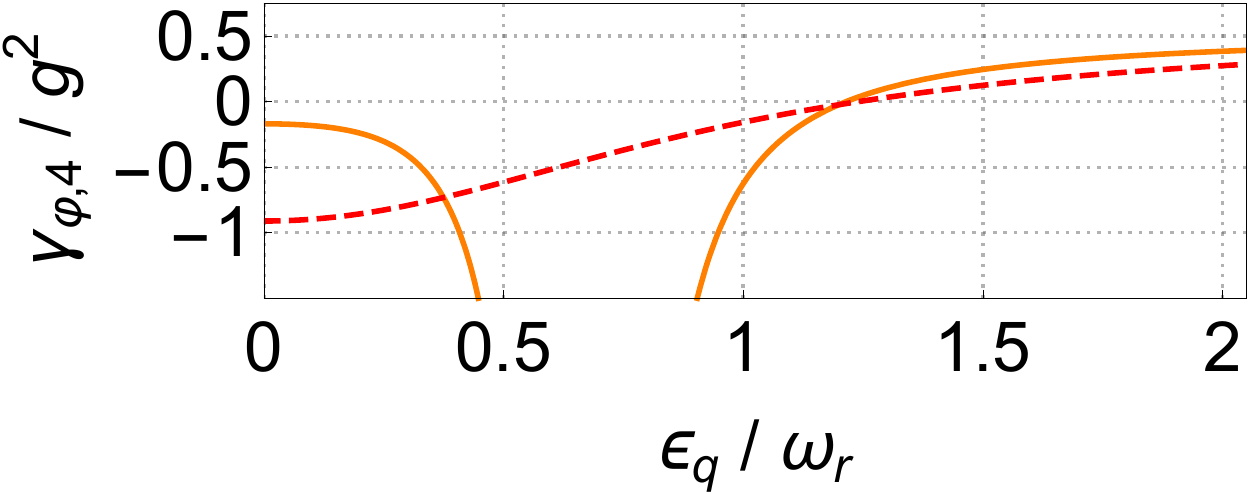} & 
				\includegraphics[width=.45\columnwidth]{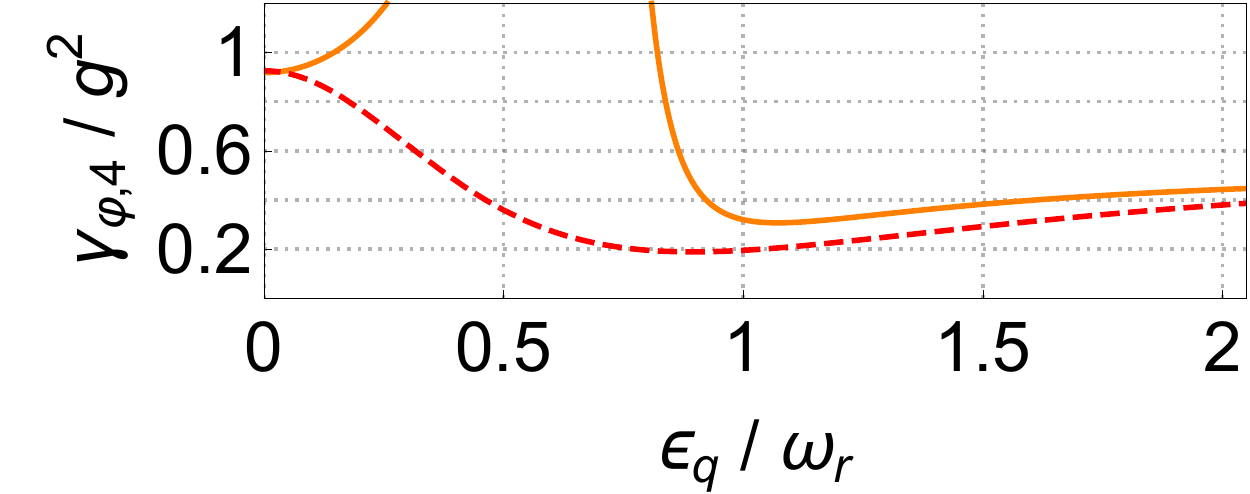}  \\
				\hline
			\begin{turn}{90}$T=0.2 \omega_{r}$\end{turn}& 	
				\includegraphics[width=.45\columnwidth]{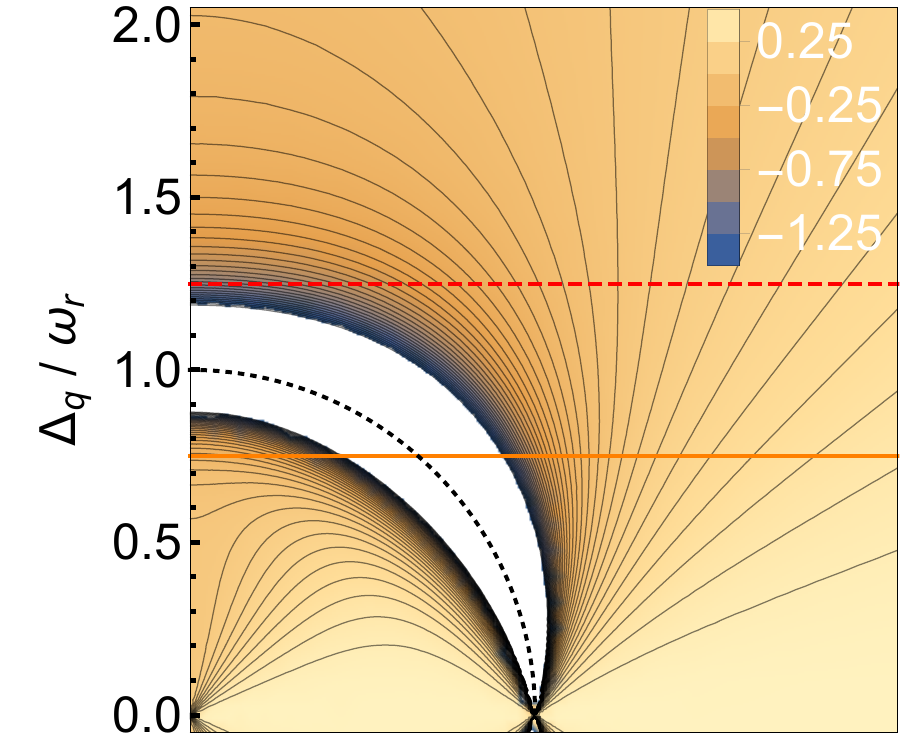}  &
				\includegraphics[width=.45\columnwidth]{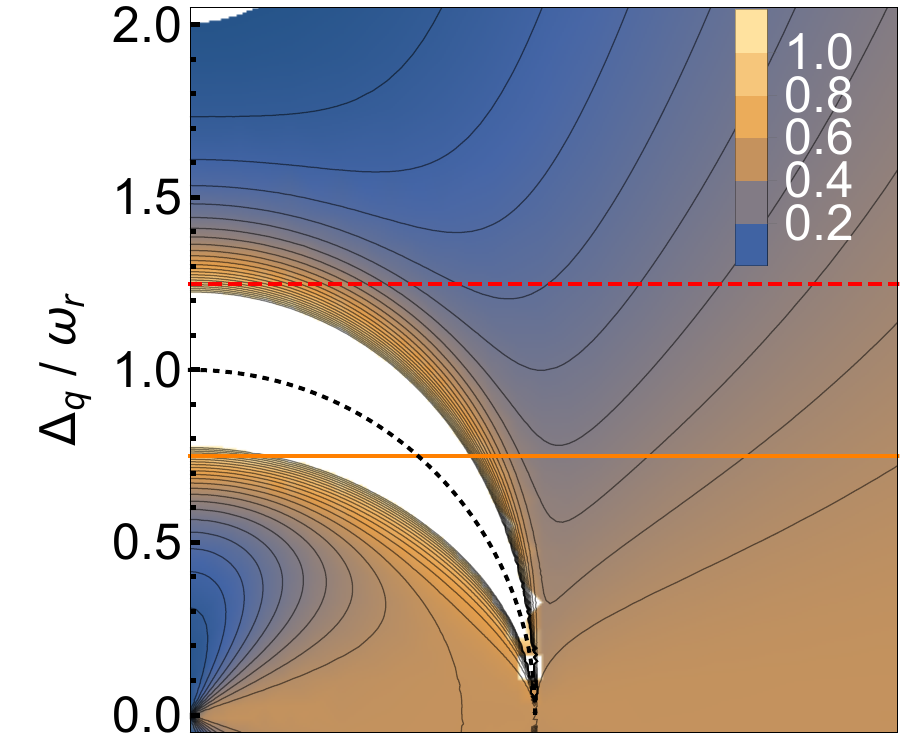} \\	 
			& 	\includegraphics[width=.45\columnwidth]{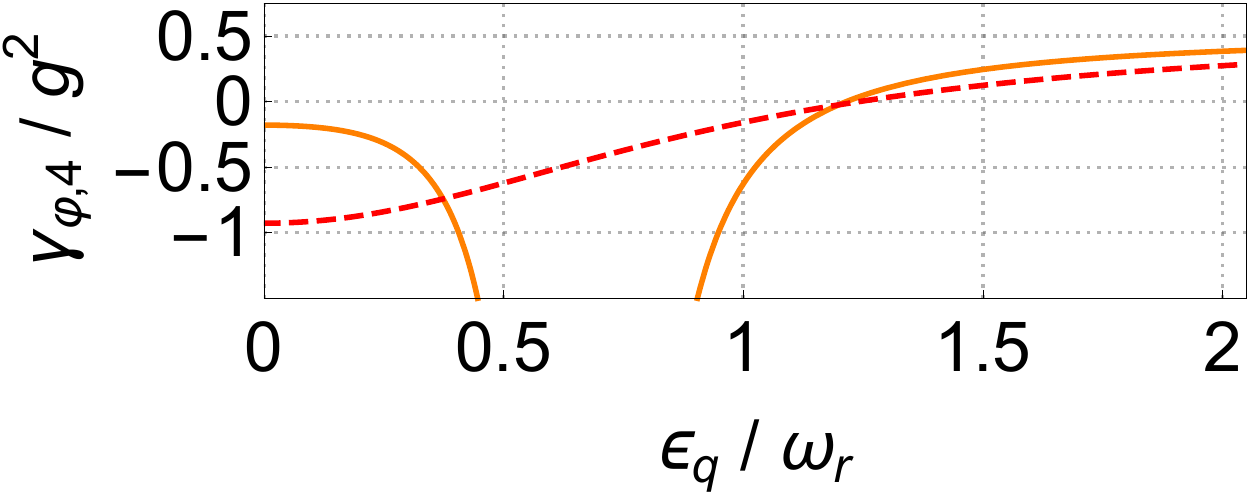} & 
				\includegraphics[width=.45\columnwidth]{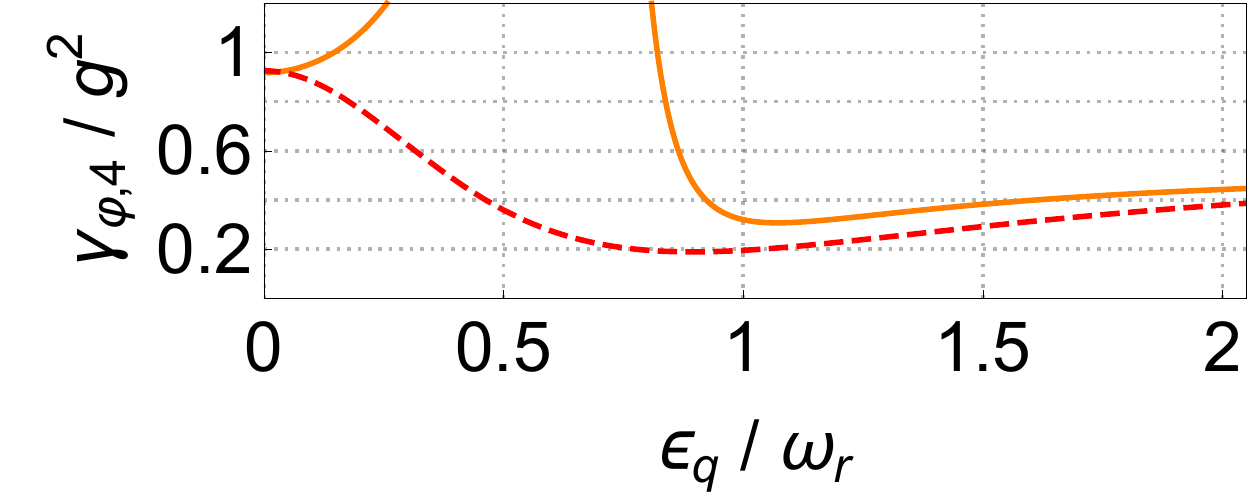}  \\
			\end{tabular} 
			\caption{(Color online) Plots of the effective qubit dephasing rate $\gamma_{\varphi,4}$ from~\eqn{eq:gamma} for an ohmic bath spectral function $C(\omega) = \omega (n_{\text{th}}(\omega)+\theta(\omega))$ and zero frequency noise of amplitude $C(0) = 1$.
				We show the contributions to the rates for the resonator unpopulated, $\alpha= 0$ (left), and for a large resonator field amplitude, $\alpha \gg 1$ (right), for the two situations of zero temperature, $T=0$ (top), and non-zero temperature, $T=0.2 \omega_{r}$ (bottom).
				Horizontal lines in the contour plots indicate the values of $\Delta_{q}$ used in the line plots below.
				The thin dotted line in the contour plots indicates the resonance condition $\omega_{q} = \omega_{r}$.
			}
			\label{fig:gamma0}
		\end{center}
	\end{figure}
	
	Figure~\ref{fig:gamma} shows  the correlated qubit relaxation and excitation rates $\gamma_{\downarrow / \uparrow,4}$. We show the rates in the two limits of zero resonator field, $\alpha=0$ and very large resonator amplitude, $\abss\alpha\gg1$, both for zero and non-zero temperature $T$.
	In Figure~\ref{fig:gamma0} we  show the effective qubit dephasing rate $\gamma_{\varphi,4}$ for the same parameters.
	Similar to the effective resonator rates $\kappa_{\pm,4}$ shown in the main text, the qubit rates are not generally well behaved close to resonance between qubit and resonator, as is expected from the perturbation theory. 
	Far from resonance however, all rates are positive and finite and therefore lead to CP dynamics in the qubit master equation.
	As before we can define an effective qubit temperature $T_{q}$ via the ratio of its relaxation and excitation rates, $\gamma_{\uparrow} / \gamma_{\downarrow} = \exp{\{ -\omega_{q} / k_{B} T_{q} \}}$, now depending on the bath temperature and the resonator field amplitude $\alpha$. 

\section{Phonon spectral functions from microscopic considerations\label{App:Phonons}}
\renewcommand{\thefigure}{\ref{App:Phonons}\arabic{figure}}
\renewcommand{\thetable}{\ref{App:Phonons}\arabic{table}}
\setcounter{figure}{0}
\setcounter{table}{0}
	
	Following the discussion in Ref.~\onlinecite{Stace:PRL:2005} we calculate expressions for the spectral function weights $\mathsf{F}$ and $\mathsf{P}$ in \eqn{eq:SpecFunc} assuming piezoelectric interactions between the phonons and the DQD, based on a simplified geometrical model of the semiconductor heterostructure.
	We find
	\begin{align}
		\mathsf{F_{piezo}} = \frac{P_1^2 \hbar d}{2\mu_1 c_n^2 \omega_r}  , \quad 
		\mathsf{P_{piezo}} = \frac{ P_1^2 \hbar}{2 \mu_3 c_p^3} ,
	\end{align}
	where $\mu_3$ is the three-dimensional mass density of the material, and $\mu_1 = \pi a^2 \mu_3 $ is the one-dimensional equivalent for a wire of radius $a$.
	Here we assumed that the interactions of substrate phonons with the electronic state in the DQD is mediated by the piezoelectricity of the wire material, while the phonon modes are defined by the bulk properties of the substrate. 
	With the DQD length $d=120$~nm, the wire radius $a = 25$~nm, the speed of sound in the wire $c_n = 4\times10^3$~m/s and in the substrate $c_p = 5\times10^3$~m/s~\cite{Fate:JAP:1975}, we find
	\begin{align}
		\mathsf{F_{piezo}} = 0.85  ,\quad		\mathsf{P_{piezo}} = 0.16.
	\end{align}
	
	Given that these estimates do not include any of the  complexities of the real experimental system (e.g.\ nearby surface gates, shear coupling between the 1D wire and the bulk substrate, finite length of wire etc.) 
	these values are in reasonable qualitative agreement with the values used in Fig.~\ref{fig:gain} in the main text, where $\mathsf{F}$ and $\mathsf{P}$ are treated as free parameters.
	
	The remaining microscopic parameter values used are summarized in Table~\ref{tab:Phonons}, where we have corrected for angular averaging in 3D piezoelectric constants compared to 1D compounds.
	\begin{table}[htb]
		\begin{center}
		\begin{tabular}{c|c|c}
				& InAs (1D Wire) & SiN (3D Substrate) \\
			\hline
			mass density $\mu_3$ & $5.7\times10^3$~kg/m$^3$ & $3.2\times10^3$~kg/m$^3$ \\
			piezoelectric constant $P\hbar$ & $0.725$~eV/nm & n.a. \\
		\end{tabular}
		\caption{Microscopic parameters for the calculation of the phonon spectral functions, partially extracted from Ref.~\onlinecite{Adachi:2005}.}
		\label{tab:Phonons}
		\end{center}
	\end{table}		

\subsection{Effect of higher temperature\label{App:HighTemp}}

\begin{figure}[ht]
		\begin{center}
			\includegraphics[width=\columnwidth]{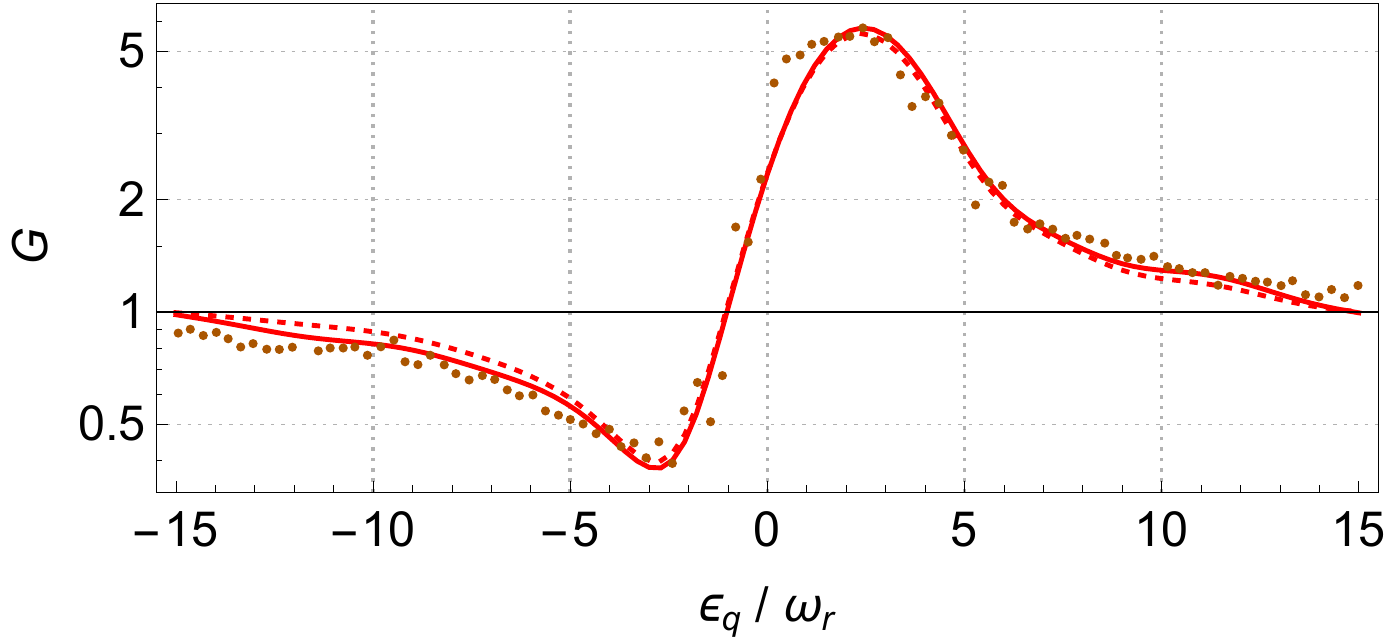}  
			\caption{(Color online) Effect of raising the phonon bath temperature from $3$~K to $9$~K. 
				We plot the power gain in the microwave resonator $G$ as a function of DQD asymmetry $\epsilon_q$. 
				The solid red line is for the elevated temperature, $k_BT/\omega_r=23.4$ (corresponding to $T=9$~K), the dotted line is for comparison at the lower temperature shown in Fig.~\ref{fig:gain} (i.e.\ $k_BT/\omega_r=7.8$, corresponding to $T=3$~K, from \cite{Gullans:PRL:2015}). All other parameters are the same for both curves.
				Shown are calculations including all  additional terms in the full master equation, Eq.~\eqref{eq:4thAll}.
				Parameters used in the calculations: $ \omega_{d}/\omega_{r}= 1$, $g/\omega_{r}=0.0125$, $\Delta_{q}/\omega_{r} = 3$, $\kappa/\omega_{r}=52 \times10^{-6}$, $\Gamma/\omega_{r}=0.34$, $\mathsf F=2.9$, $\mathsf P = 0.25$, and $\mathsf w = 1.7$, as in Fig.~\ref{fig:gain}.
			}
			\label{fig:GainHigh}
		\end{center}
	\end{figure}

	To illustrate the effect of a raising the bath temperature, we compare in Fig.~\ref{fig:GainHigh} the microwave power gain $G$ at a bath temperature of $T=3$~K (dashed red) and at $T=9$~K (solid red). 
	These calculations take into account all additional dissipators in Eq.~\eqref{eq:4thAll}. 
	We see that the main effect is a broadening of the gain profile, leading to a better fit of the theory to experimental data (orange dots) in the region $\epsilon_q / \omega_r < -5 $.
 
\end{document}